\documentclass[twocolumn,showpacs,floatfix,superscriptaddress, showkeys, prc]{revtex4-2}%

\usepackage{mathrsfs}
\usepackage{graphicx}
\usepackage{dcolumn}
\usepackage{bm}
\usepackage[colorlinks,linkcolor=blue,citecolor=blue]{hyperref}
\usepackage{amsmath, amssymb}
\usepackage{CJK}
\usepackage{multirow}
\usepackage[section]{placeins}

\newcommand{\ivec}[1]{\vec{#1}}
\newcommand{\svec}[1]{\boldsymbol{#1}}
\newcommand{\scr}[1]{{\mathscr #1}}

\newcommand{\cals}[1]{{\mathcal #1}}
\newcommand{\ff}[1]{\frac{1}{#1}}

\newcommand{\lrlc}[1]{\left|#1\right>}
\newcommand{\lrcl}[1]{\left<#1\right|}
\newcommand{\lrs}[1]{\left[#1\right]}

\newcommand{\Lrb}[1]{\left\{#1\right\}}

\usepackage{ulem}

\newcommand{\delete}{\bgroup\markoverwith{\textcolor{red}{\rule[0.5ex]{2pt}{1pt}}}\ULon}

\begin{document}
\title{Relativistic Hartree-Fock-Bogoliubov model for axially deformed nuclei}

\author{Jing Geng}
\affiliation{School of Nuclear Science and Technology, Lanzhou University, Lanzhou 730000, China}
\author{Wen Hui Long}\email{longwh@lzu.edu.cn}
\affiliation{School of Nuclear Science and Technology, Lanzhou University, Lanzhou 730000, China}
\affiliation{Joint Department for Nuclear Physics, Lanzhou University and Institute of Modern Physics, CAS, Lanzhou 730000, China}
\affiliation{Frontier Science Center for Rare isotope, Lanzhou University, Lanzhou 730000, China}

\begin{abstract}

\noindent\textbf{Background:} At the forefront of nuclear science, unstable nuclei are of special significance, in which the deformation and pairing correlations play an essential role in determining the nuclear structure. Under the mean field approach, the pairing and deformation effects have been uniformly considered within the deformed relativistic Hartree-Bogoliubov (D-RHB) model. However, due to the limitation of the Hartree approach, the important ingredient of nuclear force, such as the tensor force that contributes via the Fock diagram, is missing in the D-RHB model. Recently, the axially deformed relativistic Hartree-Fock (D-RHF) model was established for deformed nuclei, in which the $\pi$-pseudovector ($\pi$-PV) can get the tensor force effects into account naturally. While limited by the BCS pairing, it cannot be safely applied to describe unstable nuclei.

\noindent\textbf{Purpose:} The aim of this work is to develop the axially deformed relativistic Hartree-Fock-Bogoliubov (D-RHFB) model for the reliable description of wide-range unstable nuclei, by utilizing the spherical Dirac Woods-Saxon (DWS) base.

\noindent\textbf{Method:} Staring from the Lagrangian density that foots on the meson-propagated picture of nuclear force, the full Hamiltonian, that contains both mean field and pairing contributions, is derived by quantizing the Dirac spinor field in the Bogoliubov quasi-particle space, and the expectation with respect to the Bogoliubov ground state gives the full energy functional. As an extension of the D-RHF model, the degree of freedom associated with the $\rho$-tensor ($\rho$-T) coupling is implemented, and incorporating with the Bogoliubov scheme the finite-range Gogny force D1S is utilized as the pairing force. Moreover, qualitative analysis on the nature of the $\pi$-PV and $\rho$-T couplings are presented for better understanding their enhancements on the deformation effects.

\noindent\textbf{Results:} Space convergence related to the spherical DWS base is confirmed for the D-RHFB model by taking light nucleus $^{24}$Mg and mid-heavy one $^{156}$Sm as candidates. Compared to light nuclei, extraordinary more negative energy states are necessitated to keep the expansion completeness on the spherical DWS base for mid-heavy and heavy nuclei, due to the enhanced correlations between the expansion components with large $\kappa$-quantity as indicated by the nature of the $\pi$-PV and $\rho$-T couplings. Furthermore, because of the enhanced deformation effects by the $\pi$-PV and $\rho$-T couplings, the RHF Lagrangian PKA1 presents deeper bound ground state for $^{24}$Mg than the other selected Lagrangians, in addition to predicting a fairly deep bound local minimum with large oblate deformation.

\noindent\textbf{Conclusions:} Axially deformed relativistic Hartree-Fock-Bogoliubov model has been established with confirmed convergence checks. The effects of the $\pi$-PV and $\rho$-T couplings, coupled with nuclear deformation, are analyzed in both light and mid-heavy nuclear systems, which are expected to be manifested in wide-range unstable nuclei.

\end{abstract}

\maketitle


\section{Introductions}

The worldwide developments of the new generation radioactive-ion-beam (RIB) facilities, including the Cooler Storage Ring (CSR) at the Heavy Ion Research Facility in Lanzhou (HIRFL) in China \cite{Zhan2010NPA834.694c}, the RIKEN Radioactive Ion Beam Factory (RIBF) in Japan \cite{Motobayashi2010NPA834.707c}, the Rare Isotope Science Project (RISP) in Korea \cite{Tshoo2013BIMA317.242}, the Facility for Antiproton and Ion Research (FAIR) in Germany \cite{Sturm2010NPA834.682c}, the Second Generation System On-Line Production of Radioactive Ions (SPIRAL2) at GANIL in France \cite{Gales2010NPA834.717c}, the Facility for Rare Isotope Beams (FRIB) in the USA \cite{Thoennessen2010NPA834.688c} etc., have greatly enriched nuclear science, and the boundary of the nuclear chart is being extended continuously. One of the common targets of these new generation RIBs is to explore the unstable nuclei (also referred as exotic nuclei) far away from the $\beta$-stability line \cite{Jonson2004PhyRep389.1, Tanihata1995PPNP35.505, Jensen2004RMP76.215, Casten2000PPNP45.S171, Ershov2010JPG:NP37.064026}.

Exploring from the stable to unstable regions of the nuclear chart, plenty of novel phenomena have been discovered in unstable nuclei, for instance, the quenching of traditional magic shells and the emergences of new ones \cite{Hoffman2008PRL100.152502, Simon1999PRL83.496, Motobayashi1995PLB346.9, TshooPRL109.022501, Ozawa2000PRL84.5493, Kanungo2009PRL102.152501, Gade2006PRC74.021302, Steppenbeck2015PRL114.252501, Steppenbeck2013Nature502.207, Li2019PLB788.192}, the dilute matter distributions --- halo structures \cite{Minamisono1992PRL69.2058, Tanihata1985PRL55.2676, Schwab1995ZPA350.283, Meng1998PRL80.460, Long2010PRC81.031302(R)}, etc. Accompanying with the novel phenomena, the weak-bound mechanism, that accounts for the stabilization of unstable nuclei, challenges our understanding on nuclear physics from both experimental and theoretical aspects. On the other hand, the rapid neutron-caption process ($r$-process), that accounts for the origin of heavy elements in the universe \cite{Burbidge1957RMP29.547, Qian2007PR442.237, Langanke2003RMP75.819}, gets large amount of unstable nuclei involved, which calls for precise understanding on the structures, decays and reactions of those nuclei. Not only for nuclear physics, unstable nuclei are also of special significance in relevant field such as astrophysics, which calls for continuous efforts to devote. Despite rapid development of modern nuclear facilities and advanced nuclear detectors, many of the unstable nuclei are not synthesized experimentally, and the pioneering theoretical exploration can be rather helpful.

In contrast to stable ones, the one- and/or two-nucleon separation energies of unstable nuclei can be as small as $1\sim 2$ MeV or even less, due to the extreme imbalanced neutron-proton ratio. For such weakly bound nuclear systems, nucleons can be gradually scattered into the continuum by such as pairing correlations, and the deduced continuum effects shall be treated carefully. In fact, as one of the crucial weak-bound mechanisms, pairing correlations play an important role not only in stabilizing unstable nuclei but also in developing the novel nuclear phenomena such as the halo structures \cite{Meng1996PRL77.3963, Meng1998PRL80.460, Meng2002PRC65.041302R, Meng2006Prog.Part.Nucl.Phys57.470, T.Niksic2011Prog.Part.Nucl.Phys66.519, Vretenar2005PhyRep409.101, Long2010PRC81.031302(R)}. Compared to the BCS method \cite{Bardeen1957PR108.1175}, which is widely used to describe the pairing correlations in stable nuclei, the Bogoliubov transformation \cite{Bogoliubov1958DAN119.244, Bogoliubov1959DAN124.1011, Valatin1961PR122.1012} owns the advantage of providing unified treatments of both pairing and continuum effects \cite{Meng2006Prog.Part.Nucl.Phys57.470, T.Niksic2011Prog.Part.Nucl.Phys66.519, Vretenar2005PhyRep409.101}, which is essential for the reliable description of  unstable nuclei.

On the other hand, as revealed by many experimental and theoretical studies, most of nuclei in nuclear chart, except a few nearby magic numbers, are of the deformed shapes deviating from the spherical symmetry for both ground and exited states. As we learn from the textbook, nuclear structures can be changed systematically with the deformation. Benefited from the notable progresses achieved on the laser spectroscopy at RIB facilities, more and more novel nuclear phenomena which couple with the deformation effects were discovered, such as the island of inversion \cite{Alburger1964PR136.B916, Crawford2019PRL122.052501, Ahn2019PRL123.212501, Lenzi2010PRC82.054301}, shape coexistence \cite{Garrett2019PRL123.142502, Gaudefroy2009PRL102.092501, Togashi2016PRL117.172502}, super- and hyper-deformed configurations \cite{Oberstedt2007PRL99.042502, Ichikawa2011PRL107.112501}, etc. Not only stable ones, many of unstable nuclei can be also deformed. As indicated in Ref. \cite{Xia2018ADNDT121.1}, the deformation affects can be significant in deciding the boundary of nuclear chart, which the new generation RIB facilities focus on. For the reliable description of unstable nuclei, it becomes necessary to incorporate the deformation effects, in addition to the careful treatment of the weak-bound mechanism, such as the pairing correlations.

In this work, we restrict ourselves under the relativistic scheme. As one of the most successful nuclear models, the relativistic mean field (RMF) theory \cite{Walecka1974Ann.Phys83.491, Serot1986ANP16.1} (also referred as covariant density functional theory), which contains only the Hartree diagram of the meson-propagated nuclear force, has achieved many successes in describing various nuclear phenomena \cite{Reinhard1989Rep.Prog.Phys52.439, Ring1996Prog.Part.Nucl.Phys37.193, Bender2003Rev.Mod.Phys75.121, Vretenar2005PhyRep409.101, Meng2006Prog.Part.Nucl.Phys57.470, T.Niksic2011Prog.Part.Nucl.Phys66.519, Liang2015PRep137.129, Shen2019PPNP109.103713}. Aiming at the reliable description of unstable nuclei, continuous efforts were devoted to the extension of the RMF model --- the axially deformed relativistic Hartree-Bogoliubov model \cite{Lalazissis1999PRC60.014310, Stoitsov1998PRC58.2086, Stoitsov2000PRC61.034311, Zhou2010PRC82.011301, Li2012PRC85.024312, Chen2012PRC85.067301}, referred as D-RHB model here, by incorporating the effects of the deformation and pairing correlations which are treated within the Bogoliubov scheme. It shall be reminded that it is not an easy task to solve the derived RHB equation directly in coordinate space, which is a partial differential equation, even an integral one if utilizing the finite range pairing force.

At the early stage, the axially deformed harmonic oscillator (HO) base was introduced to expand the Bogoliubov spinors and the local mean fields  \cite{Lalazissis1999PRC60.014310, Lalazissis1999NPA650.133, Lalazissis1999PRC60.051302, Afanasjev1999PRC60.051303, Tian2009PRC80.024313}. It was further extended to describe the triaxially deformed nuclei by performing the expansions over the three-dimensional Cartesian HO base \cite{Niksic2010PRC81.054318}. However, due to the nature of the HO potential, the D-RHB model with the HO base met the difficulty in providing appropriate asymptotic behaviors of the wave functions, particularly for the unstable nuclei of dilute matter distributions. Such difficulty might be overcome by considering fairly large amount of oscillator shells, while leading to huge numerical cost. Another alterative recipe is to perform the expansions over the transformed HO base that can modify the unphysical asymptotic properties \cite{Stoitsov1998PRC58.2086, Stoitsov2000PRC61.034311}. Although the HO base owns the advantage of simplicity, lots of efforts were devoted to developing the D-RHB model by expanding the Bogoliubov spinors on the complete set made of the solutions of the Dirac equation with a spherical Dirac Woods-Saxon (DWS) potential, which is referred as the DWS base \cite{Zhou2003PRC68.034323, Li2012PRC85.024312, Chen2012PRC85.067301}. In contrast to the HO base, the DWS base owns the advantage in providing appropriate asymptotic behaviors of the wave functions because of the nature of the Woods-Saxon type potential \cite{Woods1954PhysRev95.577}, and such advantage is essential for the reliable description of unstable nuclei. The D-RHB model with the spherical DWS base has achieved great successes in describing the axially deformed nuclei \cite{Zhou2010PRC82.011301, Li2012PRC85.024312, Chen2012PRC85.067301}, as well as in providing a global nuclear mass table \cite{Zhang2020PRC102.024314, Zhang2021PRC104.L021301}.

Despite the successes achieved by the RMF models, one may notice that only the Hartree terms of the meson-propagated nuclear force are considered, and the Fock diagrams are excluded just for simplicity. Not only for providing a complete picture of the meson-propagated nuclear force, the presence of the Fock terms can also take the important ingredient of nuclear force into account, such as the tensor force component carried by the $\pi$ and $\rho$-tensor couplings \cite{Long2008EPL82.12001, Wang2013PRC87.047301, Li2016PLB753.97, Jiang2015PRC91.034326, Zong2018CPC42.024101}, which contribute mainly via the Fock diagram. Based on the density-dependent relativistic Hartree-Fock (RHF) theory \cite{Long2006PLB640.150, Long2007PRC76.034314} and its extension --- the relativistic Hartree-Fock-Bogoliubov (RHFB) theory \cite{Long2010PRC81.024308}, significant improvements due to the Fock terms have been found on the self-consistent description of nuclear shell evolutions \cite{Long2008EPL82.12001, Long2009PLB680.428, Wang2013PRC87.047301, Li2016PLB753.97}, the spin and isospin excitations \cite{Liang2008PRL101.122502, Liang2009PRC79.064316, Liang2012PRC85.064302, Niu2013PLB723.172, Niu2017PRC95.044301}, symmetry energy \cite{Sun2008PRC78.065805, Long2012PRC85.025806, Zhao2015JPG42.095101}, new magicity \cite{Li2016PLB753.97, Li2014PLB732.169, Li2019PLB788.192, Liu2020PLB806.135524}, pseudo-spin symmetry \cite{Long2006PLB639.242, Liang2010EPJ44.119, Long2010PRC81.031302(R), Geng2019PRC100.051301R} and novel nuclear phenomena \cite{Long2010PRC81.031302(R), Lu2013PRC87.034311, Li2019PLB788.192}, etc. Not only for the $\pi$ and $\rho$-tensor couplings, it was illustrated that the tensor force components can be naturally manifested in the Fock terms of other meson-exchange channels \cite{Jiang2015PRC91.025802, Jiang2015PRC91.034326, Zong2018CPC42.024101}. This is quite meaningful for the reliable description of unstable nuclei, in which the tensor force can play a significant role not only in determining the shell evolution but also in describing the excitation modes \cite{Bai2010PRL105.072501, Minato2013PRL110.122501, Wang2020PRC101.064306}.

In contrast to the simple Hartree terms, the Fock terms contribute complicated non-local mean fields, which makes the RHF calculations quite time consuming and limits its extensive applications in deformed nuclei \cite{Ebran2011PRC83.064323}. Recently, by expanding the Dirac spinor on the spherical DWS base, the axially deformed relativistic Hartree-Fock model \cite{Geng2020PRC101.064302} was established, here referred as the D-RHF model, in which the pairing correlations are treated by using the BCS method. Taking the deformed $^{20}$Ne as an example, it is found that the $\pi$-pseudo-vector coupling, coupled with the deformation effects, presents notable improvement in reproducing the binding energy of $^{20}$Ne, and the carried tensor force components essentially change the evolution behavior of single-particle (s.p.) spectra with respect to the deformation \cite{Geng2020PRC101.064302}. This encourages us much to further improve the description of the pairing effects by incorporating with the Bogoliubov transformation.

Aiming at reliable description of wide-range unstable nuclei, the axially deformed relativistic Hartree-Fock-Bogoliubov model, here referred as the D-RHFB model, is developed in this work. The paper is organized as follows. In Sec. \ref{sec:General Formalism}, the complete theoretical framework is briefly introduced, and the D-RHFB frame is established by quantizing the Dirac spinor field in the Bogoliubov quasi-particle space, in which the Bogobliubov quasi-particle spinors are expanded on the spherical DWS base, similar as we did in developing the D-RHF model. Afterwards, by taking the deformed light nucleus $^{24}$Mg and mid-heavy one $^{156}$Sm as the candidates, the space truncations related with the spherical DWS base are discussed in Sec. \ref{sec:RESULTS AND DISCUSSIONS}, and in particular we focus on the role of the $\rho$-tensor couplings in describing the structure properties of the selected nuclei. Finally, we give a summary in Sec. \ref{sec:SUMMARY}. In addition to the Appendix in Ref. \cite{Geng2020PRC101.064302}, the detailed formulas of the $\rho$-tensor and $\rho$-vector-tensor couplings are list in Appendix \ref{sec:APP-A}, and Appendix \ref{sec:APP-B} and \ref{sec:APP-C} show the details of pairing potentials with Gogny-type pairing force and microscopic cent-of-mass corrections, respectively.

\section{General Formalism}\label{sec:General Formalism}

To provide a complete understanding on the theoretical framework, here we briefly recall the general formalism of the relativistic Hartree-Fock (RHF) theory at first. In addition to the D-RHF model in Ref. \cite{Geng2020PRC101.064302}, the degrees of freedom associated with the $\rho$-tensor ($\rho$-T) coupling is implemented in this work. Afterwards, the relativistic Hartree-Fock-Bogoliubov model for axially deformed nuclei will be introduced in details, referred as the D-RHFB model, in which the Bogoliubov quasi-particle wave functions are expanded on the spherical Dirac Woods-Saxon (DWS) base \cite{Zhou2003PRC68.034323}.

\subsection{RHF Lagrangian and Hamiltonian}

Under the meson-exchange diagram of nuclear force, the Lagrangian density for the nuclear systems, the starting point of the theory, can be constructed by considering the degrees of freedom associated with nucleon field ($\psi$), meson and photon ($A^\mu$) fields. For the meson fields, which propagate the nucleon-nucleon interaction, two isoscalar mesons $\sigma$ and $\omega^\mu$, and two isovectors ones $\ivec\rho^\mu$ and $\ivec\pi$ of the following quantum numbers $(I^\pi, \tau)$ are considered,
\begin{equation}
  \sigma(0^+,0), \hspace{1em}\omega^\mu (1^-,0), \hspace{1em}\ivec\rho^\mu(1^-,1), \hspace{1em} \ivec \pi(0^-,1),
\end{equation}
where $I$, $\pi$ and $\tau$ are the spin, parity and isopin of the selected mesons, respectively. Here and in the following, the arrows are used to denote the isovectors and the bold types for the space vectors. Thus, the Lagrangian density of nuclear system can be expressed as,
\begin{equation}\label{eq:Lagrangian}
    \scr L = \scr L_N + \scr L_\sigma + \scr L_\omega + \scr L_\rho + \scr L_\pi + \scr L_{A} + \scr L_{I},
\end{equation}
in which the Lagrangians $\scr L_\phi$ ($\phi = N,\sigma,\omega,\rho,\pi$ and $A$) of the nucleon, meson and photon fields read as,
\begin{subequations}
\begin{align}
    \scr L_N =& \bar\psi\left( i\gamma^\mu\partial_\mu -M \right) \psi, \\
    \scr L_\sigma = & +\ff2 \partial^\mu \sigma \partial_\mu\sigma - \ff2 m_\sigma^2 \sigma^2, \\
    \scr L_\omega = & -\ff4 \Omega^{\mu\nu} \Omega_{\mu\nu} + \ff2 m_\omega^2 \omega_\mu \omega^\mu, \\
    \scr L_\rho = & -\ff4 \vec{R}_{\mu\nu} \cdot \vec{R}^{\mu\nu} + \ff2 m_\rho^2 \vec{\rho}^\mu \cdot \vec{\rho}_\mu, \\
    \scr L_\pi =&  +\ff2 \partial_\mu \vec\pi \cdot \partial^\mu \vec \pi - \ff2 m_\pi^2 \vec\pi \cdot\vec\pi, \\
    \scr L_A  =&  -\ff4 F^{\mu\nu} F_{\mu\nu},
\end{align}
\end{subequations}
with the field tensors $\Omega^{\mu\nu}\equiv \partial^\mu\omega^\nu - \partial^\nu \omega^\mu, \vec R^{\mu\nu}\equiv \partial^\mu \vec\rho^\nu - \partial^\nu\vec\rho^\mu$, and $F^{\mu\nu}\equiv \partial^\mu A^\nu - \partial^\nu A^\mu$. Considering the Lorentz scalar ($\sigma$-S), vector ($\omega$-V, $\rho$-V and $A$-V), tensor ($\rho$-T) and pseudo-vector ($\pi$-PV) couplings, the Lagrangian density $\scr L_I$, that describes the interactions between nucleon and meson/photon fields, can be written as,
\begin{equation}
  \begin{split}
    \scr L_I = &-g_\sigma\big(\bar\psi\psi\big) \sigma - g_\omega \big(\bar\psi\gamma^\mu \psi\big)\omega_\mu \\
    & - g_\rho \big(\bar\psi\gamma^\mu \vec\tau\psi\big) \cdot \vec \rho_\mu + \frac{f_\rho}{2M}\big( \bar\psi\sigma^{\mu\nu}\ivec\tau \psi\big) \cdot\partial_\nu \vec\rho_\mu\\
    & - \frac{f_\pi}{m_\pi}\big( \bar\psi\gamma_5 \gamma^\mu\vec\tau\psi\big)\cdot \partial_\mu \ivec\pi - e\Big(\bar\psi\gamma^\mu \frac{1-\tau}{2}\psi\Big) A_\mu.
  \end{split}
\end{equation}
In the above expressions, $M$ and $m_\phi$ are the masses of nucleon and mesons, and $g_\phi$ ($\phi = \sigma$, $\omega$ and $\rho$) and $f_{\phi'}$ ($\phi' = \rho$ and $\pi$) represent various meson-nucleon coupling strengths. For the isospin projection $\tau$, the convention that $\tau\lrlc{n}=\lrlc{n}$ and $\tau\lrlc{p}=-\lrlc{p}$ is used in this work.

From the Lagrangian density (\ref{eq:Lagrangian}), one can obtain the Hamiltonian via the Legendre transformation. After neglecting the retardation effects, namely ignoring the time-component of the four-momentum carried by the mesons and photon, and substituting the form solution of the meson and photon field equations, the Hamiltonian can be derived as \cite{Bouyssy1987PRC36.380},
\begin{equation}\label{eq:Hamiltonian}
    H = T + \sum_\phi  V_\phi,
\end{equation}
where the kinetic energy ($ T$) and potential energy ($ V_\phi$) parts read as,
\begin{align}
     T =& \int d\svec r \bar\psi(\svec r) \left(-i\svec\gamma \cdot\svec \nabla + M\right) \psi(\svec r), \label{eq:kinetic-O}\\
     V_\phi = & \ff2 \int d\svec r d\svec r' \bar\psi(\svec r) \bar\psi(\svec r') \Gamma_\phi D_\phi(\svec r-\svec r') \psi(\svec r') \psi(\svec r).\label{eq:potential-O}
\end{align}
In the above expressions, $\phi$ represents various two-body interaction channels, namely the $\sigma$-S, $\omega$-V, $\rho$-V, $\rho$-T, $\rho$-vector-tensor ($\rho$-VT), $\pi$-PV and $A$-V couplings, and the relevant vertex $\Gamma_\phi( x, x')$ read as,
\begin{subequations}
\begin{align}
    \Gamma_{\sigma\text{-S}} \equiv & -g_\sigma(x) g_\sigma(x'), \\
    \Gamma_{\omega\text{-V}} \equiv & \left(g_\omega \gamma_\mu\right)_{x} \left( g_\omega\gamma^\mu \right)_{x'}, \\
    \Gamma_{\rho\text{-V}} \equiv & \left( g_\rho \gamma_\mu \ivec\tau \right)_x \cdot \left( g_\rho \gamma^\mu \ivec\tau \right)_{x'}, \\
    \Gamma_{\rho\text{-T}} \equiv & \frac{1}{4M^2} \left( f_\rho \sigma_{\nu k} \vec\tau \partial^k \right)_{x} \cdot \left( f_\rho \sigma^{\nu l} \vec\tau \partial_l\right)_{x'}, \\
    \Gamma_{\rho\text{-VT}} \equiv & \frac{1}{2M} \left( f_\rho \sigma^{k\nu} \vec\tau \partial_k \right)_{x} \cdot \left( g_\rho \gamma_\nu \vec\tau \right)_{x'} \nonumber \\
    & + \frac{1}{2M}  \left( g_\rho \gamma_\nu \vec\tau \right)_{x} \cdot \left( f_\rho \sigma^{k\nu} \vec\tau \partial_k\right)_{x'}, \\
    \Gamma_{\pi\text{-PV}} \equiv & \frac{-1}{m_\pi^2} \left( f_\pi \vec\tau \gamma_5 \gamma_\mu \partial^\mu \right)_{x} \cdot \left( f_\pi \vec\tau \gamma_5 \gamma_\nu \partial^\nu \right)_{x'}, \\
    \Gamma_{A\text{-V}} \equiv & \frac{e^2}{4} \left( \gamma_\mu (1- \tau) \right)_{x} \left( \gamma^\mu (1- \tau) \right)_{x'}.
\end{align}
\end{subequations}
After neglecting the retardation effects, the propagators $D_\phi(\svec r-\svec r')$ of the meson and photon field in $V_\phi$ are of the Yukawa form as,
\begin{equation}\label{eq:Yukawa}
    D_\phi = \frac{1}{4\pi} \frac{e^{-m_\phi\left| \svec r-\svec r' \right|}}{\left|\svec r-\svec r'\right|}, \hspace{2em} D_{A} = \frac{1}{4\pi} \frac{1}{\left|\svec r-\svec r'\right|}.
\end{equation}

\subsection{Quantization of the Hamiltonian}
In this work, we intend to derive the full energy functional from the Hamiltonian (\ref{eq:Hamiltonian}) by considering its expectation with respect to the Bogoliubov ground state, which may deduce both contributions from the mean field and pairing correlations. While it is not so straightforward to quantize the Dirac spinor field $\psi$ in the Bogoliubov quasi-particle space, as one did in deriving the RHF energy functional \cite{Bouyssy1987PRC36.380, Geng2020PRC101.064302}.

Under the RHF approach, the Dirac spinor field $\psi$ can be quantized as,
\begin{subequations}\label{eq:expansionHF}
\begin{align}
  \psi(x) = & \sum_{l} \psi_l(x)  c_l, &
  \bar\psi(x) = & \sum_{l} \bar\psi_{l}(x)  c_l^\dag,
\end{align}
\end{subequations}
where $x=(t,\svec r)$, the creation and annihilation operators $c_l^\dag$ and $c_l$ in the Hartree-Fock (HF) space are defined by the positive energy solutions of the Dirac equation, and $\psi_l(x)$ is the s.p. wave function. In quantizing the Dirac spinor $\psi$, i.e., Eq. (\ref{eq:expansionHF}), the contributions from the negative energy solutions of the Dirac equation are ignored to keep consistence with the mean field approach, namely the no-sea approximation. Thus, the HF ground state $\lrlc{\text{HF}}$ can be defined as,
\begin{equation}\label{eq:HFground}
  \lrlc{\text{HF}} \equiv  \prod_{l=1}^A c_l^\dag \lrlc{-},
\end{equation}
where $\lrlc{-}$ is the vacuum state. The expectation of the Hamiltonian (\ref{eq:Hamiltonian}) with respect to $\lrlc{\text{HF}}$ leads to the RHF energy functional, whereas the pairing effects in open-shell nuclei are considered for instance by the BCS method \cite{Geng2020PRC101.064302}.

The Bogoliubov scheme is another efficient way to describe the pairing correlations. From the Bogoliubov transformation, that defines the relation between the HF particle and Bogoliubov quasi-particle spaces as,
\begin{equation}\label{eq:Bogoliubov}
    \begin{pmatrix} \beta_k \\[0.5em] \beta_k^\dag \end{pmatrix} = \sum_l \begin{pmatrix} U_{lk}^* & V_{lk}^* \\[0.5em] V_{lk} & U_{lk} \end{pmatrix} \begin{pmatrix} c_l \\[0.5em] c_l^\dag \end{pmatrix},
\end{equation}
where $\beta_k$ and $\beta_k^\dag$ are the annihilation and creation operators of the Bogoliubov quasi-particle, respectively, one may establish the following specific relationship between the HF and HFB wave functions as,
\begin{align}
     \psi_k^V(x) = &\sum_l V_{lk}\psi_l(x), & \bar\psi_k^V(x) =& \sum_l V_{lk}^* \bar\psi_l(x), \\
     \psi_{\bar k}^U(x) = &\sum_l U_{lk} \psi_{\bar l}(x), & \bar\psi_{\bar k}^U(x) = &\sum_l U_{lk}^* \bar\psi_{\bar l}(x),
\end{align}
where $\psi^V$ and $\psi^U$ denote the $V$- and $U$-components of the Bogoliubov quasi-particle spinor, respectively. In the above expressions, the index $l$ and $k$ denote the HF s.p. states and the Bogoliubov quasi-particle states, respectively, and $\bar l$ and $\bar k$ hold for the relevant time-reversal partners. Thus, in terms of $\beta_k$ and $\beta_k^\dag$, the Dirac spinor field $\psi$ can be expanded as,
\begin{subequations}\label{eq:expansionHFB}
\begin{align}
  \psi(x) = & \sum_k \big(\psi_{\bar k}^{U}(\svec r) e^{-i\varepsilon_k t} \beta_k + \psi_k^V(\svec r) e^{+i\varepsilon_k t} \beta_k^\dag\big), \\
  \bar\psi(x) = & \sum_k \big(\bar\psi_k^V(\svec r) e^{-i\varepsilon_k t} \beta_k + \bar\psi_{\bar k}^{U}(\svec r) e^{+i\varepsilon_k t}\beta_k^\dag\big),
\end{align}
\end{subequations}
in which $\varepsilon_k$ is the quasi-particle energy, and the explicitly introduced time-reversal partners are necessitated to form the Cooper pairs. One may notice that it is not trivial from Eq. (\ref{eq:expansionHF}) to the above expansions. This is due to the fact that the relation between the HF ground state (\ref{eq:HFground}) and HFB one is not as straightforward as one expects. For the HFB ground state $\lrlc{\text{HFB}}$, it shall fulfill the condition \cite{Ring1980Springer-Verlag},
\begin{equation}\label{eq:HFBground}
    \beta_k\lrlc{\text{HFB}} = 0.
\end{equation}
With such condition, it is hard to have unique form similar as the HF one $\lrlc{\text{HF}}$.

Since the expansion (\ref{eq:expansionHFB}) is deduced from the quantization (\ref{eq:expansionHF}) and the Bogoliubov transformation (\ref{eq:Bogoliubov}), the quasi-particle operators $\beta_k$ and $\beta_k^\dag$ also obey the anticommutation relation of fermions. With the quantization form (\ref{eq:expansionHFB}) for the Dirac spinor field $\psi$, the kinetic energy (\ref{eq:kinetic-O}) and potential energy (\ref{eq:potential-O}) terms in the Hamiltonian can be expressed as,
\begin{widetext}
\begin{align}
  T = & \sum_{kk'}\int d\svec r \bar\psi_k^V(\svec r)(-i\svec\gamma\cdot\svec\nabla + M)\psi_{k'}^V (\svec r) \beta_k \beta_{k'}^\dag, \label{eq:Ham-k}\\
  V_\phi = & \ff2 \sum_{k_1k_2 k_2'k_1'} \int d\svec r d\svec r'  \Big[\bar\psi_{k_1}^V(\svec r)
  \bar\psi_{k_2}^V(\svec r')\Gamma_\phi (\svec r, \svec r') D_{\phi}(\svec r-\svec r') \psi_{k_2'}^V(\svec r')  \psi_{k_1'}^V(\svec r) \beta_{k_1} \beta_{k_2} \beta_{k_2'}^\dag \beta_{k_1'}^\dag  \nonumber\\
  &\hspace{7.25em}+ \bar\psi_{k_1}^V(\svec r)\bar\psi_{\bar k_2}^U(\svec r')\Gamma_\phi(\svec r, \svec r') D_{\phi}(\svec r-\svec r')\psi_{\bar k_2'}^U(\svec r')\psi_{k_1'}^V(\svec r) \beta_{k_1} \beta_{k_2}^\dag \beta_{k_2'}\beta_{k_1'}^\dag\Big]\label{eq:Ham-v},
\end{align}
\end{widetext}
in which the terms with zero expectation referring to $\lrlc{\text{HFB}}$ are omitted. Obviously, the first term in $V_\phi$ correspond to the contribution of the mean field, and the second one accounts for the pairing correlations.

Thus, in stead of deriving an explicit form of $\lrlc{\text{HFB}}$ beforehand, one can obtain the full energy functional directly from the expectation of the Hamiltonian (\ref{eq:Hamiltonian}) with respect to $\lrlc{\text{HFB}}$ as,
\begin{equation}\label{eq:Full-E}
\begin{split}
  E = & \lrcl{\text{HFB}} H\lrlc{\text{HFB}}  \\
    = & E^{\text{kin.}} + \sum_{\phi} \big( E_{\phi}^D + E_\phi^E + E_\phi^{pp}\big),
\end{split}
\end{equation}
where the kinetic energy $E^{\text{kin.}}$, the Hartree potential energy $E_\phi^D$ and Fock one $E_\phi^E$, and the pairing energy $E_\phi^{pp}$ read as,
\begin{subequations}\label{eq:Full-E-d}
\begin{align}
    E^{\text{kin.}} = & \sum_{k}\int d\svec r \bar\psi_k^V(\svec r)(-i\svec\gamma\cdot\svec\nabla + M)\psi_{k}^V(\svec r),  \label{eq:kinetic}\\
    E_\phi^{\text{D}} = & + \ff2 \sum_{k k'}  \int d\svec r d\svec r' \bar\psi_{k}^V(\svec r) \bar\psi_{k'}^V(\svec r') \nonumber \\
    &\hspace{3em}\times\Gamma_\phi D_{\phi}(\svec r-\svec r') \psi_{k'}^V(\svec r')  \psi_{k}^V(\svec r), \label{eq:potential-H}\\
    E_\phi^{\text{E}} = &  - \ff2 \sum_{k k'}  \int d\svec r d\svec r' \bar\psi_{k}^V(\svec r) \bar\psi_{k'}^V(\svec r')\nonumber \\
    &\hspace{3em}\times\Gamma_\phi D_{\phi}(\svec r-\svec r') \psi_{k}^V(\svec r') \psi_{k'}^V(\svec r), \label{eq:potential-F}\\
    E_{\phi}^{pp} =& + \ff2 \sum_{k k'}  \int d\svec r d\svec r' \bar\psi_{k}^V(\svec r)\bar\psi_{\bar k}^U(\svec r')\nonumber \\
    &\hspace{3em}\times\Gamma_\phi D_{\phi}(\svec r-\svec r')\psi_{\bar k'}^U(\svec r')\psi_{k'}^V(\svec r)\label{eq:potential-P}.
\end{align}
\end{subequations}
Notice that the expectation of the quantized two-body interaction ${V}_\phi$ (\ref{eq:Ham-v}) does give the Hartree and Fock potential energies, namely Eqs. (\ref{eq:potential-H}) and (\ref{eq:potential-F}), while the pairing energy (\ref{eq:potential-P}) does not contain commutative antisymmetric contributions.

In order to get reasonable description of the pairing effects, people often utilize the phenomenological pairing force, such as the zero-range delta force \cite{Meng1998NPA635.3}, finite-range Gogny force \cite{Meng1998NPA635.3, Long2010PRC81.024308}, and separable pairing force \cite{Tian2009PLB676.44}, etc. In this work, we choose the finite-range Gogny force D1S \cite{Berger1984NPA428.23} as the pairing force. Specifically, the term $\Gamma_\phi D_\phi(\svec r-\svec r')$ in $E_{\phi}^{pp}$ is replaced by $(\gamma_0)_{x}(\gamma_0)_{x'} V_{\text{Gogny}}^{pp}(\svec r-\svec r')$, see the following context for details.

With the obtained energy functional (\ref{eq:Full-E}), one can derive the RHFB equation for the Bogoliubov quasi-particles following the variational principle. While, one may also notice that only the $V$-components of the quasi-particle spinors remain in the kinetic and potential energies, see Eqs. (\ref{eq:kinetic}-\ref{eq:potential-F}). If performing the variation with respect to the $V$-component, only part of the RHFB equation can be deduced. In fact, the $U$- and $V$-components shall fulfill certain condition due to the unitary of the Bogoliubov transformation \cite{Ring1980Springer-Verlag}. Thus to obtain the full RHFB equation, it requires to perform the variation of the energy functional with respect to the generalized density matrix $\scr R$ as,
\begin{equation}
   \scr R\equiv  \begin{pmatrix} \rho & \kappa \\ -\kappa^* & 1-\rho^* \end{pmatrix},
\end{equation}
where $\rho$ is the density matrix and $\kappa$ is the pairing tensor \cite{Ring1980Springer-Verlag}. Following the procedure detailed in Ref. \cite{Jean-Paul1986MITpress}, one may obtain the quasi-particle Hamiltonian $\cals H$ as,
\begin{equation}
    \cals H = \frac{\partial E}{\partial \scr R} 
    = \begin{pmatrix} h & \Delta \\[0.5em] -\Delta^* & -h^*  \end{pmatrix}, 
\end{equation}
where $h=\partial E/\partial\rho$ and $\Delta=\partial E/\partial\kappa$ are respectively the s.p. Hamiltonian and pairing potential which will be detailed in the following context.

For spherical nuclei, the deduced RHFB equation is an integro-differential equation because of the non-local Fock terms and pairing potentials \cite{Long2010PRC81.024308}. Such equation is hard to solve in coordinate space. For axially deformed nuclei, the RHFB equation becomes an integral partial-differential equation, leading to even more complicated numerical problem. In this work, we will expand the $U$- and $V$-components of quasi-particle wave functions on the spherical Dirac Woods-Saxon (DWS) base \cite{Zhou2003PRC68.034323}, similar as we did in Refs. \cite{Long2010PRC81.024308, Geng2020PRC101.064302}.

Not only for the convenience of solving the RHFB equation, it is shown in Ref. \cite{Geng2020PRC101.064302} that one can avoid the singularity at $\svec r = \svec r'$ in the propagators (\ref{eq:Yukawa}) by introducing the expansion on the spherical DWS base, since only finite terms among the propagator decompositions can contribute with a given space truncation, see Eqs. (31-32) in Ref. \cite{Geng2020PRC101.064302}. This is one of the reason why we do not utilize the axially deformed harmonic oscillator base to expanded the quasi-particle wave functions, although it has simple analytic form. Not only due to that, the spherical DWS base can provide appropriate asymptotical behaviors of the wave functions. It is not so significant for the stable nuclei, while can be essential when people extend from the stable to unstable region of the nuclear chart. For weakly bound unstable nuclei, the continuum effects can be involved and thus the appropriate asymptotical behaviors of the wave functions are significant for the reliable description \cite{Meng1998NPA635.3,Meng2006Prog.Part.Nucl.Phys57.470,Meng1996PRL77.3963,Meng1998PRL80.460}.

\subsection{RHFB energy functional and eigenvalue equations with spherical DWS base}
From the spherical to axial symmetry, the s.p./quasi-particle angular momentum $j$ is not a conserved physical quantity anymore, and its projection $m$ remains as a good quantum number. In addition to the axial symmetry, another restriction, the reflection symmetry with respect to the $z=0$ plane is also introduced for the axially deformed nuclei in this work, which indicates conserved parity $\pi$ for s.p./quasi-particle states. In the following, we utilize the index $i=(\nu\pi m)$ to denote the quasi-particle orbits of axially deformed nuclei, $\nu$ for the index of the orbits in the $(\pi m)$-block.

Using the spherical DWS base, the $U$- and $V$-components of the quasi-particle wave functions $\psi^U$ and $\psi^V$ can be expanded as
\begin{equation}\label{equ:C}
    \psi_{\nu\pi m}^U = \sum_{a} C_{a,i}^U \psi_{am}, \hspace{2em}
    \psi_{\nu\pi m}^V = \sum_{a} C_{a,i}^V \psi_{am},
\end{equation}
where the coefficients $C_{a,i}^U$ and $C_{a,i}^V$ are restricted as real number, and the index $a = (n\kappa)$, together with the angular momentum projection $m$, define the states in the spherical DWS base, with $\kappa = \pm (j+1/2)$ and $j=l\mp1/2$, $l$ for orbital angular momentum and $n$ for the principle number. For the spherical DWS base, the wave function $\psi_{n\kappa m}$ can be explicitly expressed as
\begin{equation}\label{eq:nkappam}
    \psi_{n\kappa m} = \frac{1}{r} \begin{pmatrix} G_{n\kappa}(r) \Omega_{\kappa m}(\vartheta,\varphi) \\[0.5em] iF_{n\kappa}(r) \Omega_{-\kappa m}(\vartheta,\varphi) \end{pmatrix},
\end{equation}
where $\Omega_{\kappa m}$ is the spinor spherical harmonics \cite{Varshalovich1988World-Scientific}.

Similar as we did in Ref. \cite{Geng2020PRC101.064302}, the expansions of $\psi_{\nu \pi m}^U$ and $\psi_{\nu\pi m}^V$ are further abbreviated as
\begin{subequations}\label{eq:expansion-wav}
\begin{align}
    \psi_{\nu\pi m}^V =& \sum_\kappa \frac{1}{r} \begin{pmatrix} \cals G_{i\kappa}^V \Omega_{\kappa m} \\[0.5em] i\cals F_{i\kappa}^V \Omega_{-\kappa m} \end{pmatrix}, \\
    \psi_{\nu\pi m}^U =& \sum_\kappa \frac{1}{r} \begin{pmatrix} \cals G_{i\kappa}^U \Omega_{\kappa m} \\[0.5em] i\cals F_{i\kappa}^U \Omega_{-\kappa m} \end{pmatrix},
\end{align}
\end{subequations}
where the quantities $\cals G$ and $\cals F$ contain the superposition of radial wave functions $G$ and $F$ as,
\begin{subequations}
\begin{align}
    \cals G_{i\kappa}^V = &\sum_n C_{n\kappa,i}^V G_{n\kappa}, & \cals F_{i\kappa}^V = &\sum_n C_{n\kappa,i}^V F_{n\kappa}, \\
    \cals G_{i\kappa}^U =& \sum_n C_{n\kappa,i}^U G_{n\kappa}, & \cals F_{i\kappa}^U =& \sum_n C_{n\kappa,i}^U F_{n\kappa}.
\end{align}
\end{subequations}

Employing the expansions (\ref{eq:expansion-wav}), the kinetic energy functional (\ref{eq:kinetic}) can be explicitly expressed as,
\begin{equation}\label{eq:ENE-Kin}
\begin{split}
    E^{\text{kin.}} = \sum_{i\kappa} \int dr \Big\{& \cals F_{i\kappa}^V \Big[ \frac{d\cals G_{i\kappa}^V}{dr} + \frac{\kappa}{r} \cals G_{i\kappa}^V - M \cals F_{i\kappa}^V \Big] \\
    -& \cals G_{i\kappa}^V \Big[ \frac{d\cals F_{i\kappa}^V}{dr} -\frac{\kappa}{r} \cals F_{i\kappa}^V - M \cals G_{i\kappa}^V \Big] \Big\}.
\end{split}
\end{equation}

For the potential energies, there exist the Hartree and Fock terms. For the Hartree terms (\ref{eq:potential-H}), the contributions from various meson-nucelon coupling channels can be derived as,
\begin{subequations}\label{eq:ENE-Hartree}
\begin{align}
    E_{\sigma\text{-S}}^D =& \frac{2\pi}{2} \int r^2dr \sum_{\lambda_d} \Sigma_{S,\sigma\text{-S}}^{\lambda_d}(r) \rho_s^{\lambda_d}(r), \\
    E_{\omega\text{-V}}^D =& \frac{2\pi}{2} \int r^2dr \sum_{\lambda_d} \Sigma_{0,\omega\text{-V}}^{\lambda_d}(r) \rho_b^{\lambda_d}(r), \\
    E_{\rho\text{-V}}^D =& \frac{2\pi}{2} \int r^2dr \sum_{\lambda_d} \Sigma_{0,\rho\text{-V}}^{\lambda_d}(r) \rho_{b,3}^{\lambda_d}(r),\\
    E_{\rho\text{-T}}^D =& \frac{2\pi}{2} \int r^2dr \sum_{\lambda_d\mu} \Sigma_{T,\rho\text{-T}}^{\lambda_d\mu}(r) \rho_{T,3}^{\lambda_d\mu}(r),\\
    E_{\rho\text{-VT}}^D = & \frac{2\pi}{2} \int r^2dr \Big[\sum_{\lambda_d\mu}\Sigma_{T,\rho\text{-TV}}^{\lambda_d\mu}(r) \rho_{T,3}^{\lambda_d\mu}(r) \nonumber \\
    & \hspace{4em}+ \sum_{\lambda_d}\Sigma_{0,\rho\text{-VT}}^{\lambda_d}(r) \rho_{b,3}^{\lambda_d}(r) \Big].
\end{align}
\end{subequations}
In the above expressions, $\Sigma_{S,\phi}$, $\Sigma_{0,\phi}$ and $\Sigma_{T,\phi}$ represent respectively the local scalar, vector and tensor self-energies from various channel $\phi$, including the contributions of the Hartree terms and rearrangement term due to the density dependence of the meson-nucleon coupling strengths \cite{Long2006PLB640.150, Long2007PRC76.034314}. For the $A$-V couplings, the expressions can be obtained similarly as the $\omega$-V and $\rho$-V ones. The details can be found in the Appendix A 2 of Ref. \cite{Geng2020PRC101.064302}. In Appendix \ref{subsec:Hartree-rhot}, the details of the $\rho$-T and $\rho$-VT couplings are implemented, namely $\Sigma_{\text{T},\rho\text{-T}}^{\lambda_d\mu}$, $\Sigma_{T, \rho\text{-TV}}^{\lambda_d\mu}$, $\Sigma_{0, \rho\text{-VT}}^{\lambda_d}$, as well as the tensor density $\rho_{T,3}^{\lambda_d\mu}$.

The Fock terms of the two-body interactions present rather complicated contributions to the energy functional. Even though, the Fock energy functional $E_\phi^E$, i.e., Eq. (\ref{eq:potential-F}), can be uniformly described as,
\begin{equation}\label{eq:ENE-Fock}
\begin{split}
    E_\phi^E =& \ff2 \int drdr'\sum_{i\kappa_1\kappa_2} \begin{pmatrix} \cals G_{i\kappa_1}^V & \cals F_{i\kappa_1}^V  \end{pmatrix}_r  \\
    &\hspace{2em}\times \begin{pmatrix} Y_{G,\pi m}^{\kappa_1,\kappa_2;\phi} & Y_{F,\pi m}^{\kappa_1,\kappa_2;\phi} \\[0.5em] X_{G,\pi m}^{\kappa_1,\kappa_2;\phi} & X_{F,\pi m}^{\kappa_1,\kappa_2;\phi} \end{pmatrix}_{r,r'} \begin{pmatrix} \cals G_{i\kappa_{2}}^V \\[0.5em] \cals F_{i\kappa_2}^V\end{pmatrix}_{r'},
\end{split}
\end{equation}
where $\phi$ represents the coupling channels $\sigma$-S, $\omega$-V, $\rho$-V, $\rho$-T, $\rho$-VT, $\pi$-PV and $A$-V. The details of the $\rho$-T and $\rho$-VT couplings are given in Appendix \ref{subsec:Fock-rhot} and \ref{subsec:Contact-rhot}, and the others can be found in Appendix A 3 of Ref. \cite{Geng2020PRC101.064302}. For the nuclei with odd neutron and/or proton numbers, the blocking effects shall be considered, and one needs to replace the $V$-components by the $U$-components for the blocked orbits in dealing with the kinetic energy and potential energy terms.

For the contributions from the pairing correlations, i.e., Eq. (\ref{eq:potential-P}), we adopt the Gogny force D1S as the pairing force, due to the advantage as mentioned in Ref. \cite{Geng2020PRC101.064302} that the finite-range nature of Gogny force can lead to a natural convergence with the configuration space in evaluating the pairing effects. The Gogny-type pairing force read as
\begin{equation}\label{eq:Gogny}
  \begin{split}
    V^{pp}_{\text{Gogny}}(\svec r-\svec r') = & \sum_{i=1,2} \exp\Big(\frac{(\svec r-\svec r')^2}{\mu_i^2}\Big) \\
    &\hspace{-2em}\times \big(W_i + B_i P^\sigma - H_i P^\tau - M_i P^\sigma P^\tau\big),
  \end{split}
\end{equation}
with the parameters $\mu_i$, $W_i$, $B_i$, $H_i$, $M_i$ $(i=1,2)$ as the finite range part of the Gogny force, and $P^\sigma$ and $P^\tau$ are the spin and isospin exchange operators, respectively. Employing the expansions (\ref{equ:C}), the pairing energy can be derived as
\begin{equation}\label{eq:ENE-pairing}
    E^{pp} = \ff2 \sum_{m\kappa_1\kappa_2} \int dr dr' \sum_{\sigma\sigma'}^{\pm} K_{m\kappa_1\kappa_2}^{\sigma\sigma'}(r,r') \Delta_{m\kappa_1\kappa_2}^{\sigma\sigma'}(r,r'),
\end{equation}
where the pairing potential $\Delta^{\sigma\sigma'}$  is given by the variation with respect to the pairing tensor components $K^{\sigma\sigma'}$ \cite{Jean-Paul1986MITpress},
\begin{equation}\label{eq:PairingP}
  \Delta_{m\kappa_1 \kappa_2}^{\sigma\sigma'} =  \sum_{m'\kappa_1'\kappa_2'} \Pi_{m\kappa_1\kappa_2,m'\kappa_1'\kappa_2'}^{\sigma\sigma'} K_{m'\kappa_1'\kappa_2'}^{\sigma\sigma'}.
\end{equation}
In the above expressions, $\Pi^{\sigma\sigma'}$ is the combination of the C-G coefficients and the radial part of the Gogny force, and $\sigma,\ \sigma'=\pm$ correspond to the upper ($+$) and lower ($-$) components of $\psi_{\nu\pi m}^{U,V}$ in Eqs. (\ref{eq:expansion-wav}). In deriving pairing energy, the $L$-$S$ coupling scheme is used \cite{Meng1998NPA635.3}. In order to fully evaluate the pairing effects, the contributions of all $J$ ($\svec J = \svec L+\svec S$) components are considered in this work, rather than only $J=0$ component as in Ref.\cite{Geng2020PRC101.064302}. The details are given in Appendix \ref{sec:APP-B}.

Since the quasi-particle spinors $\psi^U$ and $\psi^V$ are expanded on the spherical DWS base, the RHFB equation for the Bogoliubov quasi-particles turns to be an eigenvalue equation as,
\begin{equation}\label{eq:HFB-E}
    \sum_{a'} \begin{pmatrix} -h_{aa'}^{i} +\lambda & \Delta_{aa'}^i \\[0.5em] \Delta_{aa'}^{i} & h_{aa'}^{i}-\lambda \end{pmatrix} \begin{pmatrix} C_{a',i}^U \\[0.5em] C_{a',i}^V \end{pmatrix} = \varepsilon_i \begin{pmatrix} C_{a,i}^U \\[0.5em] C_{a,i}^V \end{pmatrix},
\end{equation}
where the chemical potential $\lambda$ is introduced to conserve the particle number on the average. In the above equation, the s.p. Hamiltonian $h_{aa'}^i$ contains three parts,
\begin{equation}
    h_{aa'}^i = h_{aa'}^{m,\text{kin.}} + h_{aa'}^{m,\text{D}} + h_{aa'}^{m,\text{E}},
\end{equation}
which correspond to the kinetic, local mean-field and non-local mean-field terms, respectively. The kinetic energy term $h_{aa'}^{m,\text{kin.}}$ reads as
\begin{equation}
  \begin{split}
    h_{aa'}^{m,\text{kin}.} = \int dr \Big\{& -G_a \Big[ \frac{dF_{a'}}{dr} -\frac{\kappa}{r} F_{a'} - MG_{a'} \Big]  \\
    & + F_a \Big[ \frac{dG_{a'}}{dr} + \frac{\kappa}{r} G_{a'} - MF_{a'} \Big] \Big\},
  \end{split}
\end{equation}
in which $\kappa = \kappa'$. For the local mean-field term $h_{aa'}^{m,\text{D}}$ that contains the Hartree mean field and rearrangement term, it can be expressed as,
\begin{equation}
\begin{split}
    h_{aa'}^{m,D} = \int dr \sum_{\lambda_d}&(-1)^{m+\ff2} \begin{pmatrix} G_a & F_a \end{pmatrix}_r  \\
    &\times \begin{pmatrix} \Sigma_+^{\lambda_d} & \Pi_+^{\lambda_d}  \\[0.5em] \Pi_-^{\lambda_d}& \Sigma_-^{\lambda_d}   \end{pmatrix}_r \begin{pmatrix} G_{a'} \\[0.5em] F_{a'} \end{pmatrix}_r,
\end{split}
\end{equation}
where the local self-energy terms $\Sigma_\pm$ and $\Pi_\pm$ read as,
\begin{align}
    \Sigma_{\pm}^{\lambda_d} & \Big( \sum_{\phi'} \Sigma_{0,\phi'}^{\lambda_d} \pm \Sigma_{S,\sigma\text{-S}}^{\lambda_d}\Big) \scr D_{\kappa m,\kappa'm}^{\lambda_d0}, \\
    \Pi_{\pm}^{\lambda_d} =&\sum_\mu \left(\Sigma_{T,\rho\text{-T}}^{\lambda_d\mu} + \Sigma_{T,\rho\text{-VT}}^{\lambda_d\mu}\right)\cals Q_{\pm\kappa m,\mp\kappa'm}^{\lambda_d\mu\sigma}.
\end{align}
In the above expressions, $\phi'$ corresponds to the vector ($\omega$-V, $\rho$-V and $A$-V) and $\rho$-VT coupling channels, and the rearrangement terms $\Sigma_R$ due to the density dependencies of the coupling strengths. The details of $\scr D$ and $\cals Q$ symbols are given in Appendix \ref{sec:APP-A-symbols}.

For the non-local term $h_{aa'}^{m,\text{E}}$ that contains the contributions of the Fock terms, it can be written in a compact form as,
\begin{equation}
  \begin{split}
    h_{aa'}^{m,\text{E}} = \sum_{\phi} \int &drdr' \begin{pmatrix} G_a & F_a \end{pmatrix}_r   \\
    & \times \begin{pmatrix} Y_{G,\pi m}^{\kappa\kappa',\phi} & Y_{F,\pi m}^{\kappa\kappa',\phi} \\ X_{G,\pi m}^{\kappa\kappa',\phi} & X_{F,\pi m}^{\kappa\kappa',\phi} \end{pmatrix}_{r,r'} \begin{pmatrix} G_{a'} \\ F_{a'}\end{pmatrix}_{r'},
  \end{split}
\end{equation}
in which the details of the non-local mean fields $X_G$, $X_F$, $Y_G$ and $Y_F$ can be found in Appendix \ref{subsec:Fock-rhot} and \ref{subsec:Contact-rhot}, and Appendix A 2 of Ref. \cite{Geng2020PRC101.064302}. Similar as the non-local Fock term $h_{aa'}^{m,\text{E}}$, using the derived pairing potential (\ref{eq:PairingP}), the pairing term $\Delta_{aa'}^{m}$ can be written as
\begin{equation}
  \begin{split}
    \Delta_{aa'}^{m} =  \int dr&dr'\begin{pmatrix}
    G_{a} & F_{a} \end{pmatrix}_r  \\
    & \times\begin{pmatrix}\Delta_{m\kappa\kappa' }^{++} & \Delta_{m\kappa\kappa' }^{+-}\\[0.5em] \Delta_{m\kappa\kappa' }^{-+} & \Delta_{m\kappa\kappa'}^{--}   \end{pmatrix}_{r,r'} \begin{pmatrix} G_{a'} \\[0.5em] F_{a'} \end{pmatrix}_{r'},
  \end{split}
\end{equation}
where the terms $\Delta_{m\kappa\kappa'}^{\pm\pm}$ are given in Eq. (\ref{eq:PairingP}).

In addition to the energy functional (\ref{eq:Full-E}), the corrections on the center-of-mass (CoM) motion is introduced as usual. The CoM corrections are evaluated in a microscopic way \cite{Bender2000EPJA7.467,Long2004PRC69.034319} as,
\begin{equation}\label{eq:CoM}
    E_{\text{c.m.}} = -\frac{1}{2AM} \lrcl{\text{HFB}} \hat{\svec P}_{\text{c.m.}}^2 \lrlc{\text{HFB}},
\end{equation}
where $A$ is the nuclear mass number and $\svec P_{\text{c.m.}} = \sum_i \svec p_i$ is the CoM momentum. It shall be mentioned that the CoM corrections are not involved in the variation of the energy functional. For completeness, we also give the details of the CoM corrections in Appendix \ref{sec:APP-C}.

\subsection{Analysis of the coupling channels}\label{sec:channels}
In order to have a better understanding on the deformation effects, particularly the evolution of the shell structure with respect to the deformation, it is worthwhile to perform a qualitative analysis on the nature of the meson-nucleon coupling channels. In the following, we focus on the $\pi$-PV and $\rho$-T coupling channels, and the analysis is carried out using the momentum representation for convenience.

In the momentum space, the energy functional contributed by the Fock diagrams of the $\pi$-PV and $\rho$-T couplings can be expressed as,
\begin{align}
    E_{\pi\text{-PV}} =& -\ff2 \frac{1}{(2\pi)^6} \frac{f_\pi^2}{m_\pi^2} \int d\svec pd\svec p' \cals T_{\tau\tau'} \bar u(\svec p) \bar u(\svec p') \nonumber \\
    &\times \big( \gamma_5 \svec\gamma \cdot \svec q \big) \big( \gamma_5 \svec\gamma \cdot \svec q \big)' \frac{1}{m_\pi^2 +\svec q^2} u(\svec p) u(\svec p'), \label{eq:Pi-PV-E}\\
    E_{\rho\text{-T}} =& -\ff2 \frac{1}{(2\pi)^6} \frac{f_\rho^2}{(2M)^2} \int d\svec p d\svec p' \cals T_{\tau\tau'} \bar u(\svec p) \bar u(\svec p') \nonumber \\
    &\times \big( \sigma_{i\nu} q^i \big) \big( \sigma^{k\nu}q_k \big)' \frac{1}{m_\rho^2 +\svec q^2} u(\svec p) u(\svec p'),\label{eq:rho-T-E}
\end{align}
where $\cals T_{\tau\tau'}=2-\delta_{\tau\tau'}$, and the momentum transfer $\svec q = \svec p-\svec p'$. In the above contributions, the zero-range parts are cancelled by adding the following contact terms,
\begin{align}
    E_{\pi\text{-PV}}^\delta = +\ff6 \frac{1}{(2\pi)^6} & \frac{f_\pi^2}{m_\pi^2} \int d\svec pd\svec p' \cals T_{\tau\tau'} \bar u(\svec p) \bar u(\svec p') \nonumber \\
     &\times \big( \gamma_5\svec\gamma \big) \cdot \big( \gamma_5\svec\gamma\big)' u(\svec p) u(\svec p'),\label{eq:Pi-PV-C}\\
    E_{\rho\text{-T}}^\delta = +\ff6 \frac{1}{(2\pi)^6} & \frac{f_\rho^2}{(2M)^2} \int d\svec pd\svec p' \cals T_{\tau\tau'} \bar u(\svec p)\bar u(\svec p') \nonumber \\
     &\times \big(\sigma_{\mu i}\big)\big(\sigma^{\mu i}\big)' u(\svec p) u(\svec p'). \label{eq:rho-T-C}
\end{align}

For the $\pi$-PV coupling, the non-relativistic reduction leads to the vertex of the tensor force component as,
\begin{align}
    V_{T}^\pi =& (\svec\sigma\cdot\svec q)(\svec\sigma'\cdot\svec q) - \ff3 \left(\svec\sigma\cdot\svec\sigma'\right) \svec q^2 \nonumber \\
    =&\frac{4}{3} S^2 q^2 \sqrt{\frac{4\pi}{5}}Y_{20}(\vartheta,\varphi), \label{eq:Pi-PV}
\end{align}
where the total spin $\svec S = \svec s+\svec s'$, and $\vartheta$ denotes the angle between $\svec S$ and transferring momentum $\svec q$. It shall be mentioned that in the above expressions, the term $(\svec\sigma\cdot\svec q)(\svec\sigma'\cdot\svec q)$ originates from the $\pi$-PV coupling (\ref{eq:rho-T-E}), whereas the term $\left(\svec\sigma\cdot\svec\sigma'\right) \svec q^2$ does not simply correspond to the contact term (\ref{eq:Pi-PV-C}) because
\begin{equation}
  \frac{(\svec\sigma\cdot\svec\sigma')\svec q^2}{m_\pi^2 +\svec q^2} = (\svec\sigma\cdot\svec\sigma') - (\svec\sigma\cdot\svec\sigma') \frac{m_\pi^2}{m_\pi^2 +\svec q^2},
\end{equation}
in which the first term in the right-hand side corresponds to the contact term (\ref{eq:Pi-PV-C}), and the second term is introduced to have the tensor-type vertex (\ref{eq:Pi-PV}).

For the time component ($\mu=0$) of the $\rho$-T coupling (\ref{eq:rho-T-E}), one can obtain similar expression as Eq. (\ref{eq:Pi-PV}) but with opposite sign, following the procedure as deriving the $V_T^\pi$ of the $\pi$-PV coupling. It shall be stressed that the derived expression corresponds to the coupling between the upper and lower components of Dirac spinor, in contrast to the $\pi$-PV coupling.

For the space components of the $\rho$-T coupling, the situation is more complicated, and the vertex $\big(\sigma_{jl} q^l\big) \big(\sigma'^{jk} q_k\big)$ can be expressed as,
\begin{equation}
    -\begin{pmatrix}\big( \svec\sigma\times \svec q \big)\cdot\big(\svec\sigma'\times \svec q\big) & 0 \\[0.5em] 0 & \big(\svec\sigma\times \svec q \big)\cdot\big(\svec\sigma'\times \svec q\big)   \end{pmatrix}.
\end{equation}
Similarly as deriving Eq. (\ref{eq:Pi-PV}), one can deduce the following results for the space component of the $\rho$-T coupling as,
\begin{equation}
\begin{split}
-\Big[\left( \svec\sigma\times \svec q \right)\cdot\left( \svec\sigma'\times \svec q \right) &- \frac{2}{3}\left( \svec\sigma\cdot\svec\sigma'\right) \svec q^2\Big]  \\
    =&+\frac{4}{3} S^2 q^2 \sqrt{\frac{4\pi}{5}} Y_{20}(\vartheta,\varphi).
\end{split}\label{eq:rho-T}
\end{equation}
Notice that for the term $2/3(\svec\sigma\cdot\svec\sigma')\svec q^2$, the factor $2$ originates from the contact term (\ref{eq:rho-T-C}).

It can be seen that the $\pi$-PV and the space component of the $\rho$-T couplings lead to similar results, see Eqs. (\ref{eq:Pi-PV}) and (\ref{eq:rho-T}). However, it does not correspond to the similar tensor effects, because $\vartheta$ and $\varphi$ represent the angles between the vectors $\svec S$ and $\svec q$. Qualitatively, considering $\svec s = \ff2\svec\sigma$ and $ \svec s ' = \ff2\svec\sigma'$, both Eqs. (\ref{eq:Pi-PV}) and (\ref{eq:rho-T}) lead to opposite contributions for the cases of $\svec s = \svec s'$ and $\svec s = -\svec s'$. In addition, the ones of the $\pi$-PV coupling [Eq. (\ref{eq:Pi-PV})] are opposite to those of the $\rho$-T space components [Eq. (\ref{eq:rho-T})], being consistent with the opposite tensor effects. It shall be stressed that such analysis is rather rough, and the situation in realistic nuclei can be much more complicated.

Despite the tensor effects, the expressions (\ref{eq:Pi-PV}) and (\ref{eq:rho-T}) can still help us to understand the effects of the $\pi$-PV and $\rho$-T couplings in finite nuclei, at least qualitatively. For the convenience of  discussions, we show the two-body interaction using the $L$-$S$ coupling scheme as,
\begin{equation}\label{eq:LS-scheme}
 \lrcl{(\textstyle{\ff2\ff2})S,(l_1l_2)L; JM} V_\phi \lrlc{(\textstyle{\ff2\ff2})S',(l_1'l_2')L'; J'M'},
\end{equation}
in which $V_\phi$ represents the coupling channels, and $\ff2$ and $l$ are respectively the spin and orbital angular momentum of the expansion terms of the wave functions $\psi_{\nu\pi m}^V$ and $\psi_{\nu\pi m}^U$, see Eq. (\ref{eq:expansion-wav}). Because of the spherical harmonics function $Y_{20}$ in Eqs. (\ref{eq:Pi-PV}, \ref{eq:rho-T}), it can be deduced that $L = |L'-2|, \cdots, L'+2$. Qualitatively speaking, compared with the scalar ($\sigma$-S) and vector ($\omega$-V, $\rho$-V and $A$-V) couplings, more expansion terms can be involved, which may lead to tight correlations between the $\pi$-PV/$\rho$-T couplings and the deformation effects. In fact, notable enhancement with the respect to the deformation from the $\pi$-PV coupling has been manifested in $^{20}$Ne \cite{Geng2020PRC101.064302}.

\section{RESULTS AND DISCUSSIONS}\label{sec:RESULTS AND DISCUSSIONS}

For the newly developed D-RHFB model in this work, we implement the $\rho$-T and $\rho$-VT couplings, in addition to using Bogoliubov scheme to deal with the pairing correlations. For the first attempts, we select the light nucleus $^{24}$Mg and the mid-heavy one $^{156}$Sm as the candidates, both of which are potentially deformed. The D-RHFB calculations are performed by utilizing the RHF Lagrangians PKA1 \cite{Long2007PRC76.034314}, PKO1 \cite{Long2006PLB640.150} and PKO2 \cite{Long2008EPL82.12001}, as compared to the newly proposed RMF one DD-LZ1 \cite{Wei2020CPC44.074107}. It is worthwhile to mention that in addition to the ones considered in PKO2 and DD-LZ1, PKA1 contains the degrees of freedom associated with the $\pi$-PV and $\rho$-T couplings, and the $\pi$-PV coupling is included in PKO1.



In determining the spherical DWS base, the spherical Dirac equations are solved by setting the spherical box size as 20 fm with radial mesh step 0.1 fm. Compared to the D-RHF model with BCS pairing \cite{Geng2020PRC101.064302}, the implemented  $\rho$-T and $\rho$-VT couplings notably increase the numerical complexity, and the calculations become even more time consuming.

\subsection{Space truncations and Convergence check}
In order to provide reliable description of nuclei, particularly for the weakly bound unstable nuclei, it is necessary to perform the convergence check with respect to the space truncations for the newly developed D-RHFB model. Similar to the D-RHF model, there exist several space truncations, namely the maximum value of the angular momentum projection $m$, the expansion terms (denoted by $\lambda_p$) of the density-dependent coupling strengths, and the configuration space of the spherical DWS base, namely the $n$ and $\kappa$ quantities in Eqs. (\ref{equ:C}) and (\ref{eq:expansion-wav}). Fortunately, one does not need to care much about the pairing window because of the mentioned advantage of the finite-range Gogny pairing force before.

For the maximum value of $m$, referred as $m_{\max}$, it depends on the specific nucleus, which can be determined referring to the fundamental s.p. level scheme given by the traditional shell model. Here the $m_{\max}$-value is selected as $11/2$ for $^{24}$Mg and $17/2$ for $^{156}$Sm, both of which are large enough for selected nuclei. For the expansion terms of the density-dependent coupling strengths, see Eq. (46) in Ref. \cite{Geng2020PRC101.064302}, we consider five terms $\lambda_p=0,2,4,6,8$ for both $^{24}$Mg and $^{156}$Sm, which is accurate enough. Combined with the following discussions, it can be deduced that the numerical cost will be doubly enlarged with larger $m_{\max}$ and/or more $\lambda_p$-terms.

For the space truncation of the spherical DWS base, i.e., the cutoff on the principle number $n$ and $\kappa$-quantity, it depends on the specific $m$-value as seen from the expansions (\ref{equ:C}) or (\ref{eq:expansion-wav}). The detailed cutoff on the $\kappa$-quantity has been described in Ref. \cite{Geng2020PRC101.064302}, here recalled as follows for completeness. For the orbits with the maximum $m$-value $m_{\max}$, the number of $\kappa$-blocks in the expansion (\ref{eq:expansion-wav}) is set as $K_{m_{\max}}$, giving the maximum absolute $\kappa$ value as $k_{\max} = m_{\max} + K_{m_{\max}}-1/2$. Then for the orbits with arbitrary $m$, the $\kappa$-blocks involved in Eq. (\ref{eq:expansion-wav}) read as $|\kappa| = m+1/2, m+3/2, \cdots, k_{\max}$ and $\text{sign}(\kappa) = \pi*(-1)^{|\kappa|}$, $\pi=\pm$ for positive/negative parity. Here we selected $K_{m_{\max}}=4$ with $m_{\max}=11/2$ for $^{24}$Mg and $K_{m_{\max}}=3$ with $m_{\max}=17/2$ for $^{156}$Sm, and such selections are tested to be accurate enough.

For the cutoff over the principle number $n$ in the expansion (\ref{equ:C}), the maximum $n$-values for each $\kappa$-block are determined by the given energy cutoff. In expanding the spinors $\psi^U$ and $\psi^V$, both positive and negative energy solutions of the spherical Dirac equation shall be considered to keep the completeness of the expansion. One should not mix this with the no-sea approximation adopted in the mean field approach, which corresponds to neglecting the Dirac sea in calculating the densities/currents. The energy cutoff read as $E_\pm^C$, corresponding the positive ($+$) and negative ($-$) energy cutoff in the spherical DWS base. Namely, the states with positive (negative) energies $E$, that $E-M<E_+^C$ ($E+M>E_-^C$), are considered in the expansion (\ref{equ:C}). One may notice that it is different from Ref. \cite{Geng2020PRC101.064302}, in which the maximum $n$-values, reading as $N_D$ and $N_F$ therein, are simply fixed as the same large enough values for all $\kappa$-blocks. In general, preserving similar numerical accuracy, the energy cutoff utilized in this work leads to smaller configuration space for the spherical DWS base, and the numerical cost can be also reduced by some amount.

\begin{figure}[hbpt]
  \centering
  \includegraphics[width=0.48\textwidth]{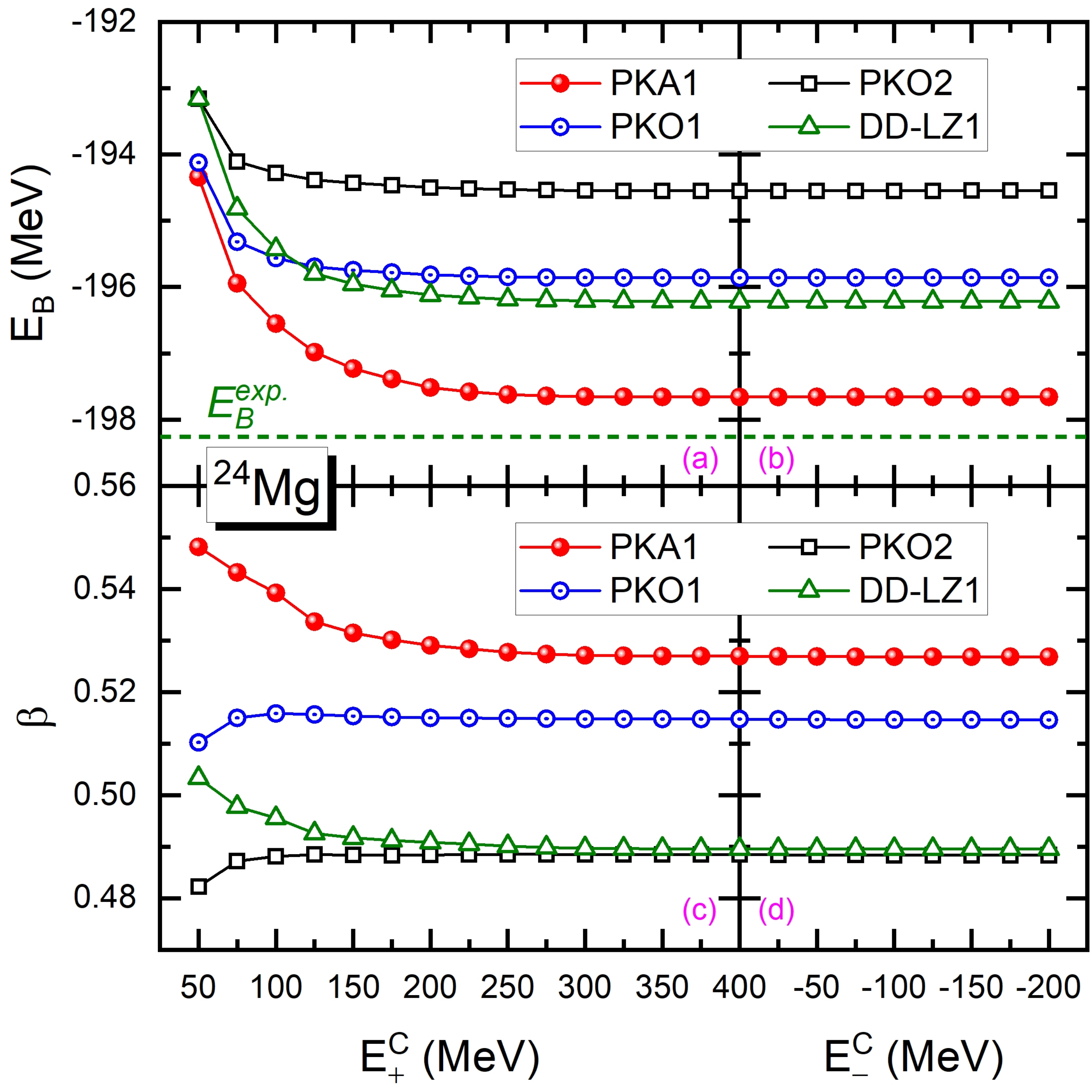}
  \caption{(Color Online) Binding energy $E_B$ (MeV) [plots (a) and (b)] and quadruple deformation $\beta$ [plots (c) and (d)] for $^{24}$Mg with respect to the positive ($+$) and negative ($-$) energy cutoff $E_\pm^C$ (MeV) in expanding the spinors $\psi^U$ and $\psi^V$. The results are calculated by PKA1, PKO1, PKO2 and DD-LZ1 with the initial deformation $\beta_0=0.4$. The marked experimental $E_B$-value reads as $-$198.26 MeV \cite{Wang2017CPC41.030003}.}\label{Fig:Mg24-EB}
\end{figure}

\begin{figure}[hbpt]
  \centering
  \includegraphics[width=0.48\textwidth]{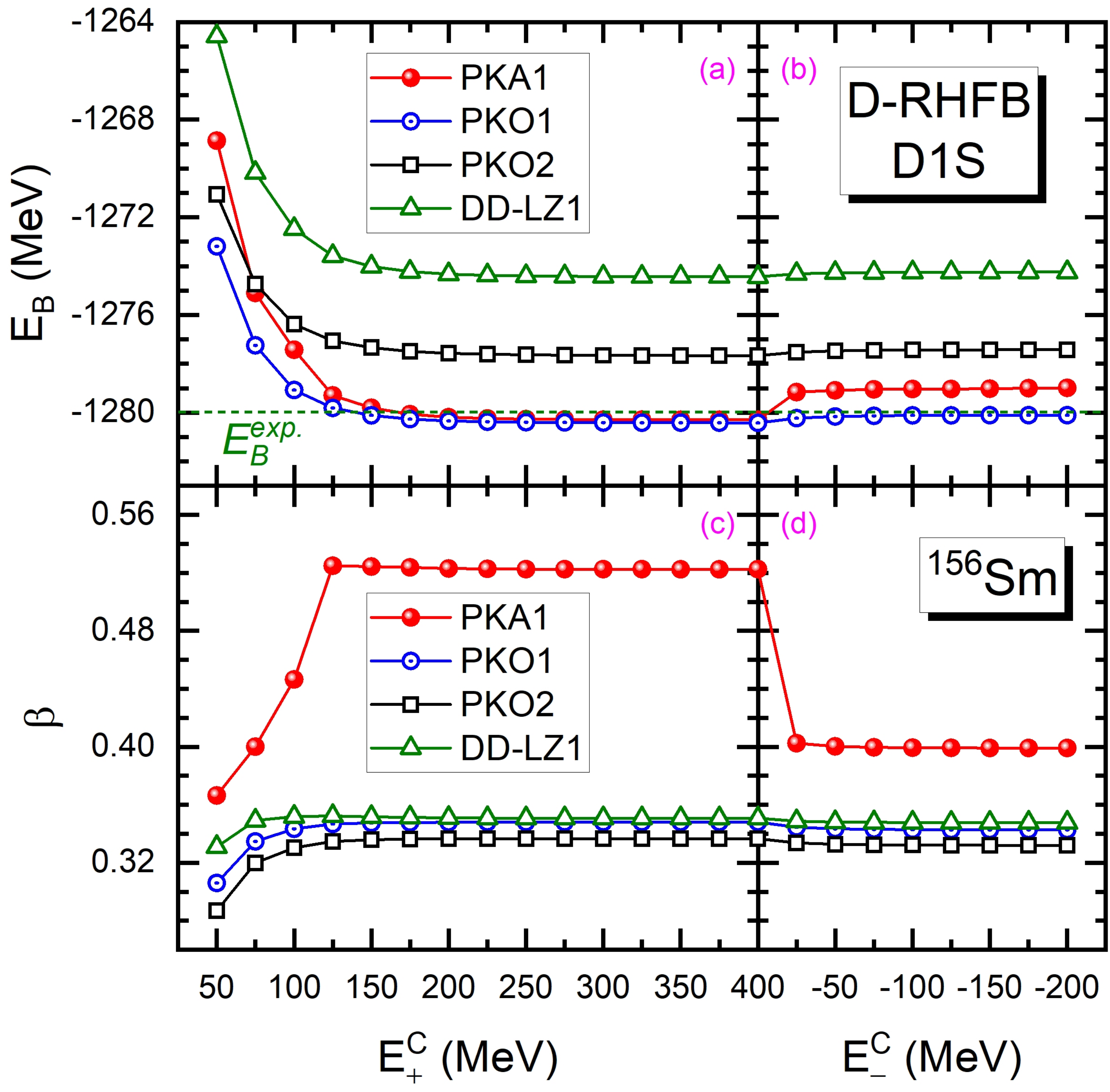}\\
  \caption{(Color Online) Same as Fig. \ref{Fig:Mg24-EB}, but for $^{156}$Sm. The marked experimental $E_B$-value reads as $-$1279.98 MeV \cite{Wang2017CPC41.030003}. }\label{Fig:Sm156}
\end{figure}

Figures \ref{Fig:Mg24-EB} and \ref{Fig:Sm156} show the convergence tests respectively for $^{24}$Mg and $^{156}$Sm, and the calculations are performed with the initial deformation $\beta_0=0.4$, using the RHF Lagrangians PKA1, PKO1 and PKO2, and the RMF one DD-LZ1. In both figures, plots (a) and (c) present the convergence with respect to $E_+^C$ respectively for the binding energy $E_B$ (MeV) and quadruple deformation $\beta$, in which $E_-^C$ is fixed as $0$ MeV, and plots (b) and (d) show the results with respect to $E_-^C$, in which $E_+^C$ is fixed as $400$ MeV. As seen from Figs. \ref{Fig:Mg24-EB} and \ref{Fig:Sm156}, both binding energy $E_B$ [plots (a)] and deformation $\beta$ [plots (c)] tend to be converged, when $E_+^C>250$ MeV for the light $^{24}$Mg and $E_+^C>200$ MeV for the mid-heavy $^{156}$Sm. With more negative energy states are involved in the expansion (\ref{equ:C}), the results for $^{24}$Mg remain almost unchanged for all the selected effective Lagrangians, see plots (b) and (d) in Fig. \ref{Fig:Mg24-EB}.

However, the situation changes for the mid-heavy $^{156}$Sm, particularly for the RHF Lagrangian PKA1. From the left panels to right panels of Fig. \ref{Fig:Sm156}, one can see slight but visual changes on both binding energy $E_B$ and quadruple deformation $\beta$ given by PKO1, PKO2 and DD-LZ1. On the contrary, notable changes are surprisingly found on the values of $E_B$ and $\beta$ given by PKA1. It is also interesting to see that the results are quickly converged with slightly more negative $E_-^C$-values.

Combined with Figs. \ref{Fig:Mg24-EB} and \ref{Fig:Sm156}, one may conclude that the convergence test is no doubt necessitated for the positive energy cutoff $E_+^C$ for both light and heavy nuclei, and $E_+^C\geqslant350$ MeV can be accurate enough.  Concerning the negative energy cutoff, negative $E_-^C$-value means unbound negative energy states of the spherical DWS base getting involved in expanding the $\psi^V$ and $\psi^U$. For the light nuclei, here $^{24}$Mg, the results remain almost unchanged with more negative $E_-^C$-value, as shown in Fig. \ref{Fig:Mg24-EB}. While for mid-heavy and heavy nuclei, it becomes necessary to have more negative energy states in the expansion (\ref{equ:C}), particularly for the RHF Lagrangian PKA1, and $E_-^C\leqslant-100$ MeV can be accurate enough for selected $^{156}$Sm. For heavier nuclei, such as the superheavy ones, a careful convergence test with respect to $E_-^C$ is deserved for reliable calculations. In fact, similar tests under the RMF scheme were discussed in Ref. \cite{Zhou2003PRC68.034323}, and similar trend with respect to $E_-^C$ were found, despite less notable than PKA1.

\subsection{Significance of the negative energy states of the DWS base for mid-heavy nucleus $^{156}$Sm}

\begin{figure*}[htbp]
  \centering
  \includegraphics[width=0.48\textwidth]{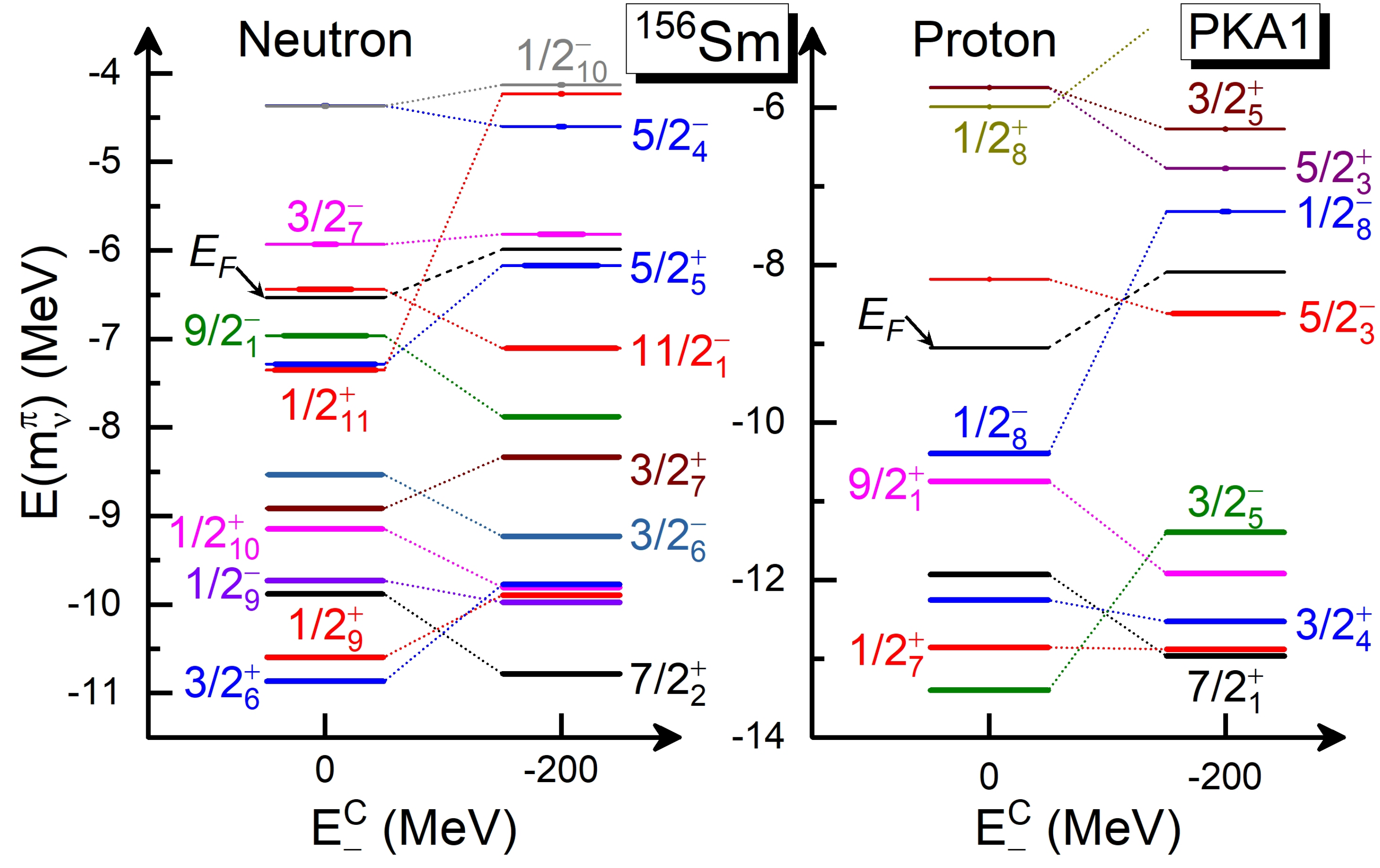}
  \includegraphics[width=0.48\textwidth]{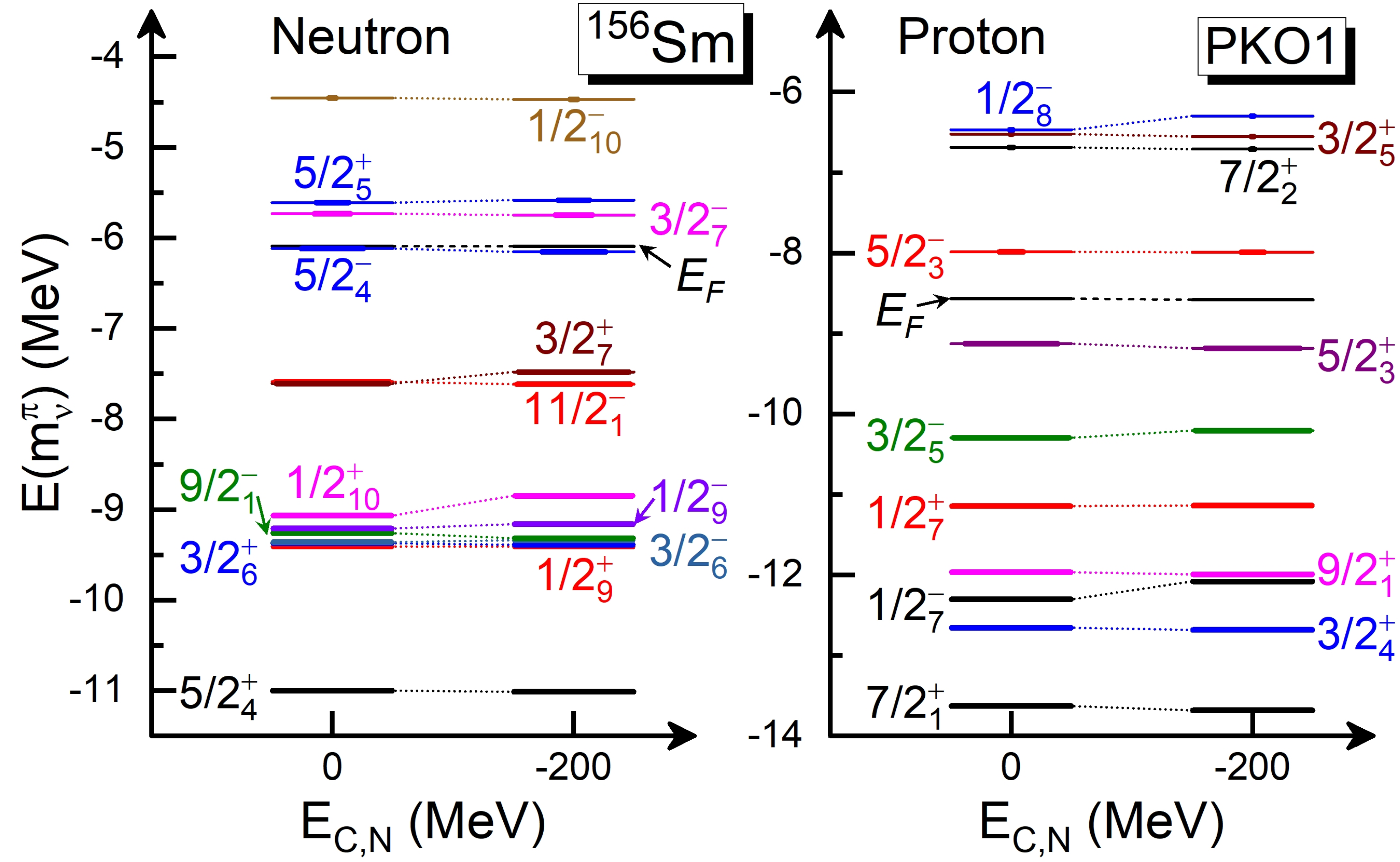}
  \caption{(Color Online) Neutron and proton spectra of $^{156}$Sm given by PKA1 (left plots) and PKO1 (right plots) with negative energy cutoff $E_-^C = 0$ MeV and $-200$ MeV, and the positive one $E_+^C=400$ MeV. The ultra thick bars represent the occupation probabilities of the orbits $m_\nu^\pi$ and $E_F$ denotes the Fermi levels. }\label{Fig:LEV-Sm152}
\end{figure*}

In order to understand the notable changes from zero $E_-^C$-value to negative ones in Fig. \ref{Fig:Sm156} on the quadruple deformation $\beta$ and binding energy $E_B$, we show the neutron and proton canonical single particle spectra of $^{156}$Sm in Fig. \ref{Fig:LEV-Sm152}, and the results are calculated by PKA1 (left plots) and PKO1 (right plots) with $E_-^C=0$ and $-200$ MeV. In Fig. \ref{Fig:LEV-Sm152}, $E_{\text{F}}$ is used to denote the Fermi levels and $m_\nu^\pi$ for the canonical single particle orbits.

Let's firstly focus on the results given by PKA1, namely the left two plots of Fig. \ref{Fig:LEV-Sm152}. For the results with $E_-^C =0$ MeV, which indicates only the bound negative energy states of the spherical DWS base are involved in expanding $\psi^V$ and $\psi^U$ [Eq. (\ref{equ:C})], it can be seen that the neutron orbit $1/2_{11}^+$ and the proton one $1/2_8^-$ are almost fully occupied. On the contrary, the neutron orbit ${11/2}_1^-$ with rather large $m$-value is only partly occupied, and the proton one $5/2_3^-$ is almost empty. With more negative energy states of the spherical DWS base involved in the expansion (\ref{equ:C}), i.e., the results with $E_-^C=-200$ MeV in Fig. \ref{Fig:LEV-Sm152}, the neutron $11/2_1^-$ and proton $5/2_3^-$ orbits become deeper bound and fully occupied, whereas few particles populate on much less bound neutron $1/2_{11}^+$ and proton $1/2_8^-$ orbits. Similar as $11/2_1^-$, the neutron orbit $9/2_1^-$ also becomes deeper bound and fully occupied from zero $E_-^C$ value to $-200$ MeV.

Compared to the notable changes given by PKA1 from zero $E_-^C$ value to $-200$ MeV, the results given by PKO1 are not significantly changed, see the right two plots in Fig. \ref{Fig:LEV-Sm152}. Even though, with $E_-^C=-200$ MeV, both PKA1 and PKO1 give similarly populated neutron/proton orbits around the Fermi levels, while with different ordering.  In order to better understand the systematics, Table \ref{Tab:Q2-Sm152} shows the quadruple momenta $Q_2$ (fm$^{-2}$) of the critical neutron (upper panel) and proton (lower panel) orbits, whose occupations given by PKA1 are remarkably changed from zero $E_-^C$ value to $-200$ MeV. As seen from Table \ref{Tab:Q2-Sm152}, the neutron orbits $9/2_1^-$ and $11/2_1^-$ carry evident oblate nature with rather negative $Q_2$-values, and the others are of prolate distributions. Combined with the results in the left two plots of Fig. \ref{Fig:LEV-Sm152} and Table \ref{Tab:Q2-Sm152}, it is obvious that the empty prolate neutron orbit $1/2_{11}^+$ and proton one $1/2_8^-$, as well as the full-occupied oblate neutron orbits $9/2_1^-$ and $11/2_1^-$ and proton one $5/2_3^-$, account for reduced quadruple deformation $\beta$ of $^{156}$Sm in the PKA1 results, see Fig. \ref{Fig:Sm156}. It is also worthwhile to mention that the $Q_2$-values given by PKA1 for the occupied prolate neutron orbit $5/2_5^+$ and proton one $5/2_3^-$ are substantially reduced as well, being consistent with the deformation reduction from zero $E_-^C$-value to $-200$ MeV. In contrast to that, the results given by PKO1 remain nearly unchanged, similar as the systematics shown in Fig. \ref{Fig:Sm156}.

\begin{table}[htbp]
\caption{Quadruple momentum $Q_{2}$ (fm$^{-2}$) of neutron ($N$) orbits $9/2_{1}^-$, $11/2_{1}^-$, $1/2_{11}^+$ and $5/2_5^+$, and proton ($P$) ones $5/2_{3}^-$ and $1/2_{8}^-$. These results are calculated with PKA1 and PKO1 by selecting the negative energy cutoff $E_-^C$ respectively as $0$ MeV and $-200$ MeV and $E_+^C=400$ MeV. } \renewcommand{\arraystretch}{1.5}\setlength{\tabcolsep}{0.7em}\label{Tab:Q2-Sm152}
\begin{tabular}{c|c|rr|rr}  \hline\hline
 \multicolumn{2}{c|}{} & \multicolumn{2}{c|}{PKO1} & \multicolumn{2}{c}{PKA1}  \\ \hline
\multicolumn{2}{c|}{$E_-^C$}           &   0~~~~       &   $-$200~~&      0~~~~ &    $-$200~~\\ \hline
\multirow{4}{*}{$N$}& 9/2$_{1}^-$  &  $-$6.184     &  $-$6.155 &  $-$5.405  &  $-$5.980 \\ 
&11/2$_{1}^-$ & $-$22.940     & $-$22.913 & $-$24.216  & $-$24.107 \\ 
&1/2$_{11}^+$ &    57.140     &    56.775 &    53.972  &    64.870 \\ 
&5/2$_{5}^+$  &    32.564     &    32.230 &    41.796  &    34.599 \\ \hline \hline
\multirow{2}{*}{$P$}&5/2$_{3}^-$  &    23.941     &    23.715 &    35.173  &    27.843 \\ 
&1/2$_{8}^-$  &    41.475     &    40.874 &    54.807  &    48.624 \\ \hline\hline
\end{tabular}
\end{table}

As we mentioned before, the negative energy states of the spherical DWS base shall be considered to keep the expansion completeness of the wave functions $\psi^U$ and $\psi^V$, although the relevant expansion coefficients are rather small. As seen from Fig. \ref{Fig:Sm156}, it can be concluded that the results given by PKA1 seem more sensitive to the negative energy cutoff $E_-^C$, as compared to the other selected effective Lagrangians. In fact, such notable model difference can be understood qualitatively from the role of the $\rho$-T coupling. According to the analysis in Sec. \ref{sec:channels}, the $\rho$-T coupling, as well as the $\pi$-PV coupling, can get more expansion terms of the wave functions to contribute in the two-body interactions, particularly for the ones with large $|\kappa|$-values whose couplings lead to large $L$/$L'$ values. For the mentioned neutron orbits $9/2_1^-$ and $11/2_1^-$, which are dominated by the large $|\kappa|$-value blocks, it is not enough to keep the completeness with $E_-^C=0$ MeV, because only several bound negative energy states of the DWS base can contribute to the expansion (\ref{equ:C}) for the relevant $\kappa$-blocks. Compared to the RMF Lagrangians, as well as the RHF ones PKO$i$ ($i=1,2,3$), the strong $\rho$-T coupling carried by PKA1 enhances the correlations between the terms with large $|\kappa|$-values, and thus the completeness of the relevant $\kappa$-blocks is essential for the reliable description with PKA1. It is also worthwhile to remind that the $\pi$-PV couplings in PKO1 and PKO3 \cite{Long2008EPL82.12001} are much weaker than the $\rho$-T coupling in PKA1, which may interpret the fact that PKO1 gives similar systematics with respect to $E_-^C$-value as PKO2 and DD-LZ1.

\subsection{Description of light nucleus $^{24}$Mg}

After the convergence check, we performed the D-RHFB calculations for the light nucleus $^{24}$Mg, using the RHF Lagrangians PKA1, PKO1 and PKO2, and the RMF one DD-LZ1. Table \ref{Tab:Bulk-24Mg} shows the binding energy $E_B$ (MeV), quadruple deformation $\beta$ and charge radius $r_c$ (fm) for $^{24}$Mg. The local minima are obtained by the self-consistent D-RHFB calculations with different initial deformation, in which the energy cutoff of the spherical DWS base are selected as $E_+^C = 350$ MeV and $E_-^C=-100$ MeV with $m_{\max}=11/2$ and $K_{m_{\max}}=4$.

From Table \ref{Tab:Bulk-24Mg}, it is found that all the selected effective Lagrangians present rather large prolate deformation ($\beta\sim 0.5$) for the ground state of $^{24}$Mg. For the binding energy $E_B$ of the ground state, it is well reproduced by the selected effective Lagrangians, among which PKA1 shows the best agreement with the data \cite{Wang2017CPC41.030003}, and less good by the new developed RMF Lagrangian DD-LZ1. Besides, one can also find some systematics that may be related with the role played by the $\pi$-PV and $\rho$-T couplings. Specifically, from PKO2 to PKO1 that contains the $\pi$-PV coupling, both binding energy $E_B$ and deformation $\beta$ increase much, and to PKA1 that takes both $\pi$-PV and $\rho$-T couplings into account, the values of $E_B$ and $\beta$ are further enlarged. Compared to the RHF Lagrangians, DD-LZ1 gives similar deformation as PKO2, but with larger $E_B$-value than PKO1 and PKO2, which might be due to improved in-medium balance between the nuclear attractions and repulsions as revealed in Refs. \cite{Geng2019PRC100.051301R, Wei2020CPC44.074107}.

\begin{table}[htbp]
\caption{Binding energy $E_B$ (MeV), quadruple deformation including the total ($\beta$), neutron ($\beta_n$) and proton ($\beta_p$) ones, and charge radius $r_c$ (fm) at various local minima of $^{24}$Mg, calculated by PKA1, PKO1, PKO2 and DD-LZ1. The experimental $E_B$-value reads as $-$198.26 MeV \cite{nudatWang2017CPC41.030003}. } \renewcommand{\arraystretch}{1.5}\setlength{\tabcolsep}{0.4em}\label{Tab:Bulk-24Mg}
\begin{tabular}{l|r|rrr|r}  \hline\hline
                        &   $E_B$               &  $\beta$  & $\beta_n$ & $\beta_p$ &  $r_c$  \\ \hline
\multirow{3}{*}{PKA1}   &   \textbf{$-$197.656} &     0.527 &     0.522 &     0.532 &  2.990  \\
                        &   $-$193.580          &  $-$0.766 &  $-$0.758 &  $-$0.774 &  3.188  \\
                        &   $-$190.569          &  $-$0.306 &  $-$0.302 &  $-$0.311 &  3.004  \\ \hline
\multirow{2}{*}{PKO1}   &   \textbf{$-$195.857} &     0.515 &     0.509 &     0.520 &  2.972  \\
                        &   $-$190.364          &  $-$0.267 &  $-$0.263 &  $-$0.271 &  2.935 \\ \hline
\multirow{2}{*}{PKO2}   &   \textbf{$-$194.544} &     0.488 &     0.483 &     0.494 &  2.959  \\
                        &   $-$190.089          &  $-$0.185 &  $-$0.183 &  $-$0.188 &  2.902 \\ \hline
\multirow{2}{*}{DD-LZ1} &   \textbf{$-$196.214} &     0.489 &     0.484 &     0.495 &  2.961  \\
                        &   $-$189.847          &  $-$0.248 &  $-$0.245 &  $-$0.252 &  2.948 \\ \hline\hline
\end{tabular}
\end{table}

Moreover, all the selected Lagrangians give the local minimum at $\beta\sim(-0.18, -0.31)$ with similar binding energies for $^{24}$Mg. Different from the others, PKA1 provides an additional minimum with rather large oblate deformation $\beta = -0.766$, which corresponds to the second minimum in the PKA1 results. In fact, not shown in Table \ref{Tab:Bulk-24Mg}, the shape constrained calculations with PKO1 also present a weak local minimum at similar oblate deformation as PKA1, but with rather high-lying binding energy. Similar minimum has been also found in the calculations of $^{20}$Ne with PKO1 and PKO3 \cite{Geng2020PRC101.064302}. Combined with the systematics from PKO2 to PKO1 and further to PKA1 in describing the ground state of $^{24}$Mg, the second minimum with large oblate deformation given by PKA1 can be taken as another evidence that the $\rho$-T coupling, as well as the $\pi$-PV coupling \cite{Geng2020PRC101.064302}, enhances the deformation effects.

\begin{figure}[hbpt]
  \centering
  \includegraphics[width=0.48\textwidth]{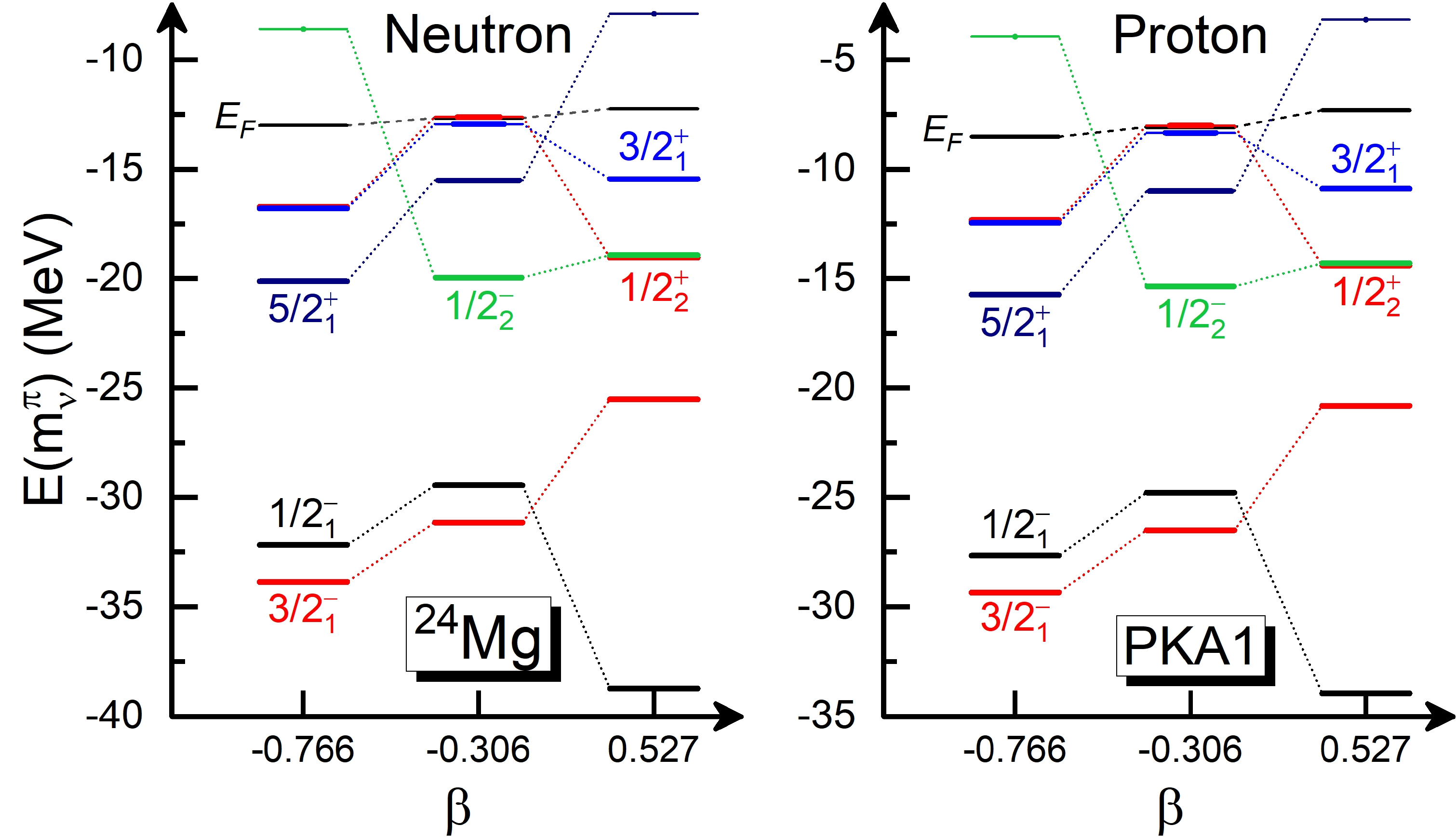}
  \caption{(Color Online) Neutron (left panel) and proton (right panel) spectra of the local minimum (denoted by $\beta$) of $^{24}$Mg, calculated by PKA1 with $E_+^C = 350$ MeV and $E_-^C=-100$ MeV. The ultra thick bars represent the occupation probabilities of the orbits $m_\nu^\pi$ and $E_F$ denotes the Fermi levels. }\label{Fig:LEV-Mg24}
\end{figure}

As an implemented illustration, Fig. \ref{Fig:LEV-Mg24} shows the canonical neutron (left panel) and proton (right panel) spectra given by PKA1 for $^{24}$Mg at the local minimum $\beta=-0.766$, $-0.306$ and $0.527$, in which the ultra thick bars denote the occupation probabilities and $E_F$ for the Fermi levels. Before detailed discussions, it is worthwhile mentioning the general deformation behaviors on nuclear structure. Specifically, branching from the spherical $j$-orbit, here denoted by the dominant expansion terms of the spherical DWS base, the orbits with small $m$-values becomes deeper bound and those with large $m$-values tend to be raised up with enhanced prolate deformation, and vice versa for enhanced oblate deformation.

As shown in Fig. \ref{Fig:LEV-Mg24},  both neutron and proton orbits $5/2_1^+$, which are not occupied for the ground state of $^{24}$Mg, become deeply bound at the oblate local minima. This can be partly treated as the natural results of the oblate deformation. On the other hand, such effects can be enhanced by the $\pi$-PV and $\rho$-T couplings, particularly for the latter. According to the expansion (\ref{equ:C}), one can easily deduce that the orbits $5/2_1^+$ are dominated by the $d_{5/2}$-components. Following the analysis in Sec. \ref{sec:channels}, the couplings between the $d_{5/2}$-components and the others (mainly the $s_{1/2}$-, $d_{3/2}$-, $p_{1/2}$- and $p_{3/2}$-ones for $^{24}$Mg), leading to more $L$ terms as referred to the $L$-$S$ scheme (\ref{eq:LS-scheme}), can be enhanced by the $\rho$-T coupling, as well as by the $\pi$-PV one. Thus, it is not hard to understand that PKA1 gives a deeper bound second minimum with larger oblate deformation than the other selected models, due to the enhanced correlations between the intruded $5/2_1^+$ orbits and the others.

As shown in Fig. \ref{Fig:LEV-Mg24}, one may also notice that both neutron and proton orbits $1/2_2^+$ and $3/2_1^+$ are nearly degenerated at the oblate minima. For a qualitative understanding, Table \ref{Tab:XC-24Mg} gives the proportions of the dominant components given by PKA1 in expanding the neutron orbits $1/2_2^+$ and $3/2_1^+$, as well as the $Q_2$ values (fm$^{-2}$). It can be seen that for the ground state ($\beta=0.527$) the $1d_{5/2}$-components play the dominant role for both neutron orbits $1/2_2^+$ and $3/2_1^+$, being consistent with the large prolate deformation. It may also partly interpret the fact that PKA1 presents deeper bound ground state for $^{24}$Mg than the others, as indicated by the nature of the $\pi$-PV and $\rho$-T couplings. From the prolate to oblate minima, the proportions of the $1d_{5/2}$-components tend to be gradually reduced, particular from $\beta=-0.306$ to $\beta=-0.776$. In contrast to that, the $2s_{1/2}$ proportion for the orbit $1/2_2^+$ and the $1d_{3/2}$ one for the orbit $3/2_1^+$ increase much with the enhanced oblate deformation.

\begin{table}[htbp]
\caption{Proportions (in percentage) of the main expansion components of neutron orbits $1/2_2^+$ and $3/2_1^+$ at the local minima (denoted by $\beta$) of $^{24}$Mg, as well as the quadruple momentum $Q_2$ (fm$^{-2}$). The results are calculated by PKA1 with $E_+^C = 350$ MeV and $E_-^C=-100$ MeV.} \renewcommand{\arraystretch}{1.8}\setlength{\tabcolsep}{0.5em}\label{Tab:XC-24Mg}
\begin{tabular}{r|rrr|rrr}  \hline\hline
         &  \multicolumn{3}{c|}{$1/2_2^+$}  & \multicolumn{3}{c}{$3/2_1^+$} \\
$\beta$~~~&    $2s_{1/2}$  &  $1d_{5/2}$ &   $Q_2$~ &  $1d_{3/2}$  &  $1d_{5/2}$  &  $Q_2$~ \\ \hline
0.527    &   17.8\%  &  75.0\%  &     12.15  &   9.9\%  &  86.8\%  &     6.12 \\
$-$0.306 &   34.2\%  &  63.6\%  &   $-$5.33  &  19.8\%  &  79.2\%  &  $-$4.40 \\
$-$0.766 &   63.2\%  &  27.3\%  &   $-$9.23  &  52.7\%  &  43.0\%  &  $-$8.62 \\ \hline\hline
\end{tabular}
\end{table}

For the ground state of $^{24}$Mg, as shown in Fig. \ref{Fig:LEV-Mg24}, there exist fairly large gaps between the neutron/proton orbits $1/2_2^+$ and $3/2_1^+$, which is consistent with the deformation effects since both orbits are dominated by the $1d_{5/2}$-components, namely branching from the spherical $1d_{5/2}$ orbit. When nucleus becomes oblately deformed, it is expected that the ordering of the orbits $1/2_2^+$ and $3/2_1^+$ inverses and the distance between enlarges with enhanced oblate deformation, if the orbits are still dominated by the $1d_{5/2}$-components. However, as shown in Fig. \ref{Fig:LEV-Mg24} and Table \ref{Tab:XC-24Mg}, the orbits $1/2_2^+$ and $3/2_1^+$ given by PKA1 are nearly degenerated at the oblate minima, and consistently the $2s_{1/2}$ and $1d_{3/2}$ proportions, respectively in the orbits $1/2_2^+$ and $3/2_1^+$, become the dominant ones. As well known, the spherical $2s_{1/2}$ and $1d_{3/2}$ orbits form the pseudo-spin partners, which are expected to be degenerated around the Fermi levels. Thus, the degenerated $1/2_2^+$ and $3/2_1^+$ orbits at the oblate minima of $^{24}$Mg might be taken as the qualitative manifestation of the pseudo-spin symmetry in deformed nuclei, which seems to compete with the deformation effects.

\section{SUMMARY}\label{sec:SUMMARY}

In this work, the axially deformed relativistic Hartree-Fock-Bogoliubov (D-RHFB) model is established, and the spherical Dirac Woods-Saxon (DWS) base is utilized in expanding the Bogoliubov quasi-particle spinors and solving the RHFB equations. It is shown that starting from the Lagrangian density, the expectation of the derived Hamiltonian with respect to the Bogoliubov ground state can give the full energy functional, that contains both contributions from the Hartree-Fock mean fields and pairing correlations. Following such procedure, there do not exist antisymmetric terms in the pairing energy. In addition, the degree of freedom associated with the $\rho$-T coupling, automatically the $\rho$-VT one, is implemented, and the analysis on the nature of the $\pi$-PV and $\rho$-T are presented to provide a qualitative understanding for their enhancement on the deformation effects. For practical applications, the finite-range Gogny-type force is utilized as the pairing force.

For the reliable description, the space truncations related to the spherical DWS base are carefully verified by taking the light nucleus $^{24}$Mg and mid-heavy one $^{156}$Sm as the candidates. It is found that for light nuclei the completeness of the expansion on the spherical DWS base can be achieved by considering less negative energy states in the base than for the mid-heavy and heavy ones, in which the correlations between the expansion components of the DWS base with large $\kappa$-quantity are enhanced due to the strong $\rho$-T couplings carried by the RHF Lagrangian PKA1. This indicates the necessity of a careful convergence test regarding the completeness of the expansion in the future applications of the D-RHFB model in heavy and superheavy nuclei.

Taking deformed $^{24}$Mg as an example, the roles played by the $\rho$-T and $\pi$-PV couplings, which can couple with deformation effects tightly, are discussed qualitatively, as well as the pseudo-spin symmetry. Besides deeper bound ground state than those given by the other effective Lagrangians, PKA1 that contains the $\rho$-T coupling predicts a fairly deep bound local minimum with large oblate deformation for $^{24}$Mg. As indicated from the nature of the $\rho$-T and $\pi$-PV couplings, this may result from the enhanced correlations between the orbits $5/2_1^+$ and the others. For perspectives, the effects of $\pi$-PV and $\rho$-T couplings, coupled with nuclear deformation, deserve to be studied systematically in the future applications of the D-RHFB model for the wide-range unstable nuclei.

\section{ACKNOWLEDGMENTS}

This work was partly supported by the Strategic Priority Research Program of Chinese Academy of Sciences (Grant No. XDB34000000) and by the Fundamental Research Funds for the Central Universities (lzujbky-2021--sp41 and lzujbky-2021--sp36). The authors also want to thank the computation resources provided by the Supercomputing Center of Lanzhou University and the Southern Nuclear Science Computing Center.


\appendix

\begin{widetext}
\section{Detailed formulas for the $\rho$-T and $\rho$-VT couplings} \label{sec:APP-A}
\subsection{Compounded symbols}\label{sec:APP-A-symbols}

In this work, the density dependences of the meson-nucleon coupling strengths are introduced to evaluate the nuclear in-medium effects \cite{Long2006PLB640.150, Long2007PRC76.034314, Geng2019PRC100.051301R}. Under the axial symmetry, together with the reflection symmetry with respect to the $z=0$ plane, the density-dependent coupling strengths can be decomposed as,
\begin{equation}
    g_\phi(\rho_b) = \sqrt{2\pi} \sum_{\lambda_p}g_\phi^{\lambda_p}(r) Y_{\lambda_p0}(\vartheta,\varphi),
\end{equation}
in which $\lambda_p$ shall be even, and $g_\phi$ here represents the coupling strengths $g_\rho$ and $f_\rho$. Similar as the $\pi$-PV coupling, there exist gradients over the propagator in the $\rho$-T and $\rho$-VT coupling channels,
\begin{align}
  \svec \nabla_{\svec r} D_{\rho}(\svec r,\svec r') =& -m_\rho \sum_{\lambda_y\mu_y} \sum_{\lambda_1}^{\lambda_y\pm 1}(-1)^{\mu_y} C_{\lambda_y010}^{\lambda_10} S_{\lambda_y\lambda_1}(r,r') \svec Y_{\lambda_y\mu_y}^{\lambda_1}(\svec\Omega)Y_{\lambda_y-\mu_y}(\svec\Omega'),\\
  \svec\nabla_{\svec r} \svec\nabla_{\svec r'} D_\rho(\svec r, \svec r')=  & + m_\rho^2 \sum_{\lambda_y\mu_y} \sum_{\lambda_1\lambda_2}^{\lambda_y\pm 1} (-1)^{\mu_y} C_{\lambda_y010}^{\lambda_10} C_{\lambda_y010}^{\lambda_20} \cals V_{\lambda_y}^{\lambda_1\lambda_2}(r, r') \svec Y_{\lambda_y\mu_y}^{\lambda_1}(\svec\Omega) \svec Y_{\lambda_y-\mu_y}^{\lambda_2}(\svec\Omega'),
\end{align}
where $\svec Y_{\lambda\mu}^{\lambda'}$ represents the vector spherical harmonics \cite{Varshalovich1988World-Scientific}, and the radial parts read as,
\begin{align}
    \cals V_{\lambda_y}^{\lambda_1\lambda_2}(r,r') =& -R_{\lambda_1\lambda_2}(r,r') + \frac{1}{m_\rho^2 r^2} \delta(r-r'), \label{eq:V}\\
    R_{\lambda_1\lambda_2}(r,r') =& \sqrt{\frac{1}{rr'}} \Big[I_{\lambda_1+\ff2}(m_\rho r) K_{\lambda_2+\ff2}(m_\rho r') \theta(r'-r) + K_{\lambda_1+\ff2}(m_\rho r) I_{\lambda_2+\ff2}(m_\rho r')\theta(r-r')\Big], \\
    \cals S_{\lambda_y\lambda_1}(r,r') =& \sqrt{\frac{1}{rr'}} \Big[ I_{\lambda_1+\ff2}(m_\rho r) K_{\lambda_y+\ff2}(m_\rho r') \theta(r'-r) - K_{\lambda_1+\ff2}(m_\rho r) I_{\lambda_y+\ff2}(m_\rho r')\theta(r-r') \Big],
\end{align}
where $I$ and $K$ are the modified Bessel functions.

In practice, we performed the integrations with respect to the angle variables $\svec\Omega = \left(\vartheta,\varphi\right)$, since the expansion of the wave functions $\psi_{\nu\pi m}$ is carried on the spherical Dirac Woods-Saxon (DWS) base. Such integration contains the Harmonic functions deduced from the decompositions of the propagators and the coupling strengths, and the couplings between the spherical Dirac spinors,
\begin{equation}
    \sqrt{2\pi} \int d\svec\Omega Y_{\lambda_d\mu_d}(\svec\Omega) Y_{\lambda_\mu -\mu_y}(\svec\Omega) Y_{\lambda_p0} (\svec\Omega) = \frac{1}{\sqrt2} \hat\lambda_d \hat\lambda_y \hat\lambda_p^{-1} C_{\lambda_d0\lambda_y0}^{\lambda_p0} C_{\lambda_d\mu\lambda_y-\mu}^{\lambda_p0} \equiv (-1)^\mu \Theta_{\lambda_d\lambda_p}^{\lambda_y\mu},
\end{equation}
where $\mu = \mu_d=\mu_y$, and $(\lambda_d\mu_d)$, $(\lambda_y\mu_y)$ and $\lambda_p$ denote the terms originating from the couplings between the Dirac spinors, the decompositions of the propagators and the coupling strengths, respectively.

During the derivations of the $\rho$-T and $\rho$-VT couplings using the spherical DWS base, it is convenient to introduce the following symbols, namely $\scr D$, $\bar{\scr D}$, $\cals Q$ and $\bar{\cals Q}$, to abbreviate the complicated expressions,
\begin{align}
    \scr D_{\kappa_1m_1;\kappa_2m_2}^{\lambda\mu} \equiv & \frac{1}{\sqrt2} \hat j_1 \hat j_2 \hat\lambda^{-1} C_{j_1\ff2 j_2-\ff2}^{L0} C_{j_1-m_1 j_2m_2}^{\lambda\mu}, \\
    \bar{\scr D}_{\kappa_1m_1;\kappa_2m_2}^{\lambda\bar\mu} \equiv& (-1)^{\kappa_1} \scr D_{\kappa_1-m_1;\kappa_2m_2}^{\lambda\bar \mu}, \\
    \cals Q_{\kappa_1m_1;\kappa_2m_2}^{\lambda\mu\sigma} \equiv & (-1)^{j_1+l_1-\ff2}\sqrt{3}\hat j_1\hat j_2\hat l_1\hat l_2\sum_{J} C_{l_10l_20}^{\lambda0}C_{\lambda\mu1\sigma}^{JM} \begin{Bmatrix}j_1 & j_2 & J \\ l_1 & l_2 & \lambda \\ \ff2 & \ff2 & 1\end{Bmatrix} C_{j_1-m_1j_2m_2}^{JM},\\
  \bar{\cals Q}_{\kappa_1m_1;\kappa_2m_2}^{\lambda\bar\mu\sigma} \equiv & (-1)^{\kappa_1}\cals Q_{\kappa_1-m_1;\kappa_2m_2}^{\lambda\bar\mu\sigma}.
\end{align}
In the symbols $\scr D$ and $\bar{\scr D}$, $\mu =m_2 -m_1$ and $\bar\mu = m_2+m_1$, whereas for the symbols $\cals Q$ and $\bar{\cals Q}$, one can find $\mu+\sigma = m_2-m_1$ and $\bar\mu+\sigma = m_2 + m_1$.

\subsection{Energy functionals and self-energies of the Hartree terms}\label{subsec:Hartree-rhot}

In the Hartree diagrams of the $\rho$-T and $\rho$-VT couplings, the expansion terms of the local densities, including the baryonic $\rho_b^{\lambda_d}$ and tensor $\rho_T^{\lambda_d\mu}$, can be expressed as,
\begin{align}
  \rho_b^{\lambda_d}(r) = & \sum_i v_i^2 \sum_{\kappa\kappa'} (-1)^{m+\ff2} \scr D_{\kappa m,\kappa'm}^{\lambda_d0} \Big[ \frac{\cals G_{i\kappa}^V(r) \cals G_{i\kappa'}^V(r)}{2\pi r^2} + \frac{\cals F_{i\kappa}^V(r) \cals F_{i\kappa'}^V(r)}{2\pi r^2}\Big],\\
    \rho_{T}^{\lambda_d\mu}(r) =& \sum_{i} v_i^2\sum_{\kappa \kappa' } (-1)^{m+\ff2} \Big[\cals Q_{\kappa m,-\kappa'm}^{\lambda_d\mu\sigma} \frac{\cals G_{i\kappa }^V(r) \cals F_{i\kappa'}^V(r)}{2\pi r^2} + \cals Q_{-\kappa m,\kappa'm}^{\lambda_d\mu\sigma}\frac{\cals F_{i\kappa}^V(r) \cals G_{i\kappa'}^V(r)}{2\pi r^2}\Big],
\end{align}
in which $v_i^2$ represent the degeneracy of the orbit $i$, in general equal to $2$ for axially deformed nuclei, and $\mu+\sigma=0$.

It shall be noticed that the $\rho$-VT coupling in fact contains the $\rho$-vector-tensor and $\rho$-tensor-vector parts, which are specified as the $\rho$-VT and $\rho$-TV in the following context, respectively. The self-energies contributed by the Hartree terms of the $\rho$-T, $\rho$-VT and $\rho$-TV couplings can be derived as,
\begin{align}
    \Sigma_{T,\rho\text{-T}}^{\lambda_d\mu}(r) =& \frac{m_\rho^2}{4M^2}\sum_{\lambda_p} f_\rho^{\lambda_p}(r) \int r'^2dr' \sum_{\lambda_y}\sum_{\lambda_1\lambda_2}^{\lambda_y\pm1} C_{\lambda_y010}^{\lambda_10} C_{\lambda_1\mu1\sigma}^{\lambda_y0} \Theta_{\lambda_d\lambda_p}^{\lambda_1\mu} \cals V_{\lambda_y}^{\lambda_1\lambda_2}(r,r')S_{\text{T}}^{\lambda_y\lambda_2}(r') , \label{eq:app-RT-H-self}\\
    \Sigma_{T,\rho\text{-TV}}^{\lambda_d\mu}(r) =& \frac{m_\rho}{2M}\sum_{\lambda_p} f_\rho^{\lambda_p}(r) \int r'^2dr' \sum_{\lambda_y}\sum_{\lambda_1}^{\lambda_y\pm1} C_{\lambda_y010}^{\lambda_10} C_{\lambda_1\mu1\sigma}^{\lambda_y0} \Theta_{\lambda_d\lambda_p}^{\lambda_1\mu} S_{\lambda_y\lambda_1}(r,r') S_{\text{V}}^{\lambda_y}(r'), \label{eq:app-RTV-H-self}\\
    \Sigma_{0,\rho\text{-VT}}^{\lambda_d}(r) = & \frac{m_\rho}{2M} \sum_{\lambda_p} g_\rho^{\lambda_p}(r) \int r'^2dr' \sum_{\lambda_y} \sum_{\lambda_1}^{\lambda_y\pm1} \Theta_{\lambda_d\lambda_p}^{\lambda_y0}  S_{\lambda_y\lambda_1}(r',r)S_{\text{T}}^{\lambda_y\lambda_1}(r') \label{eq:app-RVT-H-self},
\end{align}
where the source terms $S_{\text{T}}^{\lambda_y\lambda_1}$ and $S_{\text{V}}^{\lambda_y}$ read as,
\begin{equation}
  S_{\text{T}}^{\lambda_y\lambda_1}(r') =  \sum_{\lambda_p'\lambda_d'\mu'}  C_{\lambda_y010}^{\lambda_10} C_{\lambda_1\mu'1\sigma}^{\lambda_y0} \Theta_{\lambda_d'\lambda_p'}^{\lambda_1\mu'} f_\rho^{\lambda_p'}(r') \rho_{T,3}^{\lambda_d'\mu'}(r'), \hspace{4em} S_{\text{V}}^{\lambda_y} = \sum_{\lambda_p' \lambda_d' }  \Theta_{\lambda_d'\lambda_p'}^{\lambda_y0} g_\rho^{\lambda_p'}(r') \rho_{b,3}^{\lambda_d'}(r')
\end{equation}
with $\rho_{b,3}^{\lambda_d'} = \rho_{b,n}^{\lambda_d'} - \rho_{b,p}^{\lambda_d'}$ and $\rho_{T,3}^{\lambda_d'\mu} = \rho_{T,n}^{\lambda_d'\mu} - \rho_{T,p}^{\lambda_d'\mu}$. In terms of the self-energies, the energy functionals contributed by the Hartree terms can be written as,
\begin{align}
    E_{\rho\text{-T}}^{D} =& +\frac{2\pi}{2} \int r^2dr \sum_{\lambda_d\mu} \Sigma_{T,\rho\text{-T}}^{\lambda_d\mu}(r) \rho_{T,3}^{\lambda_d\mu}(r), \\
    E_{\rho\text{-VT}}^{D} =& +\frac{2\pi}{2} \int r^2dr \Big[\sum_{\lambda_d\mu} \Sigma_{T,\rho\text{-TV}}^{\lambda_d\mu}(r) \rho_{T,3}^{\lambda_d\mu}(r) + \sum_{\lambda_d} \Sigma_{0,\rho\text{-VT}}^{\lambda_d}(r) \rho_{b,3}^{\lambda_d}(r)\Big].
\end{align}
Due to the density dependencies carried by the coupling strengths $g_\rho$ and $f_\rho$, the deduced rearrangement terms from the variation of the energy functional read as,
\begin{align}
\Sigma_{R,\rho\text{-T}}^{D,\lambda}(r) =& \frac{ m_\rho^2}{4M^2 r^2}\sum_{\lambda_p\lambda_d\mu} \Big[ \frac{\partial f_\rho^{\lambda_p}}{\partial \rho_b^{\lambda}} \rho_{T,3}^{\lambda_d\mu} \Big]_r \int r'^2dr' \sum_{\lambda_y}\sum_{\lambda_1\lambda_2}^{\lambda_y\pm1} C_{\lambda_y010}^{\lambda_10}C_{\lambda_1\mu1\sigma}^{\lambda_y0} \Theta_{\lambda_d\lambda_p}^{\lambda_1\mu} \cals V_{\lambda_y}^{\lambda_1\lambda_2}(r,r') S_{\text{T}}^{\lambda_y\lambda_2}(r'), \\
\Sigma_{R,\rho\text{-VT}}^{D,\lambda}(r) =& \frac{m_\rho}{2M r^2} \sum_{\lambda_p\lambda_d\mu} \Big[\frac{\partial f_\rho^{\lambda_p}}{\partial \rho_b^{\lambda}} \rho_{T,3}^{\lambda_d\mu}\Big]_r \int r'^2dr' \sum_{\lambda_y}\sum_{\lambda_1}^{\lambda_y\pm1} C_{\lambda_y010}^{\lambda_10} C_{\lambda_1\mu1\sigma}^{\lambda_y0}\Theta_{\lambda_d\lambda_p}^{\lambda_1\mu} S_{\lambda_y\lambda_1}(r,r') S_{\text{V}}^{\lambda_y}(r') \nonumber \\
&+ \frac{ m_\rho}{2M r^2} \sum_{\lambda_p\lambda_d} \Big[ \frac{\partial g_\rho^{\lambda_p}}{\partial \rho_b^{\lambda}} \rho_{b,3}^{\lambda_d}\Big] \int r'^2dr' \sum_{\lambda_y} \sum_{\lambda_1}^{\lambda_y\pm1} \Theta_{\lambda_d\lambda_p}^{\lambda_y0} S_{\lambda_y\lambda_1}(r',r) S_{\text{T}}^{\lambda_y\lambda_1}(r').
\end{align}

\subsection{Energy functionals and self-energies from the Fock terms}\label{subsec:Fock-rhot}

In the Fock diagrams of the $\rho$-T and $\rho$-VT couplings, the expressions are rather complicated. To abbreviate the expressions, we introduce the symbols $\cals R^{\sigma\sigma'}$ with $\sigma$, $\sigma'=\pm$ to denote the non-local density terms, which appear in the non-local self-energy terms $Y_G$, $Y_F$, $X_G$ and $X_F$ as,
\begin{align}
    \cals R_{\kappa\kappa',\pi m}^{++}(r,r') =& \sum_\nu v_i^2 \cals G_{n\kappa m,\kappa}^V(r) \cals G_{\nu\pi m,\kappa'}^V(r'),  & \cals R_{\kappa\kappa',\pi m}^{+-}(r,r') =& \sum_\nu v_i^2 \cals G_{n\kappa m,\kappa}^V(r) \cals F_{\nu\pi m,\kappa'}^V(r'), \\
    \cals R_{\kappa\kappa',\pi m}^{-+}(r,r') = &\sum_\nu v_i^2 \cals F_{n\kappa m,\kappa}^V(r) \cals G_{\nu\pi m,\kappa'}^V(r'), & \cals R_{\kappa\kappa',\pi m}^{--}(r,r') =& \sum_\nu v_i^2 \cals F_{n\kappa m,\kappa}^V(r) \cals F_{\nu\pi m,\kappa'}^V(r').
\end{align}
In the above expressions, $v_i^2$ represents the degeneracy of the orbit $i = (\nu\pi m)$, similar as in the local density expressions.

\subsubsection{Non-local mean fields contributed by the $\rho$-T coupling}

For the $\rho$-T coupling, it contains the time and space components, which are referred as T-$t$ and T-$\svec s$ in the following context, respectively. Using the self-energies $Y_G$, $Y_F$, $X_G$ and $X_F$, the energy functional can be written as,
\begin{equation}
    E_{\rho\text{-T}}^{E} = \frac{1}{2} \int drdr' \sum_{i}v_i^{2} \sum_{\kappa_1\kappa_2} \begin{pmatrix} \cals G_{i\kappa_1}^V & \cals F_{i\kappa_1}^V \end{pmatrix}_r \begin{pmatrix} Y_{G,\pi m}^{\kappa_1\kappa_2,\rho\text{-T}} & Y_{F,\pi m}^{\kappa_1\kappa_2,\rho\text{-T}}\\[0.5em] X_{G,\pi m}^{\kappa_1\kappa_2,\rho\text{-T}} & X_{F,\pi m}^{\kappa_1\kappa_2,\rho\text{-T}} \end{pmatrix}_{r,r'} \begin{pmatrix} \cals G_{i\kappa_2}^V \\[0.5em] \cals F_{i\kappa_2}^V\end{pmatrix}_{r'},
\end{equation}
in which the terms $Y_G$, $Y_F$, $X_G$ and $X_F$ contain the time- and space-component contributions, e.g.,
\begin{equation}
  Y_{G,\pi m}^{\kappa_1\kappa_2,\rho\text{-T}} =  Y_{G,\pi m}^{\kappa_1\kappa_2,\text{T-}t} + Y_{G,\pi m}^{\kappa_1\kappa_2,\text{T-}\svec s}.
\end{equation}
For the time (T-$t$) component, the non-local self-energies $Y_G$, $Y_F$, $X_G$, $X_F$ are derived as,
\begin{align}
    Y_{G,\pi m}^{\kappa_1\kappa_2,\text{T-}t} = & -\dfrac{1}{2\pi} \frac{m_\rho^2}{4M^2} \sum_{\pi'm'} \cals T_{\tau\tau'} \sum_{\kappa_1'\kappa_2'}  \cals R_{\kappa_1'\kappa_2',\pi'm'}^{--} \sum_{\lambda_p\lambda_p'} f_\rho^{\lambda_p}(r) f_{\rho}^{\lambda_p'}(r') \sum_{\lambda_y} \sum_{\lambda_1\lambda_2}^{\lambda_y\pm1} \cals V_{\lambda_y}^{\lambda_1\lambda_2}(r,r') \hat{\cals A}_{-\kappa_1'-\kappa_2'm';+\kappa_1+\kappa_2m}^{\lambda_p\lambda_p',\lambda_y\lambda_1\lambda_2},\\
    Y_{F,\pi m}^{\kappa_1\kappa_2,\text{T-}t} = & -\dfrac{1}{2\pi} \frac{m_\rho^2}{4M^2} \sum_{\pi'm'} \cals T_{\tau\tau'} \sum_{\kappa_1'\kappa_2'} \cals R_{\kappa_1'\kappa_2',\pi'm'}^{-+} \sum_{\lambda_p\lambda_p'} f_\rho^{\lambda_p}(r) f_{\rho}^{\lambda_p'}(r') \sum_{\lambda_y} \sum_{\lambda_1\lambda_2}^{\lambda_y\pm1} \cals V_{\lambda_y}^{\lambda_1\lambda_2}(r,r') \hat{\cals A}_{-\kappa_1'+\kappa_2'm';+\kappa_1-\kappa_2m}^{\lambda_p\lambda_p',\lambda_y\lambda_1\lambda_2},\\
    X_{G,\pi m}^{\kappa_1\kappa_2,\text{T-}t} = & -\dfrac{1}{2\pi} \frac{m_\rho^2}{4M^2} \sum_{\pi'm'} \cals T_{\tau\tau'} \sum_{\kappa_1'\kappa_2'} \cals R_{\kappa_1'\kappa_2',\pi'm'}^{+-} \sum_{\lambda_p\lambda_p'} f_\rho^{\lambda_p}(r) f_{\rho}^{\lambda_p'}(r') \sum_{\lambda_y} \sum_{\lambda_1\lambda_2}^{\lambda_y\pm1} \cals V_{\lambda_y}^{\lambda_1\lambda_2}(r,r') \hat{\cals A}_{+\kappa_1'-\kappa_2'm';-\kappa_1+\kappa_2m}^{\lambda_p\lambda_p',\lambda_y\lambda_1\lambda_2},\\
    X_{F,\pi m}^{\kappa_1\kappa_2,\text{T-}t} = & -\dfrac{1}{2\pi} \frac{m_\rho^2}{4M^2} \sum_{\pi'm'} \cals T_{\tau\tau'} \sum_{\kappa_1'\kappa_2'} \cals R_{\kappa_1'\kappa_2',\pi'm'}^{++} \sum_{\lambda_p\lambda_p'} f_\rho^{\lambda_p}(r) f_{\rho}^{\lambda_p'}(r') \sum_{\lambda_y} \sum_{\lambda_1\lambda_2}^{\lambda_y\pm1} \cals V_{\lambda_y}^{\lambda_1\lambda_2}(r,r') \hat{\cals A}_{+\kappa_1'+\kappa_2'm';-\kappa_1-\kappa_2m}^{\lambda_p\lambda_p',\lambda_y\lambda_1\lambda_2}.
\end{align}
In the above expressions, $\cals T_{\tau\tau'} = 2-\delta_{\tau\tau'}$ is the isospin factor, $\tau$ for $(\pi m)$ and $\tau'$ for $(\pi'm')$. For abbreviation, we introduce the symbols $\widehat{\cals A}$ to denote the combinations of the C-G coefficients given by various couplings,
\begin{equation}
    \widehat{\cals A}_{\kappa_1'\kappa_2'm';\kappa_1\kappa_2m}^{\lambda_p\lambda_p',\lambda_y\lambda_1\lambda_2} \equiv \ff2\Big\{\sum_{\sigma\sigma'} \cals P_{\kappa_1'm',\kappa_1m}^{\lambda_p,\lambda_y\lambda_1;\mu\sigma} \cals P_{\kappa_2'm',\kappa_2m}^{\lambda_p';\lambda_y\lambda_2;\mu'\sigma'} + \sum_{\sigma\sigma'}\bar{\cals P}_{\kappa_1'm',\kappa_1m}^{\lambda_p,\lambda_y\lambda_1;\bar\mu\sigma} \bar{\cals P}_{\kappa_2'm',\kappa_2m}^{\lambda_p',\lambda_y\lambda_2;\bar\mu'\sigma'}\Big\},
\end{equation}
in which the symbols $\cals P$ and $\bar{\cals P}$ read as,
\begin{equation}
  \cals P_{\kappa_1'm',\kappa_1m}^{\lambda_p;\lambda_y\lambda_1;\mu \sigma } \equiv \sum_{\lambda_d} C_{\lambda_y010}^{\lambda_10} C_{\lambda_1\mu 1\sigma }^{\lambda_y\mu_y} \cals Q_{\kappa_1'm',\kappa_1m}^{\lambda_d\mu \sigma } \Theta_{\lambda_d\lambda_p}^{\lambda_1\mu }, \hspace{2em}
  \bar{\cals P}_{\kappa_1'm',\kappa_1m}^{\lambda_p,\lambda_y\lambda_1;\bar\mu \sigma } \equiv \sum_{\lambda_d} C_{\lambda_y010}^{\lambda_10} C_{\lambda_1\bar \mu 1\sigma }^{\lambda_y \mu_y} \bar{\cals Q}_{\kappa_1'm',\kappa_1m}^{\lambda_d \bar\mu  \sigma } \Theta_{\lambda_d\lambda_p}^{\lambda_1 \bar\mu }.\label{eq:app-T-BP}
\end{equation}
Remind that in the symbol $\cals P$, $\mu+\sigma = m-m' = \mu_y$, and for the symbol $\bar{\cals P}$, one has $\bar \mu+\sigma = m+m' = \mu_y$. Following these relations, it can be deduced that finite expansion terms related to $\mu_y$ for the propagators are selected by given $m$ and $m'$. In addition, finite $(\lambda_y\lambda_1)$ terms are selected by the symbol $\Theta$ and C-G coefficient $C_{\lambda_1\mu1\sigma}^{\lambda_y \mu_y}$, as well. Thus, with given space truncations on the spherical DWS base and expanding the coupling strengths, which give finite $\lambda_d$ and $\lambda_p$ values, finite expansion terms of the propagators contribute to the energy functional, which avoids the singularity at $\svec r=\svec r'$ in the propagators.

For the space component (T-$\svec s$), the non-local self-energies $Y_G$, $Y_F$, $X_G$ and $X_F$ can be written as,
\begin{align}
    Y_{G,\pi m}^{\kappa_1\kappa_2,\text{T-}\svec s} = &+\frac{1}{2\pi} \frac{m_\rho^2}{2 M^2} \sum_{\pi'm'} \cals T_{\tau\tau'}  \sum_{\kappa_1'\kappa_2'} \cals R_{\kappa_1'\kappa_2',\pi'm'}^{++} \sum_{\lambda_p\lambda_p'} f_\rho^{\lambda_p}(r) f_\rho^{\lambda_{p'}}(r') \sum_{\lambda_y} \sum_{\lambda_1\lambda_2}^{\lambda_y\pm1} \cals V_{\lambda_y}^{\lambda_1\lambda_2} \widehat{\cals F}_{+\kappa_1'+\kappa_2'm',+\kappa_1+\kappa_2m}^{\lambda_p\lambda_p';\lambda_y\lambda_1\lambda_2},  \\
    Y_{F,\pi m}^{\kappa_1\kappa_2,\text{T-}\svec s} = &-\frac{1}{2\pi} \frac{m_\rho^2}{2 M^2} \sum_{\pi'm'} \cals T_{\tau\tau'}  \sum_{\kappa_1'\kappa_2'} \cals R_{\kappa_1'\kappa_2',\pi'm'}^{+-} \sum_{\lambda_p\lambda_p'} f_\rho^{\lambda_p}(r) f_\rho^{\lambda_{p'}}(r') \sum_{\lambda_y} \sum_{\lambda_1\lambda_2}^{\lambda_y\pm1} \cals V_{\lambda_y}^{\lambda_1\lambda_2} \widehat{\cals F}_{+\kappa_1'-\kappa_2'm',+\kappa_1-\kappa_2m}^{\lambda_p\lambda_p';\lambda_y\lambda_1\lambda_2}, \\
    X_{G,\pi m}^{\kappa_1\kappa_2,\text{T-}\svec s} = &-\frac{1}{2\pi} \frac{m_\rho^2}{2 M^2} \sum_{\pi'm'} \cals T_{\tau\tau'}  \sum_{\kappa_1'\kappa_2'} \cals R_{\kappa_1'\kappa_2',\pi'm'}^{-+} \sum_{\lambda_p\lambda_p'} f_\rho^{\lambda_p}(r) f_\rho^{\lambda_{p'}}(r') \sum_{\lambda_y} \sum_{\lambda_1\lambda_2}^{\lambda_y\pm1} \cals V_{\lambda_y}^{\lambda_1\lambda_2} \widehat{\cals F}_{-\kappa_1'+\kappa_2'm',-\kappa_1+\kappa_2m}^{\lambda_p\lambda_p';\lambda_y\lambda_1\lambda_2}, \\
    X_{F,\pi m}^{\kappa_1\kappa_2,\text{T-}\svec s} = &+\frac{1}{2\pi} \frac{m_\rho^2}{2 M^2} \sum_{\pi'm'} \cals T_{\tau\tau'}  \sum_{\kappa_1'\kappa_2'} \cals R_{\kappa_1'\kappa_2',\pi'm'}^{--} \sum_{\lambda_p\lambda_p'} f_\rho^{\lambda_p}(r) f_\rho^{\lambda_{p'}}(r') \sum_{\lambda_y} \sum_{\lambda_1\lambda_2}^{\lambda_y\pm1} \cals V_{\lambda_y}^{\lambda_1\lambda_2} \widehat{\cals F}_{-\kappa_1'-\kappa_2'm',-\kappa_1-\kappa_2m}^{\lambda_p\lambda_p';\lambda_y\lambda_1\lambda_2}.
\end{align}
For abbreviations, the introduced symbols $\widehat{\cals F}$ read as
\begin{equation}
    \widehat{\cals F}_{\kappa_1'\kappa_2'm',\kappa_1\kappa_2m}^{\lambda_p\lambda_p';\lambda_y\lambda_1\lambda_2} \equiv \ff2\Big\{ \sum_{\delta} \sum_{\sigma \sigma '} \cals N_{\kappa_1'm',\kappa_1m}^{\lambda_p,\lambda_y\lambda_1;\mu \sigma ,\delta} \cals N_{\kappa_2'm',\kappa_2m}^{\lambda_p',\lambda_y\lambda_2;\mu '\sigma ',\delta} + \sum_{\delta}\sum_{\sigma \sigma '} \bar{\cals N}_{\kappa_1'm',\kappa_1m}^{\lambda_p,\lambda_y\lambda_1;\bar\mu \sigma,\delta} \bar{\cals N}_{\kappa_2'm',\kappa_2m}^{\lambda_p',\lambda_y\lambda_2;\bar\mu' \sigma',\delta}\Big\},
\end{equation}
where the symbols $\cals N$ and $\bar{\cals N}$ are of the following forms,
\begin{equation}\label{eq:app-T-BN}
\begin{split}
    \cals N_{\kappa_1'm',\kappa_1m}^{\lambda_p,\lambda_y\lambda_1;\mu \sigma,\delta} \equiv & \sum_{\lambda_d} C_{\lambda_y010}^{\lambda_10} C_{\lambda_1\mu  1\nu}^{\lambda_y\mu_y} C_{1\delta 1\nu}^{1\sigma } \cals Q_{\kappa_1'm',\kappa_1m}^{\lambda_d\mu \sigma }\Theta_{\lambda_d\lambda_p}^{\lambda_1\mu }, \\
    \bar{\cals N}_{\kappa_1'm',\kappa_1m}^{\lambda_p,\lambda_y\lambda_1;\bar\mu \sigma ,\delta} \equiv & \sum_{\lambda_d} C_{\lambda_y010}^{\lambda_10} C_{\lambda_1 \bar \mu  1\nu}^{\lambda_y \mu_y} C_{1\delta 1\nu}^{1 \sigma } \bar{\cals Q}_{\kappa_1'm',\kappa_1m}^{\lambda_d \bar \mu  \sigma } \Theta_{\lambda_d\lambda_p}^{\lambda_1\bar\mu }.
\end{split}
\end{equation}
Remind that in the symbols $\cals N$, $\mu + \sigma = m-m'$, and for the symbols $\bar{\cals N}$ one may have $\bar u+\sigma = m+m'$.

Similar as the energy functional, the rearrangement terms $\Sigma_{R,\rho\text{-T}}^{E,\lambda_d}$ contributed by the Fock terms of the $\rho$-T coupling can be expressed formally as,
\begin{equation}
    \Sigma_{R,\rho\text{-T}}^{E,\lambda_d}(r) = \frac{1}{2\pi r^2}\int dr' \sum_{i} v_i^2 \sum_{\kappa_1\kappa_2} \begin{pmatrix}\cals G_{i\kappa_1}^V & \cals F_{i\kappa_1}^V \end{pmatrix}_r \begin{pmatrix} P_{G,\pi m,\lambda_d}^{\kappa_1\kappa_2, \rho\text{-T} } & P_{F,\pi m,\lambda_d}^{\kappa_1\kappa_2, \rho\text{-T}} \\[0.5em] Q_{G,\pi m,\lambda_d}^{\kappa_1\kappa_2, \rho\text{-T} } & Q_{F,\pi m,\lambda_d}^{\kappa_1\kappa_2, \rho\text{-T} }\end{pmatrix}_{r,r'} \begin{pmatrix} \cals G_{i\kappa_2}^V \\[0.5em] \cals F_{i\kappa_2}^V \end{pmatrix}_{r'},
\end{equation}
in which the $P$ and $Q$ terms contain the time- and space-component contributions, e.g.,
\begin{equation}
  P_{G,\pi m,\lambda_d}^{\kappa_1\kappa_2, \rho\text{-T} } =  P_{G,\pi m,\lambda_d}^{\kappa_1\kappa_2, \text{T-}t } + P_{G,\pi m,\lambda_d}^{\kappa_1\kappa_2, \text{T-}\svec s }.
\end{equation}
Concerning the time-component contributions, the terms $P$ and $Q$ read as,
\begin{align}
    P_{G,\pi m,\lambda_d}^{\kappa_1\kappa_2,\text{T-}t} = & -\frac{1}{2\pi} \frac{m_\rho^2}{4M^2} \sum_{\pi'm'} \cals T_{\tau\tau'} \sum_{\kappa_1'\kappa_2'} \cals R_{\kappa_1'\kappa_2',\pi'm'}^{--} \sum_{\lambda_p\lambda_p'} \frac{\partial f_\rho^{\lambda_p}(r)}{\partial \rho_b^{\lambda_d}(r)} f_{\rho}^{\lambda_p'}(r') \sum_{\lambda_y} \sum_{\lambda_1\lambda_2}^{\lambda_y\pm1} \cals V_{\lambda_y}^{\lambda_1\lambda_2}(r,r') \widehat{\cals A}_{-\kappa_1'-\kappa_2'm';+\kappa_1+\kappa_2m}^{\lambda_p\lambda_p',\lambda_y\lambda_1\lambda_2},\\
    P_{F,\pi m,\lambda_d}^{\kappa_1\kappa_2,\text{T-}t} = & -\frac{1}{2\pi} \frac{m_\rho^2}{4M^2} \sum_{\pi'm'} \cals T_{\tau\tau'} \sum_{\kappa_1'\kappa_2'} \cals R_{\kappa_1'\kappa_2',\pi'm'}^{-+} \sum_{\lambda_p\lambda_p'} \frac{\partial f_\rho^{\lambda_p}(r)}{\partial \rho_b^{\lambda_d}(r)} f_{\rho}^{\lambda_p'}(r') \sum_{\lambda_y} \sum_{\lambda_1\lambda_2}^{\lambda_y\pm1} \cals V_{\lambda_y}^{\lambda_1\lambda_2}(r,r') \widehat{\cals A}_{-\kappa_1'+\kappa_2'm';+\kappa_1-\kappa_2m}^{\lambda_p\lambda_p',\lambda_y\lambda_1\lambda_2},\\
    Q_{G,\pi m,\lambda_d}^{\kappa_1\kappa_2,\text{T-}t} = & -\frac{1}{2\pi} \frac{m_\rho^2}{4M^2} \sum_{\pi'm'} \cals T_{\tau\tau'} \sum_{\kappa_1'\kappa_2'} \cals R_{\kappa_1'\kappa_2',\pi'm'}^{+-} \sum_{\lambda_p\lambda_p'} \frac{\partial f_\rho^{\lambda_p}(r)}{\partial \rho_b^{\lambda_d}(r)} f_{\rho}^{\lambda_p'}(r') \sum_{\lambda_y} \sum_{\lambda_1\lambda_2}^{\lambda_y\pm1} \cals V_{\lambda_y}^{\lambda_1\lambda_2}(r,r') \widehat{\cals A}_{+\kappa_1'-\kappa_2'm';-\kappa_1+\kappa_2m}^{\lambda_p\lambda_p',\lambda_y\lambda_1\lambda_2},\\
    Q_{F,\pi m,\lambda_d}^{\kappa_1\kappa_2,\text{T-}t} = & -\frac{1}{2\pi} \frac{m_\rho^2}{4M^2} \sum_{\pi'm'} \cals T_{\tau\tau'} \sum_{\kappa_1'\kappa_2'} \cals R_{\kappa_1'\kappa_2',\pi'm'}^{++} \sum_{\lambda_p\lambda_p'} \frac{\partial f_\rho^{\lambda_p}(r)}{\partial \rho_b^{\lambda_d}(r)} f_{\rho}^{\lambda_p'}(r') \sum_{\lambda_y} \sum_{\lambda_1\lambda_2}^{\lambda_y\pm1} \cals V_{\lambda_y}^{\lambda_1\lambda_2}(r,r') \widehat{\cals A}_{+\kappa_1'+\kappa_2'm';-\kappa_1-\kappa_2m}^{\lambda_p\lambda_p',\lambda_y\lambda_1\lambda_2},
\end{align}
where $\cals T_{\tau\tau'} = 2-\delta_{\tau\tau'}$. For the space components, the $P$ and $Q$ terms read as,
\begin{align}
    P_{G,\pi m}^{\kappa_1\kappa_2,\text{T-}\svec s} = &+\frac{1}{2\pi}\frac{m_\rho^2}{2 M^2} \sum_{\pi'm'} \cals T_{\tau\tau'} \sum_{\kappa_1'\kappa_2'} \cals R_{\kappa_1'\kappa_2',\pi'm'}^{++} \sum_{\lambda_p\lambda_p'} \frac{\partial f_\rho^{\lambda_p}(r)}{\partial \rho_b^{\lambda_d}(r)} f_\rho^{\lambda_{p'}}(r') \sum_{\lambda_y} \sum_{\lambda_1\lambda_2}^{\lambda_y\pm1} \cals V_{\lambda_y}^{\lambda_1\lambda_2}(r,r') \widehat{\cals F}_{+\kappa_1'+\kappa_2'm',+\kappa_1+\kappa_2m}^{\lambda_p\lambda_p';\lambda_y\lambda_1\lambda_2}, \\
    P_{F,\pi m}^{\kappa_1\kappa_2,\text{T-}\svec s} = &-\frac{1}{2\pi}\frac{m_\rho^2}{2 M^2} \sum_{\pi'm'} \cals T_{\tau\tau'} \sum_{\kappa_1'\kappa_2'} \cals R_{\kappa_1'\kappa_2',\pi'm'}^{+-} \sum_{\lambda_p\lambda_p'} \frac{\partial f_\rho^{\lambda_p}(r)}{\partial \rho_b^{\lambda_d}(r)} f_\rho^{\lambda_{p'}}(r') \sum_{\lambda_y} \sum_{\lambda_1\lambda_2}^{\lambda_y\pm1} \cals V_{\lambda_y}^{\lambda_1\lambda_2}(r,r') \widehat{\cals F}_{+\kappa_1'-\kappa_2'm',+\kappa_1-\kappa_2m}^{\lambda_p\lambda_p';\lambda_y\lambda_1\lambda_2}, \\
    Q_{G,\pi m}^{\kappa_1\kappa_2,\text{T-}\svec s} = &-\frac{1}{2\pi}\frac{m_\rho^2}{2 M^2} \sum_{\pi'm'} \cals T_{\tau\tau'} \sum_{\kappa_1'\kappa_2'} \cals R_{\kappa_1'\kappa_2',\pi'm'}^{-+} \sum_{\lambda_p\lambda_p'} \frac{\partial f_\rho^{\lambda_p}(r)}{\partial \rho_b^{\lambda_d}(r)} f_\rho^{\lambda_{p'}}(r') \sum_{\lambda_y} \sum_{\lambda_1\lambda_2}^{\lambda_y\pm1} \cals V_{\lambda_y}^{\lambda_1\lambda_2}(r,r') \widehat{\cals F}_{-\kappa_1'+\kappa_2'm',-\kappa_1+\kappa_2m}^{\lambda_p\lambda_p';\lambda_y\lambda_1\lambda_2}, \\
    Q_{F,\pi m}^{\kappa_1\kappa_2,\text{T-}\svec s} = &+\frac{1}{2\pi}\frac{m_\rho^2}{2 M^2} \sum_{\pi'm'} \cals T_{\tau\tau'} \sum_{\kappa_1'\kappa_2'} \cals R_{\kappa_1'\kappa_2',\pi'm'}^{--} \sum_{\lambda_p\lambda_p'} \frac{\partial f_\rho^{\lambda_p}(r)}{\partial \rho_b^{\lambda_d}(r)} f_\rho^{\lambda_{p'}}(r') \sum_{\lambda_y} \sum_{\lambda_1\lambda_2}^{\lambda_y\pm1} \cals V_{\lambda_y}^{\lambda_1\lambda_2}(r,r') \widehat{\cals F}_{-\kappa_1'-\kappa_2'm',-\kappa_1-\kappa_2m}^{\lambda_p\lambda_p';\lambda_y\lambda_1\lambda_2}.
\end{align}

\subsubsection{Non-local mean fileds contributed by the $\rho$-VT coupling}

At the beginning, it shall be reminded that the $\rho$-VT coupling consists of two types of contributions, namely the $\rho$-VT and $\rho$-TV part as mentioned in expressing the Hartree terms. Similar as the $\rho$-T coupling, the $\rho$-VT  coupling contains the time and space components as well. Combined the contributions from the $\rho$-VT and $\rho$-TV terms, the whole contributions of the $\rho$-VT channel can be formally expressed as,
\begin{equation}
    E_{\rho\text{-VT}}^{E} = \frac{1}{2} \int drdr' \sum_{i}v_i^{2} \sum_{\kappa_1\kappa_2} \begin{pmatrix} \cals G_{i\kappa_1}^V & \cals F_{i\kappa_1}^V \end{pmatrix}_r \begin{pmatrix} Y_{G,\pi m}^{\kappa_1\kappa_2,\text{TV}} + Y_{G,\pi m}^{\kappa_1\kappa_2,\text{VT}} & Y_{F,\pi m}^{\kappa_1\kappa_2,\text{TV}} + Y_{F,\pi m}^{\kappa_1\kappa_2,\text{VT}} \\[0.5em] X_{G,\pi m}^{\kappa_1\kappa_2,\text{TV}} + X_{G,\pi m}^{\kappa_1\kappa_2,\text{VT}}  & X_{F,\pi m}^{\kappa_1\kappa_2,\text{TV}} + X_{F,\pi m}^{\kappa_1\kappa_2,\text{VT}} \end{pmatrix}_{r,r'} \begin{pmatrix} \cals G_{i\kappa_2}^V \\[0.5em] \cals F_{i\kappa_2}^V\end{pmatrix}_{r'}.
\end{equation}
in which $v_i^2$ denotes the degeneracy of orbit $i$, and TV and VT are used to mark the tensor-vector (TV) and vector-tensor (VT) parts, respectively. In the above expression, the $Y_G$, $Y_F$, $X_G$ and $X_F$ terms contain the contributions from both time ($t$) and space ($\svec s$) components, e.g.,
\begin{equation}
  Y_{G,\pi m}^{\kappa_1\kappa_2,\text{TV}} = Y_{G,\pi m}^{\kappa_1\kappa_2,\text{TV-}t} + Y_{G,\pi m}^{\kappa_1\kappa_2,\text{TV-}\svec s}, \hspace{4em} Y_{G,\pi m}^{\kappa_1\kappa_2,\text{VT}} =  Y_{G,\pi m}^{\kappa_1\kappa_2,\text{VT-}t} + Y_{G,\pi m}^{\kappa_1\kappa_2,\text{VT-}\svec s}.
\end{equation}
For the time components of the non-local self-energies $Y_G$, $Y_F$, $X_G$ and $X_F$, one can express the TV part as,
\begin{align}
    Y_{G,\pi m}^{\kappa_1\kappa_2,\text{TV-}t} = & -\frac{1}{2\pi} \frac{m_\rho}{2M} \sum_{\pi'm'} \cals T_{\tau\tau'} \sum_{\kappa_1'\kappa_2'} \cals R_{\kappa_1'\kappa_2',\pi' m'}^{-+} \sum_{\lambda_p\lambda_p'} f_\rho^{\lambda_p}(r) g_\rho^{\lambda_p'}(r') \sum_{\lambda_y}\sum_{\lambda_1}^{\lambda_y\pm1} S_{\lambda_y\lambda_1}(r,r') \widehat{\cals O}_{-\kappa_1'+\kappa_2'm',+\kappa_1+\kappa_2m}^{\lambda_p\lambda_p',\lambda_y\lambda_1}, \\
    Y_{F,\pi m}^{\kappa_1\kappa_2,\text{TV-}t} = & -\frac{1}{2\pi} \frac{m_\rho}{2M} \sum_{\pi'm'} \cals T_{\tau\tau'} \sum_{\kappa_1'\kappa_2'} \cals R_{\kappa_1'\kappa_2',\pi' m'}^{--} \sum_{\lambda_p\lambda_p'} f_\rho^{\lambda_p}(r) g_\rho^{\lambda_p'}(r') \sum_{\lambda_y}\sum_{\lambda_1}^{\lambda_y\pm1} S_{\lambda_y\lambda_1}(r,r') \widehat{\cals O}_{-\kappa_1'-\kappa_2'm',+\kappa_1-\kappa_2m}^{\lambda_p\lambda_p',\lambda_y\lambda_1}, \\
    X_{G,\pi m}^{\kappa_1\kappa_2,\text{TV-}t} = & -\frac{1}{2\pi} \frac{m_\rho}{2M} \sum_{\pi'm'} \cals T_{\tau\tau'} \sum_{\kappa_1'\kappa_2'} \cals R_{\kappa_1'\kappa_2',\pi' m'}^{++} \sum_{\lambda_p\lambda_p'} f_\rho^{\lambda_p}(r) g_\rho^{\lambda_p'}(r') \sum_{\lambda_y}\sum_{\lambda_1}^{\lambda_y\pm1} S_{\lambda_y\lambda_1}(r,r') \widehat{\cals O}_{+\kappa_1'+\kappa_2'm',-\kappa_1+\kappa_2m}^{\lambda_p\lambda_p',\lambda_y\lambda_1}, \\
    X_{F,\pi m}^{\kappa_1\kappa_2,\text{TV-}t} = & -\frac{1}{2\pi} \frac{m_\rho}{2M} \sum_{\pi'm'} \cals T_{\tau\tau'} \sum_{\kappa_1'\kappa_2'} \cals R_{\kappa_1'\kappa_2',\pi' m'}^{+-} \sum_{\lambda_p\lambda_p'} f_\rho^{\lambda_p}(r) g_\rho^{\lambda_p'}(r') \sum_{\lambda_y}\sum_{\lambda_1}^{\lambda_y\pm1} S_{\lambda_y\lambda_1}(r,r') \widehat{\cals O}_{+\kappa_1'-\kappa_2'm',-\kappa_1-\kappa_2m}^{\lambda_p\lambda_p',\lambda_y\lambda_1},
\end{align}
and the VT terms as,
\begin{align}
    Y_{G,\pi m}^{\kappa_1\kappa_2,\text{VT-}t} = & -\frac{1}{2\pi} \frac{m_\rho}{2M} \sum_{\pi'm'} \cals T_{\tau\tau'} \sum_{\kappa_1'\kappa_2'} \cals R_{\kappa_1'\kappa_2',\pi' m'}^{+-} \sum_{\lambda_p\lambda_p'} g_\rho^{\lambda_p}(r) f_\rho^{\lambda_p'}(r') \sum_{\lambda_y}\sum_{\lambda_1}^{\lambda_y\pm1} S_{\lambda_y\lambda_1}(r',r) \widehat{\cals O}_{-\kappa_2'+\kappa_1'm',+\kappa_2+\kappa_1m}^{\lambda_p\lambda_p',\lambda_y\lambda_1}, \\
    Y_{F,\pi m}^{\kappa_1\kappa_2,\text{VT-}t} = & -\frac{1}{2\pi} \frac{m_\rho}{2M} \sum_{\pi'm'} \cals T_{\tau\tau'} \sum_{\kappa_1'\kappa_2'} \cals R_{\kappa_1'\kappa_2',\pi' m'}^{++} \sum_{\lambda_p\lambda_p'} g_\rho^{\lambda_p}(r) f_\rho^{\lambda_p'}(r') \sum_{\lambda_y}\sum_{\lambda_1}^{\lambda_y\pm1} S_{\lambda_y\lambda_1}(r',r) \widehat{\cals O}_{+\kappa_2'+\kappa_1'm',-\kappa_2+\kappa_1m}^{\lambda_p\lambda_p',\lambda_y\lambda_1}, \\
    X_{G,\pi m}^{\kappa_1\kappa_2,\text{VT-}t} = & -\frac{1}{2\pi} \frac{m_\rho}{2M} \sum_{\pi'm'} \cals T_{\tau\tau'} \sum_{\kappa_1'\kappa_2'} \cals R_{\kappa_1'\kappa_2',\pi' m'}^{--} \sum_{\lambda_p\lambda_p'} g_\rho^{\lambda_p}(r) f_\rho^{\lambda_p'}(r') \sum_{\lambda_y}\sum_{\lambda_1}^{\lambda_y\pm1} S_{\lambda_y\lambda_1}(r',r) \widehat{\cals O}_{-\kappa_2'-\kappa_1'm',+\kappa_2-\kappa_1m}^{\lambda_p\lambda_p',\lambda_y\lambda_1}, \\
    X_{F,\pi m}^{\kappa_1\kappa_2,\text{VT-}t} = & -\frac{1}{2\pi} \frac{m_\rho}{2M} \sum_{\pi'm'} \cals T_{\tau\tau'} \sum_{\kappa_1'\kappa_2'} \cals R_{\kappa_1'\kappa_2',\pi' m'}^{-+} \sum_{\lambda_p\lambda_p'} g_\rho^{\lambda_p}(r) f_\rho^{\lambda_p'}(r') \sum_{\lambda_y}\sum_{\lambda_1}^{\lambda_y\pm1} S_{\lambda_y\lambda_1}(r',r) \widehat{\cals O}_{+\kappa_2'-\kappa_1'm',-\kappa_2-\kappa_1m}^{\lambda_p\lambda_p',\lambda_y\lambda_1}.
\end{align}
In the above expressions, the symbols $\widehat{\cals O}$ are introduced to denote the combinations of the following symbols,
\begin{equation}
    \widehat{\cals O}_{\kappa_1'\kappa_2',\kappa_1\kappa_2}^{\lambda_p\lambda_p',\lambda_y\lambda_1,m'm} \equiv \ff2\Big\{\sum_\sigma \cals P_{\kappa_1'm',\kappa_1m}^{\lambda_p\lambda_y\lambda_1,\mu\sigma} \cals S_{\kappa_2'm',\kappa_2m}^{\lambda_p'\lambda_y} + \sum_{\sigma} \bar{\cals P}_{\kappa_1'm',\kappa_1m}^{\lambda_p \lambda_y\lambda_1,\bar\mu\sigma} \bar{\cals S}_{\kappa_2'm',\kappa_2m}^{\lambda_p'\lambda_y}\Big\},
\end{equation}
in which the symbols $\cals P$ and $\bar{\cals P}$ have been defined as Eq. (\ref{eq:app-T-BP}), and the symbols $\cals S$ and $\bar{\cals S}$ correspond to the combination of the ones $\scr D$ ($\bar{\scr D}$) and $\Theta$ as defined before,
\begin{equation}
    \cals S_{\kappa_1'm',\kappa_1m}^{\lambda_p\lambda_y} \equiv  \sum_{\lambda_d} \scr D_{\kappa_1'm',\kappa_1m}^{\lambda_d\mu}\Theta_{\lambda_d\lambda_p}^{\lambda_y\mu}, \hspace{4em} \bar{\cals S}_{\kappa_1'm',\kappa_1m}^{\lambda_p\lambda_y} \equiv\sum_{\lambda_d} \bar{\scr D}_{\kappa_1'm',\kappa_1m}^{\lambda_d \bar\mu} \Theta_{\lambda_d \lambda_p }^{\lambda_y\bar\mu}.
\end{equation}
Remind that in the symbol $\scr D$, $\mu = m-m'$ and for the symbol $\bar{\scr D}$ one has $\bar\mu = m+m'$.

For the space component, the TV and VT parts of the self-energies $Y_G$, $Y_F$, $X_G$ and $X_F$ can be written as,
\begin{align}
    Y_{G,\pi m}^{\kappa_1\kappa_2,\text{TV-}\svec s} = & +\frac{1}{2\pi} \frac{\sqrt2m_\rho}{2M} \sum_{\pi'm'} \cals T_{\tau\tau'} \sum_{\kappa_1'\kappa_2'} \cals R_{\kappa_1'\kappa_2',\pi'm'}^{+-}  \sum_{\lambda_p\lambda_p'} f_\rho^{\lambda_p}(r) g_\rho^{\lambda_p'}(r') \sum_{\lambda_y}\sum_{\lambda_1}^{\lambda_y\pm1} S_{\lambda_y\lambda_1}(r,r') \widehat{\cals Q}_{+\kappa_1'-\kappa_2'm',+\kappa_1+\kappa_2m}^{\lambda_p\lambda_p',\lambda_y\lambda_1},\\
    Y_{F,\pi m}^{\kappa_1\kappa_2,\text{TV-}\svec s} = & -\frac{1}{2\pi} \frac{\sqrt2m_\rho}{2M} \sum_{\pi'm'} \cals T_{\tau\tau'} \sum_{\kappa_1'\kappa_2'} \cals R_{\kappa_1'\kappa_2',\pi'm'}^{++}  \sum_{\lambda_p\lambda_p'} f_\rho^{\lambda_p}(r) g_\rho^{\lambda_p'}(r') \sum_{\lambda_y}\sum_{\lambda_1}^{\lambda_y\pm1} S_{\lambda_y\lambda_1}(r,r') \widehat{\cals Q}_{+\kappa_1'+\kappa_2'm',+\kappa_1-\kappa_2m}^{\lambda_p\lambda_p',\lambda_y\lambda_1},\\
    X_{G,\pi m}^{\kappa_1\kappa_2,\text{TV-}\svec s} = & -\frac{1}{2\pi} \frac{\sqrt2m_\rho}{2M} \sum_{\pi'm'} \cals T_{\tau\tau'} \sum_{\kappa_1'\kappa_2'} \cals R_{\kappa_1'\kappa_2',\pi'm'}^{--}  \sum_{\lambda_p\lambda_p'} f_\rho^{\lambda_p}(r) g_\rho^{\lambda_p'}(r') \sum_{\lambda_y}\sum_{\lambda_1}^{\lambda_y\pm1} S_{\lambda_y\lambda_1}(r,r') \widehat{\cals Q}_{-\kappa_1'-\kappa_2'm',-\kappa_1+\kappa_2m}^{\lambda_p\lambda_p',\lambda_y\lambda_1},\\
    X_{F,\pi m}^{\kappa_1\kappa_2,\text{TV-}\svec s} = & +\frac{1}{2\pi} \frac{\sqrt2m_\rho}{2M} \sum_{\pi'm'} \cals T_{\tau\tau'} \sum_{\kappa_1'\kappa_2'} \cals R_{\kappa_1'\kappa_2',\pi'm'}^{-+}  \sum_{\lambda_p\lambda_p'} f_\rho^{\lambda_p}(r) g_\rho^{\lambda_p'}(r') \sum_{\lambda_y}\sum_{\lambda_1}^{\lambda_y\pm1} S_{\lambda_y\lambda_1}(r,r') \widehat{\cals Q}_{-\kappa_1'+\kappa_2'm',-\kappa_1-\kappa_2m}^{\lambda_p\lambda_p',\lambda_y\lambda_1}, \\
    Y_{G,\pi m}^{\kappa_1\kappa_2,\text{VT-}\svec s} = & +\frac{1}{2\pi} \frac{\sqrt2m_\rho}{2M} \sum_{\pi'm'} \cals T_{\tau\tau'} \sum_{\kappa_1'\kappa_2'} \cals R_{\kappa_1'\kappa_2',\pi'm'}^{-+}  \sum_{\lambda_p\lambda_p'} g_\rho^{\lambda_p}(r) f_\rho^{\lambda_p'}(r') \sum_{\lambda_y}\sum_{\lambda_1}^{\lambda_y\pm1} S_{\lambda_y\lambda_1}(r',r) \widehat{\cals Q}_{+\kappa_2'-\kappa_1'm',+\kappa_2+\kappa_1m}^{\lambda_p\lambda_p',\lambda_y\lambda_1},\\
    Y_{F,\pi m}^{\kappa_1\kappa_2,\text{VT-}\svec s} = & -\frac{1}{2\pi} \frac{\sqrt2m_\rho}{2M} \sum_{\pi'm'} \cals T_{\tau\tau'} \sum_{\kappa_1'\kappa_2'} \cals R_{\kappa_1'\kappa_2',\pi'm'}^{--}  \sum_{\lambda_p\lambda_p'} g_\rho^{\lambda_p}(r) f_\rho^{\lambda_p'}(r') \sum_{\lambda_y}\sum_{\lambda_1}^{\lambda_y\pm1} S_{\lambda_y\lambda_1}(r',r) \widehat{\cals Q}_{-\kappa_2'-\kappa_1'm',-\kappa_2+\kappa_1m}^{\lambda_p\lambda_p',\lambda_y\lambda_1},\\
    X_{G,\pi m}^{\kappa_1\kappa_2,\text{VT-}\svec s} = & -\frac{1}{2\pi} \frac{\sqrt2m_\rho}{2M} \sum_{\pi'm'} \cals T_{\tau\tau'} \sum_{\kappa_1'\kappa_2'} \cals R_{\kappa_1'\kappa_2',\pi'm'}^{++}  \sum_{\lambda_p\lambda_p'} g_\rho^{\lambda_p}(r) f_\rho^{\lambda_p'}(r') \sum_{\lambda_y}\sum_{\lambda_1}^{\lambda_y\pm1} S_{\lambda_y\lambda_1}(r',r) \widehat{\cals Q}_{+\kappa_2'+\kappa_1'm',+\kappa_2-\kappa_1m}^{\lambda_p\lambda_p',\lambda_y\lambda_1},\\
    X_{F,\pi m}^{\kappa_1\kappa_2,\text{VT-}\svec s} = & +\frac{1}{2\pi} \frac{\sqrt2m_\rho}{2M} \sum_{\pi'm'} \cals T_{\tau\tau'} \sum_{\kappa_1'\kappa_2'} \cals R_{\kappa_1'\kappa_2',\pi'm'}^{+-}  \sum_{\lambda_p\lambda_p'} g_\rho^{\lambda_p}(r) f_\rho^{\lambda_p'}(r') \sum_{\lambda_y}\sum_{\lambda_1}^{\lambda_y\pm1} S_{\lambda_y\lambda_1}(r',r) \widehat{\cals Q}_{-\kappa_2'+\kappa_1'm',-\kappa_2-\kappa_1m}^{\lambda_p\lambda_p',\lambda_y\lambda_1}.
\end{align}
In the above space-component contributions, the symbols $\widehat{\cals Q}$ are of the following form,
\begin{equation}
    \widehat{\cals Q}_{\kappa_1'\kappa_2'm',\kappa_1\kappa_2 m}^{\lambda_p\lambda_p',\lambda_y\lambda_1} \equiv \ff2\Big\{\sum_{\sigma\delta} \cals N_{\kappa_1'm',\kappa_1m}^{\lambda_p,\lambda_y\lambda_1;\mu \sigma,\delta} \cals M_{\kappa_2'm',\kappa_2m}^{\lambda_p'\lambda_y;\mu '\delta} + \sum_{\sigma\delta} \bar{\cals N}_{\kappa_1'm',\kappa_1m}^{\lambda_p',\lambda_y\lambda_1; \bar\mu \sigma,\delta} \bar{\cals M}_{\kappa_2'm',\kappa_2m}^{\lambda_p' \lambda_y; \bar\mu '\delta} \Big\}
\end{equation}
where the symbols $\cals N$ and $\bar{\cals N}$ are defined as Eq. (\ref{eq:app-T-BN}), and the symbols $\cals M$ and $\bar{\cals M}$ read as,
\begin{equation}
    \cals M_{\kappa_1'm',\kappa_1m}^{\lambda_p \lambda_y;\mu\delta } = \sum_{\lambda_d} \cals Q_{\kappa_1'm',\kappa_1m}^{\lambda_d \mu \delta }\Theta_{\lambda_d \lambda_p }^{\lambda_y\mu }, \hspace{4em} \bar{\cals M}_{\kappa_1'm',\kappa_1m}^{\lambda_p \lambda_y;\bar\mu  \delta } \equiv \sum_{\lambda_d} \bar{\cals Q}_{\kappa_1'm',\kappa_1m}^{\lambda_d \bar\mu \delta } \Theta_{\lambda_d \lambda_p }^{\lambda_y\bar\mu }.
\end{equation}
Remind that $\mu + \delta = m-m'$ in the symbol $\cals Q$ and $\bar\mu + \delta = m+m'$ in the symbol $\bar{\cals Q}$.

Similar as the energy functional, the rearrangement term $\Sigma_{R,\text{VT}}^{E,\lambda_d}$ contributed by the $\rho$-VT coupling can be formally expressed as,
\begin{equation}
    \Sigma_{R,\text{VT}}^{E,\lambda_d}(r) = \frac{1}{2\pi r^2}\int dr' \sum_{i} v_i^2 \sum_{\kappa_1\kappa_2} \begin{pmatrix} \cals G_{i\kappa_1}^V & \cals F_{i\kappa_1}^V \end{pmatrix}_r \begin{pmatrix}
    P_{G,\pi m,\lambda_d}^{\kappa_1\kappa_2, \text{TV} } + P_{G,\pi m,\lambda_d}^{\kappa_1\kappa_2, \text{VT} } &
    P_{F,\pi m,\lambda_d}^{\kappa_1\kappa_2, \text{TV} } + P_{F,\pi m,\lambda_d}^{\kappa_1\kappa_2, \text{VT} } \\[0.5em] Q_{G,\pi m,\lambda_d}^{\kappa_1\kappa_2, \text{TV} } + Q_{G,\pi m,\lambda_d}^{\kappa_1\kappa_2, \text{VT} } &
    Q_{F,\pi m,\lambda_d}^{\kappa_1\kappa_2, \text{TV} } + Q_{F,\pi m,\lambda_d}^{\kappa_1\kappa_2, \text{VT} }
    \end{pmatrix}_{r,r'} \begin{pmatrix} \cals G_{i\kappa_2}^V \\[0.5em] \cals F_{i\kappa_2}^V \end{pmatrix}_{r'},
\end{equation}
where the $P$ and $Q$ terms contain the contributions from both time ($t$) and space ($\svec s$) components, e.g.,
\begin{equation}
  P_{G,\pi m,\lambda_d}^{\kappa_1\kappa_2, \text{TV} } =  P_{G,\pi m,\lambda_d}^{\kappa_1\kappa_2, \text{TV-}t } + P_{G,\pi m,\lambda_d}^{\kappa_1\kappa_2, \text{TV-}\svec s }, \hspace{2em} P_{G,\pi m,\lambda_d}^{\kappa_1\kappa_2, \text{VT} } =  P_{G,\pi m,\lambda_d}^{\kappa_1\kappa_2, \text{VT-}t } + P_{G,\pi m,\lambda_d}^{\kappa_1\kappa_2, \text{VT-}\svec s }.
\end{equation}
The details expressions of the $P$ and $Q$ terms for the time component read as follows,
\begin{align}
    P_{G,\pi m,\lambda_d}^{\kappa_1\kappa_2,\text{TV-}t} = & -\frac{1}{2\pi}\frac{m_\rho}{2M} \sum_{\pi'm'} \cals T_{\tau\tau'} \sum_{\kappa_1'\kappa_2'} \cals R_{\kappa_1'\kappa_2',\pi' m'}^{-+} \sum_{\lambda_p\lambda_p'} \frac{\partial f_\rho^{\lambda_p}(r)}{\partial \rho_b^{\lambda_d}(r)} g_\rho^{\lambda_p'}(r') \sum_{\lambda_y}\sum_{\lambda_1}^{\lambda_y\pm1} S_{\lambda_y\lambda_1}(r,r') \widehat{\cals O}_{-\kappa_1'+\kappa_2'm',+\kappa_1+\kappa_2m}^{\lambda_p\lambda_p',\lambda_y\lambda_1}, \\
    P_{F,\pi m,\lambda_d}^{\kappa_1\kappa_2,\text{TV-}t} = & -\frac{1}{2\pi}\frac{m_\rho}{2M} \sum_{\pi'm'} \cals T_{\tau\tau'} \sum_{\kappa_1'\kappa_2'} \cals R_{\kappa_1'\kappa_2',\pi' m'}^{--} \sum_{\lambda_p\lambda_p'} \frac{\partial f_\rho^{\lambda_p}(r)}{\partial \rho_b^{\lambda_d}(r)} g_\rho^{\lambda_p'}(r') \sum_{\lambda_y}\sum_{\lambda_1}^{\lambda_y\pm1} S_{\lambda_y\lambda_1}(r,r') \widehat{\cals O}_{-\kappa_1'-\kappa_2'm',+\kappa_1-\kappa_2m}^{\lambda_p\lambda_p',\lambda_y\lambda_1}, \\
    Q_{G,\pi m,\lambda_d}^{\kappa_1\kappa_2,\text{TV-}t} = & -\frac{1}{2\pi}\frac{m_\rho}{2M} \sum_{\pi'm'} \cals T_{\tau\tau'} \sum_{\kappa_1'\kappa_2'} \cals R_{\kappa_1'\kappa_2',\pi' m'}^{++} \sum_{\lambda_p\lambda_p'} \frac{\partial f_\rho^{\lambda_p}(r)}{\partial \rho_b^{\lambda_d}(r)} g_\rho^{\lambda_p'}(r') \sum_{\lambda_y}\sum_{\lambda_1}^{\lambda_y\pm1} S_{\lambda_y\lambda_1}(r,r') \widehat{\cals O}_{+\kappa_1'+\kappa_2'm',-\kappa_1+\kappa_2m}^{\lambda_p\lambda_p',\lambda_y\lambda_1}, \\
    Q_{F,\pi m,\lambda_d}^{\kappa_1\kappa_2,\text{TV-}t} = & -\frac{1}{2\pi}\frac{m_\rho}{2M} \sum_{\pi'm'} \cals T_{\tau\tau'} \sum_{\kappa_1'\kappa_2'} \cals R_{\kappa_1'\kappa_2',\pi' m'}^{+-} \sum_{\lambda_p\lambda_p'} \frac{\partial f_\rho^{\lambda_p}(r)}{\partial \rho_b^{\lambda_d}(r)} g_\rho^{\lambda_p'}(r') \sum_{\lambda_y}\sum_{\lambda_1}^{\lambda_y\pm1} S_{\lambda_y\lambda_1}(r,r') \widehat{\cals O}_{+\kappa_1'-\kappa_2'm',-\kappa_1-\kappa_2m}^{\lambda_p\lambda_p',\lambda_y\lambda_1},\\
    P_{G,\pi m,\lambda_d}^{\kappa_1\kappa_2,\text{VT-}t} = & -\frac{1}{2\pi}\frac{m_\rho}{2M} \sum_{\pi'm'} \cals T_{\tau\tau'} \sum_{\kappa_1'\kappa_2'} \cals R_{\kappa_1'\kappa_2',\pi' m'}^{+-} \sum_{\lambda_p\lambda_p'} \frac{\partial g_\rho^{\lambda_p}(r)}{\partial \rho_b^{\lambda_d}(r)} f_\rho^{\lambda_p'}(r') \sum_{\lambda_y}\sum_{\lambda_1}^{\lambda_y\pm1} S_{\lambda_y\lambda_1}(r',r) \widehat{\cals O}_{-\kappa_2'+\kappa_1'm',+\kappa_2+\kappa_1m}^{\lambda_p\lambda_p',\lambda_y\lambda_1}, \\
    P_{F,\pi m,\lambda_d}^{\kappa_1\kappa_2,\text{VT-}t} = & -\frac{1}{2\pi}\frac{m_\rho}{2M} \sum_{\pi'm'} \cals T_{\tau\tau'} \sum_{\kappa_1'\kappa_2'} \cals R_{\kappa_1'\kappa_2',\pi' m'}^{++} \sum_{\lambda_p\lambda_p'} \frac{\partial g_\rho^{\lambda_p}(r)}{\partial \rho_b^{\lambda_d}(r)} f_\rho^{\lambda_p'}(r') \sum_{\lambda_y}\sum_{\lambda_1}^{\lambda_y\pm1} S_{\lambda_y\lambda_1}(r',r) \widehat{\cals O}_{+\kappa_2'+\kappa_1'm',-\kappa_2+\kappa_1m}^{\lambda_p\lambda_p',\lambda_y\lambda_1}, \\
    Q_{G,\pi m,\lambda_d}^{\kappa_1\kappa_2,\text{VT-}t} = & -\frac{1}{2\pi}\frac{m_\rho}{2M} \sum_{\pi'm'} \cals T_{\tau\tau'} \sum_{\kappa_1'\kappa_2'} \cals R_{\kappa_1'\kappa_2',\pi' m'}^{--} \sum_{\lambda_p\lambda_p'} \frac{\partial g_\rho^{\lambda_p}(r)}{\partial \rho_b^{\lambda_d}(r)} f_\rho^{\lambda_p'}(r') \sum_{\lambda_y}\sum_{\lambda_1}^{\lambda_y\pm1} S_{\lambda_y\lambda_1}(r',r) \widehat{\cals O}_{-\kappa_2'-\kappa_1'm',+\kappa_2-\kappa_1m}^{\lambda_p\lambda_p',\lambda_y\lambda_1}, \\
    Q_{F,\pi m,\lambda_d}^{\kappa_1\kappa_2,\text{VT-}t} = & -\frac{1}{2\pi}\frac{m_\rho}{2M} \sum_{\pi'm'} \cals T_{\tau\tau'} \sum_{\kappa_1'\kappa_2'} \cals R_{\kappa_1'\kappa_2',\pi' m'}^{-+} \sum_{\lambda_p\lambda_p'} \frac{\partial g_\rho^{\lambda_p}(r)}{\partial \rho_b^{\lambda_d}(r)} f_\rho^{\lambda_p'}(r') \sum_{\lambda_y}\sum_{\lambda_1}^{\lambda_y\pm1} S_{\lambda_y\lambda_1}(r',r) \widehat{\cals O}_{+\kappa_2'-\kappa_1'm',-\kappa_2-\kappa_1m}^{\lambda_p\lambda_p',\lambda_y\lambda_1}.
\end{align}
For the $P$ and $Q$ terms contributed by the space component, one may obtain the TV and VT parts as,
\begin{align}
    P_{G,\pi m}^{\kappa_1\kappa_2,\text{TV-}\svec s} = & +\frac{\sqrt2}{2\pi} \frac{m_\rho}{2M} \sum_{\pi'm'} \cals T_{\tau\tau'}  \sum_{\kappa_1'\kappa_2'} \cals R_{\kappa_1'\kappa_2',\pi'm'}^{+-} \sum_{\lambda_p\lambda_p'} \frac{\partial f_\rho^{\lambda_p}(r)}{\partial \rho_b^{\lambda_d}(r)} g_\rho^{\lambda_p'}(r') \sum_{\lambda_y}\sum_{\lambda_1}^{\lambda_y\pm1} S_{\lambda_y\lambda_1}(r,r') \widehat{\cals Q}_{+\kappa_1'-\kappa_2'm',+\kappa_1+\kappa_2m}^{\lambda_p\lambda_p',\lambda_y\lambda_1},\\
    P_{F,\pi m}^{\kappa_1\kappa_2,\text{TV-}\svec s} = & -\frac{\sqrt2}{2\pi} \frac{m_\rho}{2M} \sum_{\pi'm'} \cals T_{\tau\tau'}  \sum_{\kappa_1'\kappa_2'} \cals R_{\kappa_1'\kappa_2',\pi'm'}^{++} \sum_{\lambda_p\lambda_p'} \frac{\partial f_\rho^{\lambda_p}(r)}{\partial \rho_b^{\lambda_d}(r)} g_\rho^{\lambda_p'}(r') \sum_{\lambda_y}\sum_{\lambda_1}^{\lambda_y\pm1} S_{\lambda_y\lambda_1}(r,r') \widehat{\cals Q}_{+\kappa_1'+\kappa_2'm',+\kappa_1-\kappa_2m}^{\lambda_p\lambda_p',\lambda_y\lambda_1},\\
    Q_{G,\pi m}^{\kappa_1\kappa_2,\text{TV-}\svec s} = & -\frac{\sqrt2}{2\pi} \frac{m_\rho}{2M} \sum_{\pi'm'} \cals T_{\tau\tau'}  \sum_{\kappa_1'\kappa_2'} \cals R_{\kappa_1'\kappa_2',\pi'm'}^{--} \sum_{\lambda_p\lambda_p'} \frac{\partial f_\rho^{\lambda_p}(r)}{\partial \rho_b^{\lambda_d}(r)} g_\rho^{\lambda_p'}(r') \sum_{\lambda_y}\sum_{\lambda_1}^{\lambda_y\pm1} S_{\lambda_y\lambda_1}(r,r') \widehat{\cals Q}_{-\kappa_1'-\kappa_2'm',-\kappa_1+\kappa_2m}^{\lambda_p\lambda_p',\lambda_y\lambda_1},\\
    Q_{F,\pi m}^{\kappa_1\kappa_2,\text{TV-}\svec s} = & +\frac{\sqrt2}{2\pi} \frac{m_\rho}{2M} \sum_{\pi'm'} \cals T_{\tau\tau'}  \sum_{\kappa_1'\kappa_2'} \cals R_{\kappa_1'\kappa_2',\pi'm'}^{-+} \sum_{\lambda_p\lambda_p'} \frac{\partial f_\rho^{\lambda_p}(r)}{\partial \rho_b^{\lambda_d}(r)} g_\rho^{\lambda_p'}(r') \sum_{\lambda_y}\sum_{\lambda_1}^{\lambda_y\pm1} S_{\lambda_y\lambda_1}(r,r') \widehat{\cals Q}_{-\kappa_1'+\kappa_2'm',-\kappa_1-\kappa_2m}^{\lambda_p\lambda_p',\lambda_y\lambda_1},\\
    P_{G,\pi m}^{\kappa_1\kappa_2,\text{VT-}\svec s} = & +\frac{\sqrt2}{2\pi} \frac{m_\rho}{2M} \sum_{\pi'm'} \cals T_{\tau\tau'}  \sum_{\kappa_1'\kappa_2'} \cals R_{\kappa_1'\kappa_2',\pi'm'}^{-+} \sum_{\lambda_p\lambda_p'} \frac{\partial g_\rho^{\lambda_p}(r)}{\partial \rho_b^{\lambda_d}(r)} f_\rho^{\lambda_p'}(r') \sum_{\lambda_y}\sum_{\lambda_1}^{\lambda_y\pm1} S_{\lambda_y\lambda_1}(r',r) \widehat{\cals Q}_{+\kappa_2'-\kappa_1'm',+\kappa_2+\kappa_1m}^{\lambda_p\lambda_p',\lambda_y\lambda_1},\\
    P_{F,\pi m}^{\kappa_1\kappa_2,\text{VT-}\svec s} = & -\frac{\sqrt2}{2\pi} \frac{m_\rho}{2M} \sum_{\pi'm'} \cals T_{\tau\tau'}  \sum_{\kappa_1'\kappa_2'} \cals R_{\kappa_1'\kappa_2',\pi'm'}^{--} \sum_{\lambda_p\lambda_p'} \frac{\partial g_\rho^{\lambda_p}(r)}{\partial \rho_b^{\lambda_d}(r)} f_\rho^{\lambda_p'}(r') \sum_{\lambda_y}\sum_{\lambda_1}^{\lambda_y\pm1} S_{\lambda_y\lambda_1}(r',r) \widehat{\cals Q}_{-\kappa_2'-\kappa_1'm',-\kappa_2+\kappa_1m}^{\lambda_p\lambda_p',\lambda_y\lambda_1},\\
    Q_{G,\pi m}^{\kappa_1\kappa_2,\text{VT-}\svec s} = & -\frac{\sqrt2}{2\pi} \frac{m_\rho}{2M} \sum_{\pi'm'} \cals T_{\tau\tau'}  \sum_{\kappa_1'\kappa_2'} \cals R_{\kappa_1'\kappa_2',\pi'm'}^{++} \sum_{\lambda_p\lambda_p'} \frac{\partial g_\rho^{\lambda_p}(r)}{\partial \rho_b^{\lambda_d}(r)} f_\rho^{\lambda_p'}(r') \sum_{\lambda_y}\sum_{\lambda_1}^{\lambda_y\pm1} S_{\lambda_y\lambda_1}(r',r) \widehat{\cals Q}_{+\kappa_2'+\kappa_1'm',+\kappa_2-\kappa_1m}^{\lambda_p\lambda_p',\lambda_y\lambda_1},\\
    Q_{F,\pi m}^{\kappa_1\kappa_2,\text{VT-}\svec s} = & +\frac{\sqrt2}{2\pi} \frac{m_\rho}{2M} \sum_{\pi'm'} \cals T_{\tau\tau'}  \sum_{\kappa_1'\kappa_2'} \cals R_{\kappa_1'\kappa_2',\pi'm'}^{+-} \sum_{\lambda_p\lambda_p'} \frac{\partial g_\rho^{\lambda_p}(r)}{\partial \rho_b^{\lambda_d}(r)} f_\rho^{\lambda_p'}(r') \sum_{\lambda_y}\sum_{\lambda_1}^{\lambda_y\pm1} S_{\lambda_y\lambda_1}(r',r) \widehat{\cals Q}_{-\kappa_2'+\kappa_1'm',-\kappa_2-\kappa_1m}^{\lambda_p\lambda_p',\lambda_y\lambda_1}.
\end{align}

\subsection{Contact term related to the $\rho$-T coupling}\label{subsec:Contact-rhot}

For the contributions from the $\rho$-T coupling, there exist zero-range terms originated from the gradients of the propagators. Such zero-range contributions are partly removed by adding the so-called contact term as,
\begin{align}
  E_{\rho\text{-T},\delta}^{D} = & + \frac{1}{6}\frac{1}{4M^2} \sum_{ii'} \tau\tau' \int d\svec r d\svec r' \Big[f_\rho \bar\psi_{i}^V \sigma^{\mu\nu}\psi_{i}^V\Big]_{\svec r} \delta(\svec r-\svec r') \Big[f_\rho \bar\psi_{i'}^V\sigma_{\mu\nu}\psi_{i'}^V\Big]_{\svec r'}, \\
  E_{\rho\text{-T},\delta}^{E} = & - \frac{1}{6}\frac{1}{4M^2} \sum_{ii'}\cals T_{\tau\tau'} \int d\svec r d\svec r' \Big[f_\rho \bar\psi_{i}^V \sigma^{\mu\nu}\psi_{i'}^V\Big]_{\svec r} \delta(\svec r-\svec r') \Big[f_\rho \bar\psi_{i'}^V\sigma_{\mu\nu}\psi_{i}^V\Big]_{\svec r'}.
\end{align}
In the following, the detailed expressions of the contact term for the axially deformed nuclei will be list.

For the Hartree term $E_{\rho\text{-T},\delta}^{D}$, only the time component of the contact term present contributions. The local self-energy term can be obtained as,
\begin{equation}\label{eq:app-RTC-H-self}
    \Sigma_{\rho\text{-T},\delta}^{\lambda_d\mu}(r) = -\frac{1}{3}\frac{1}{4M^2} \sum_{\lambda_p} f_\rho^{\lambda_p}(r) \sum_{\lambda_y} \Theta_{\lambda_d\lambda_p}^{\lambda_y\mu} \sum_{\lambda_p' \lambda_d'}  f_\rho^{\lambda_p'}(r)   \Theta_{\lambda_d'\lambda_p'}^{\lambda_y\mu} \rho_{T,3}^{\lambda_d'\mu}(r).
\end{equation}
Thus, the Hartree energy functional of the contact term $E_{\rho\text{-T},\delta}^{D}$ can be expressed as,
\begin{equation}
    E_{\rho\text{-T},\delta}^{D} =  +\frac{2\pi}{2} \int r^2dr \sum_{\lambda_d\mu} \rho_{T,3}^{\lambda_d\mu}(r) \Sigma_{\rho\text{-T},\delta}^{\lambda_d\mu}(r).
\end{equation}
Moreover, the contribution to the rearrangement term from the Hartree term of the contact term reads as,
\begin{equation}
    \Sigma_{R,\rho\text{-T},\delta}^{D,\lambda}(r) =   -\frac{2\pi}{12M^2} \sum_{\lambda_p\lambda_d\mu} \Big[ \frac{\partial f_\rho^{\lambda_p}(r)}{\partial \rho_b^{\lambda}(r)} \rho_{\text{T},3}^{\lambda_d\mu}(r)\Big] \sum_{\lambda_y} \Theta_{\lambda_d\lambda_p}^{\lambda_y\mu} \sum_{\lambda_p' \lambda_d'}  f_\rho^{\lambda_p'}(r)   \Theta_{\lambda_d'\lambda_p'}^{\lambda_y\mu} \rho_{T,3}^{\lambda_d'\mu}(r).
\end{equation}

For the Fock contributions of the contact term, the expressions are much more complicated, which contains the time and space components. The non-local energy functional from the time/space components of contact term can be expressed as,
\begin{equation}
    E_{\rho\text{-T},\delta}^{E} = \frac{1}{2} \int dr \sum_{i}v_i^{2} \sum_{\kappa_1\kappa_2} \begin{pmatrix} \cals G_{i\kappa_1}^V & \cals F_{i\kappa_1}^V \end{pmatrix}_r \begin{pmatrix} Y_{G,\pi m}^{\kappa_1\kappa_2,\delta } & Y_{F,\pi m}^{\kappa_1\kappa_2,\delta }\\[0.5em] X_{G,\pi m}^{\kappa_1\kappa_2,\delta } & X_{F,\pi m}^{\kappa_1\kappa_2,\delta } \end{pmatrix}_{r} \begin{pmatrix} \cals G_{i\kappa_2}^V \\[0.5em] \cals F_{i\kappa_2}^V\end{pmatrix}_r,
\end{equation}
in which $v_i^2$ denotes the degeneracy of orbit $i$, and the self-energies $Y_G$, $Y_F$, $X_G$ and $X_F$ contain the contributions from the time ($\delta$-$t$) and space ($\delta$-$\svec s$) components, e.g.,
\begin{equation}
  Y_{G,\pi m}^{\kappa_1\kappa_2,\delta } =  Y_{G,\pi m}^{\kappa_1\kappa_2,\delta\text{-}t } +  Y_{G,\pi m}^{\kappa_1\kappa_2,\delta\text{-}\svec s }.
\end{equation}
For the time component, the self-energies read as
\begin{align}
    Y_{G,\pi m}^{\kappa_1\kappa_2,\delta \text{-}t} \equiv & +\frac{1}{6\pi r^2}\frac{1}{4 M^2} \sum_{\pi'm'} \cals T_{\tau\tau'} \sum_{\kappa_1'\kappa_2'} \cals R_{\kappa_1'\kappa_2',\pi'm'}^{--} \sum_{\lambda_p\lambda_p'} f_\rho^{\lambda_p}(r) f_\rho^{\lambda_p'}(r)  \widehat{\cals B}_{-\kappa_1'-\kappa_2'm';+\kappa_1+\kappa_2m}^{\lambda_p\lambda_p'},   \\
    Y_{F,\pi m}^{\kappa_1\kappa_2,\delta \text{-}t} \equiv & +\frac{1}{6\pi r^2}\frac{1}{4 M^2} \sum_{\pi'm'} \cals T_{\tau\tau'} \sum_{\kappa_1'\kappa_2'} \cals R_{\kappa_1'\kappa_2',\pi'm'}^{-+} \sum_{\lambda_p\lambda_p'} f_\rho^{\lambda_p}(r) f_\rho^{\lambda_p'}(r)  \widehat{\cals B}_{-\kappa_1'+\kappa_2'm';+\kappa_1-\kappa_2m}^{\lambda_p\lambda_p'}, \\
    X_{G,\pi m}^{\kappa_1\kappa_2,\delta \text{-}t} \equiv & +\frac{1}{6\pi r^2}\frac{1}{4 M^2} \sum_{\pi'm'} \cals T_{\tau\tau'} \sum_{\kappa_1'\kappa_2'} \cals R_{\kappa_1'\kappa_2',\pi'm'}^{+-} \sum_{\lambda_p\lambda_p'} f_\rho^{\lambda_p}(r) f_\rho^{\lambda_p'}(r)  \widehat{\cals B}_{+\kappa_1'-\kappa_2'm';-\kappa_1+\kappa_2m}^{\lambda_p\lambda_p'},\\
    X_{F,\pi m}^{\kappa_1\kappa_2,\delta \text{-}t} \equiv & +\frac{1}{6\pi r^2}\frac{1}{4 M^2} \sum_{\pi'm'} \cals T_{\tau\tau'} \sum_{\kappa_1'\kappa_2'} \cals R_{\kappa_1'\kappa_2',\pi'm'}^{++} \sum_{\lambda_p\lambda_p'} f_\rho^{\lambda_p}(r) f_\rho^{\lambda_p'}(r)  \widehat{\cals B}_{+\kappa_1'+\kappa_2'm';-\kappa_1-\kappa_2m}^{\lambda_p\lambda_p'}.
\end{align}
In the above expressions, the symbols $\widehat{\cals B}$ have the following form,
\begin{equation}
  \widehat{\cals B}_{\kappa_1'\kappa_2'm';\kappa_1\kappa_2 m}^{\lambda_p\lambda_p'} \equiv  \ff2 \sum_{\lambda_d\lambda_d' \lambda_y\sigma} \Big[ \cals Q_{\kappa_1'm', \kappa_1m}^{\lambda_d \mu\sigma}\cals Q_{\kappa_2'm',\kappa_2m}^{\lambda_d'\mu\sigma} \Theta_{\lambda_d\lambda_p}^{\lambda_y\mu} \Theta_{\lambda_d'\lambda_p'}^{\lambda_y\mu} + \bar{\cals Q}_{\kappa_1'm', \kappa_1m}^{\lambda_d\bar \mu\sigma}\bar{\cals Q}_{\kappa_2'm',\kappa_2m}^{\lambda_d' \bar \mu \sigma} \Theta_{\lambda_d\lambda_p}^{\lambda_y \bar \mu} \Theta_{\lambda_d'\lambda_p'}^{\lambda_y\bar \mu}\Big].
\end{equation}
For the space components, the self-energies can be obtained as,
\begin{align}
    Y_{G,\pi m}^{\kappa_1\kappa_2,\delta \text{-}\svec s} \equiv & -\frac{1}{3\pi r^2}\frac{1}{4 M^2} \sum_{\pi'm'} \cals T_{\tau\tau'}  \sum_{\kappa_1'\kappa_2'} \cals R_{\kappa_1'\kappa_2',\pi'm'}^{++} \sum_{\lambda_p\lambda_p'} f_\rho^{\lambda_p}(r) f_\rho^{\lambda_p'}(r)  \widehat{\scr B}_{+\kappa_1'+\kappa_2'm';+\kappa_1+\kappa_2 m}^{\lambda_p\lambda_p'},   \\
    Y_{F,\pi m}^{\kappa_1\kappa_2,\delta \text{-}\svec s} \equiv & +\frac{1}{3\pi r^2}\frac{1}{4 M^2} \sum_{\pi'm'} \cals T_{\tau\tau'}  \sum_{\kappa_1'\kappa_2'} \cals R_{\kappa_1'\kappa_2',\pi'm'}^{+-} \sum_{\lambda_p\lambda_p'} f_\rho^{\lambda_p}(r) f_\rho^{\lambda_p'}(r)  \widehat{\scr B}_{+\kappa_1'-\kappa_2'm';+\kappa_1-\kappa_2 m}^{\lambda_p\lambda_p'}, \\
    X_{G,\pi m}^{\kappa_1\kappa_2,\delta \text{-}\svec s} \equiv & +\frac{1}{3\pi r^2}\frac{1}{4 M^2} \sum_{\pi'm'} \cals T_{\tau\tau'}  \sum_{\kappa_1'\kappa_2'} \cals R_{\kappa_1'\kappa_2',\pi'm'}^{-+} \sum_{\lambda_p\lambda_p'} f_\rho^{\lambda_p}(r) f_\rho^{\lambda_p'}(r)  \widehat{\scr B}_{-\kappa_1'+\kappa_2'm';-\kappa_1+\kappa_2 m}^{\lambda_p\lambda_p'},\\
    X_{F,\pi m}^{\kappa_1\kappa_2,\delta \text{-}\svec s} \equiv & -\frac{1}{3\pi r^2}\frac{1}{4 M^2} \sum_{\pi'm'} \cals T_{\tau\tau'}  \sum_{\kappa_1'\kappa_2'} \cals R_{\kappa_1'\kappa_2',\pi'm'}^{--} \sum_{\lambda_p\lambda_p'} f_\rho^{\lambda_p}(r) f_\rho^{\lambda_p'}(r)  \widehat{\scr B}_{-\kappa_1'-\kappa_2'm';-\kappa_1-\kappa_2 m}^{\lambda_p\lambda_p'},
\end{align}
where the symbols $\widehat{\scr B}$ read as,
\begin{equation}
  \widehat{\scr B}_{\kappa_1'\kappa_2'm';\kappa_1\kappa_2 m}^{\lambda_p\lambda_p'} \equiv  \ff2 \sum_{\lambda_d\lambda_d' \lambda_y\sigma} \Big[ \cals Q_{\kappa_1'm', \kappa_1m}^{\lambda_d \mu\sigma}\cals Q_{\kappa_2'm',\kappa_2m}^{\lambda_d'\mu\sigma} \Theta_{\lambda_d\lambda_p}^{ \lambda_y\mu} \Theta_{\lambda_d'\lambda_p'}^{ \lambda_y\mu} + \bar{\cals Q}_{\kappa_1'm', \kappa_1m}^{\lambda_d \bar\mu\sigma}\bar{\cals Q}_{\kappa_2'm',\kappa_2m}^{\lambda_d' \bar\mu \sigma} \Theta_{\lambda_d\lambda_p}^{ \lambda _y\bar\mu} \Theta_{\lambda_d'\lambda_p'}^{ \lambda _y\bar\mu}\Big] .
\end{equation}

Concerning the Fock contribution to the rearrangement term, one can also formally express it as,
\begin{equation}
    \Sigma_{R,\rho\text{-T},\delta}^{E,\lambda_d} = \sum_{i} v_i^2 \sum_{\kappa_1\kappa_2} \begin{pmatrix} \cals G_{i\kappa_1}^V & \cals F_{i\kappa_1}^V \end{pmatrix}_r \begin{pmatrix}  P_{G,\pi m,\lambda_d}^{\kappa_1\kappa_2, \delta} & P_{F,\pi m,\lambda_d}^{\kappa_1\kappa_2, \delta} \\[0.5em] Q_{G,\pi m,\lambda_d}^{\kappa_1\kappa_2, \delta } & Q_{F,\pi m,\lambda_d}^{\kappa_1\kappa_2, \delta }\end{pmatrix}_{r} \begin{pmatrix} \cals G_{i\kappa_2}^V \\[0.5em] \cals F_{i\kappa_2}^V\end{pmatrix}_r,
\end{equation}
in which the $P$ and $Q$ terms contain the contributions of the time ($\delta$-$t$) and space ($\delta$-$\svec s$) components, e.g.,
\begin{equation}
  P_{G,\pi m,\lambda_d}^{\kappa_1\kappa_2, \delta} =  P_{G,\pi m,\lambda_d}^{\kappa_1\kappa_2, \delta\text{-}t} + P_{G,\pi m,\lambda_d}^{\kappa_1\kappa_2, \delta\text{-}\svec s}.
\end{equation}
For the time and space components, the $P$ and $Q$ terms read as:
\begin{align}
    P_{G,\pi m,\lambda_d}^{\kappa_1\kappa_2,\delta \text{-}t} \equiv & +\frac{1}{6\pi r^2}\frac{1}{4 M^2} \sum_{\pi'm'} \cals T_{\tau\tau'} \sum_{\kappa_1'\kappa_2'}\cals R_{\kappa_1'\kappa_2',\pi'm'}^{--} \sum_{\lambda_p\lambda_p'} \frac{\partial f_\rho^{\lambda_p}(r)}{\partial \rho_b^{\lambda_d}(r)} f_\rho^{\lambda_p'}(r)  \widehat{\cals B}_{-\kappa_1'-\kappa_2'm';+\kappa_1+\kappa_2m}^{\lambda_p\lambda_p'}, \\
    P_{F,\pi m,\lambda_d}^{\kappa_1\kappa_2,\delta \text{-}t} \equiv & +\frac{1}{6\pi r^2}\frac{1}{4 M^2} \sum_{\pi'm'} \cals T_{\tau\tau'} \sum_{\kappa_1'\kappa_2'}\cals R_{\kappa_1'\kappa_2',\pi'm'}^{-+} \sum_{\lambda_p\lambda_p'} \frac{\partial f_\rho^{\lambda_p}(r)}{\partial \rho_b^{\lambda_d}(r)} f_\rho^{\lambda_p'}(r)  \widehat{\cals B}_{-\kappa_1'+\kappa_2'm';+\kappa_1-\kappa_2m}^{\lambda_p\lambda_p'}, \\
    Q_{G,\pi m,\lambda_d}^{\kappa_1\kappa_2,\delta \text{-}t} \equiv & +\frac{1}{6\pi r^2}\frac{1}{4 M^2} \sum_{\pi'm'} \cals T_{\tau\tau'} \sum_{\kappa_1'\kappa_2'}\cals R_{\kappa_1'\kappa_2',\pi'm'}^{+-} \sum_{\lambda_p\lambda_p'} \frac{\partial f_\rho^{\lambda_p}(r)}{\partial \rho_b^{\lambda_d}(r)} f_\rho^{\lambda_p'}(r)  \widehat{\cals B}_{+\kappa_1'-\kappa_2'm';-\kappa_1+\kappa_2m}^{\lambda_p\lambda_p'},\\
    Q_{F,\pi m,\lambda_d}^{\kappa_1\kappa_2,\delta \text{-}t} \equiv & +\frac{1}{6\pi r^2}\frac{1}{4 M^2} \sum_{\pi'm'} \cals T_{\tau\tau'} \sum_{\kappa_1'\kappa_2'}\cals R_{\kappa_1'\kappa_2',\pi'm'}^{++} \sum_{\lambda_p\lambda_p'} \frac{\partial f_\rho^{\lambda_p}(r)}{\partial \rho_b^{\lambda_d}(r)} f_\rho^{\lambda_p'}(r)  \widehat{\cals B}_{+\kappa_1'+\kappa_2'm';-\kappa_1-\kappa_2m}^{\lambda_p\lambda_p'},\\
    P_{G,\pi m,\lambda_d}^{\kappa_1\kappa_2,\delta \text{-}\svec s} \equiv & -\frac{1}{3\pi r^2}\frac{1}{4 M^2} \sum_{\pi'm'} \cals T_{\tau\tau'}  \sum_{\kappa_1'\kappa_2'}\cals R_{\kappa_1'\kappa_2',\pi'm'}^{++} \sum_{\lambda_p\lambda_p'} \frac{\partial f_\rho^{\lambda_p}(r)}{\partial \rho_b^{\lambda_d}(r)} f_\rho^{\lambda_p'}(r)  \widehat{\scr B}_{+\kappa_1'+\kappa_2'm';+\kappa_1+\kappa_2m}^{\lambda_p\lambda_p'}, \\
    P_{F,\pi m,\lambda_d}^{\kappa_1\kappa_2,\delta \text{-}\svec s} \equiv & +\frac{1}{3\pi r^2}\frac{1}{4 M^2} \sum_{\pi'm'} \cals T_{\tau\tau'}  \sum_{\kappa_1'\kappa_2'}\cals R_{\kappa_1'\kappa_2',\pi'm'}^{+-} \sum_{\lambda_p\lambda_p'} \frac{\partial f_\rho^{\lambda_p}(r)}{\partial \rho_b^{\lambda_d}(r)} f_\rho^{\lambda_p'}(r)  \widehat{\scr B}_{+\kappa_1'-\kappa_2'm';+\kappa_1-\kappa_2m}^{\lambda_p\lambda_p'}, \\
    Q_{G,\pi m,\lambda_d}^{\kappa_1\kappa_2,\delta \text{-}\svec s} \equiv & +\frac{1}{3\pi r^2}\frac{1}{4 M^2} \sum_{\pi'm'} \cals T_{\tau\tau'}  \sum_{\kappa_1'\kappa_2'}\cals R_{\kappa_1'\kappa_2',\pi'm'}^{-+} \sum_{\lambda_p\lambda_p'} \frac{\partial f_\rho^{\lambda_p}(r)}{\partial \rho_b^{\lambda_d}(r)} f_\rho^{\lambda_p'}(r)  \widehat{\scr B}_{-\kappa_1'+\kappa_2'm';-\kappa_1+\kappa_2m}^{\lambda_p\lambda_p'},\\
    Q_{F,\pi m,\lambda_d}^{\kappa_1\kappa_2,\delta \text{-}\svec s} \equiv & -\frac{1}{3\pi r^2}\frac{1}{4 M^2} \sum_{\pi'm'} \cals T_{\tau\tau'}  \sum_{\kappa_1'\kappa_2'}\cals R_{\kappa_1'\kappa_2',\pi'm'}^{--} \sum_{\lambda_p\lambda_p'} \frac{\partial f_\rho^{\lambda_p}(r)}{\partial \rho_b^{\lambda_d}(r)} f_\rho^{\lambda_p'}(r)  \widehat{\scr B}_{-\kappa_1'-\kappa_2'm';-\kappa_1-\kappa_2m}^{\lambda_p\lambda_p'}.
\end{align}
It shall be noticed that for the results of contact terms, the non-local density $\cals R$ corresponds to $\cals R(r,r)$ because of zero range.

\section{Pairing potential with Gogny force}\label{sec:APP-B}
In this work, the finite-range Gogny force D1S is adopted as the pairing force, and the full contributions from all the $J$-components are considered in stead of only $J=0$ contributions adopted in Ref. \cite{Geng2020PRC101.064302}. Using the spherical DWS base, the pairing energy for axially deformed nuclei can be expressed as the following form,
\begin{equation}
    E_p =  \ff2 \sum_{m\kappa_1\kappa_2} \int dr dr'\sum_{\sigma\sigma'=\pm} K_{m\kappa_1\kappa_2}^{\sigma\sigma'}(r,r') \Delta_{m\kappa_1\kappa_2}^{\sigma\sigma'}(r,r'),
\end{equation}
in which $\sigma,\sigma'=\pm$ ($+$ and $-$ correspond to the $G$ and $F$ components respectively), and $K^{\sigma\sigma'}$ represent the components of the pairing tensor as,
\begin{align}
    K_{m\kappa\tilde\kappa}^{++}(r,r') =&  \sum_{\nu} v_i^2 \cals G_{i\kappa}^V(r)\cals G_{i\tilde\kappa}^U(r'), &
    K_{m\kappa\tilde\kappa}^{+-}(r,r') =&  \sum_{\nu} v_i^2 \cals G_{i\kappa}^V(r)\cals F_{i\tilde\kappa}^U(r'), \\
    K_{m\kappa\tilde\kappa}^{-+}(r,r') =&  \sum_{\nu} v_i^2 \cals F_{i\kappa}^V(r)\cals G_{i\tilde\kappa}^U(r'), &
    K_{m\kappa\tilde\kappa}^{--}(r,r') =&  \sum_{\nu} v_i^2 \cals F_{i\kappa}^V(r)\cals F_{i\tilde\kappa}^U(r').
\end{align}
Here $v_i^2$ denotes the degeneracy of the orbit $i = (\nu\pi m)$. Using the pairing tensor components $K^{\sigma\sigma'}$, the pairing potential terms $\Delta^{\sigma\sigma'}$ can be expressed uniformly as,
\begin{equation}
    \Delta_{m\kappa_1\kappa_2}^{\sigma\sigma'}(r,r') \equiv \sum_{m'\kappa_1'\kappa_2'} \Pi_{m\kappa_1\kappa_2,m'\kappa_1'\kappa_2'}^{\sigma\sigma'}(r,r') K_{m'\kappa_1'\kappa_2'}^{\sigma\sigma'}(r,r'),
\end{equation}
in which the symbols $\Pi$ correspond to the combination of the C-G coefficients and the radial part of the Gogny force,
\begin{align}
    \Pi_{m\kappa\tilde\kappa; m' \kappa'\tilde\kappa'}^{++}(r,r') = & \ff2\sum_\lambda \lrs{A_\lambda(r,r') \widehat{\cals X}_{m\kappa\tilde\kappa; m' \kappa'\tilde\kappa'}^\lambda - D_\lambda(r,r') \widehat{\cals S}_{m\kappa\tilde\kappa; m' \kappa'\tilde\kappa'}^\lambda }, \\
    \Pi_{m\kappa\tilde\kappa; m' \kappa'\tilde\kappa'}^{+-}(r,r') = & \ff2\sum_\lambda \lrs{A_\lambda(r,r') \widehat{\cals X}_{m\kappa-\tilde\kappa; m' \kappa'-\tilde\kappa'}^\lambda - D_\lambda(r,r') \widehat{\cals S}_{m\kappa-\tilde\kappa; m' \kappa'-\tilde\kappa'}^\lambda }, \\
    \Pi_{m\kappa\tilde\kappa; m' \kappa'\tilde\kappa'}^{-+}(r,r') = & \ff2\sum_\lambda \lrs{A_\lambda(r,r') \widehat{\cals X}_{m-\kappa\tilde\kappa; m' -\kappa'\tilde\kappa'}^\lambda - D_\lambda(r,r') \widehat{\cals S}_{m-\kappa\tilde\kappa; m' -\kappa'\tilde\kappa'}^\lambda }, \\
    \Pi_{m\kappa\tilde\kappa; m' \kappa'\tilde\kappa'}^{--}(r,r') = & \ff2\sum_\lambda \lrs{A_\lambda(r,r') \widehat{\cals X}_{m-\kappa-\tilde\kappa; m'- \kappa'-\tilde\kappa'}^\lambda - D_\lambda(r,r') \widehat{\cals S}_{m-\kappa-\tilde\kappa; m' -\kappa'-\tilde\kappa'}^\lambda }.
\end{align}
In the above expressions, the symbols $\widehat{\cals X}$ and $\widehat{\cals S}$ read as,
\begin{align}
  \widehat{\cals X}_{m\kappa\tilde\kappa; m' \kappa'\tilde\kappa'}^{\lambda} \equiv&  \sum_{JLS} \cals S_{ m\kappa\tilde\kappa}^{JLS} \cals X_{\kappa\tilde\kappa;\kappa'\tilde\kappa'}^{L,\lambda} \cals S_{ m'\kappa'\tilde\kappa'}^{JLS}, &
  \widehat{\cals S}_{m\kappa\tilde\kappa; m'\kappa'\tilde\kappa'}^{\lambda } \equiv & \sum_{JLS} (-1)^{S} \cals S_{m\kappa\tilde\kappa }^{JLS} \cals X_{\kappa\tilde\kappa;\kappa'\tilde\kappa'}^{L,\lambda} \cals S_{m'\kappa'\tilde\kappa'}^{JLS},
\end{align}
with the symbols $\cals X$ and $\cals S$ as,
\begin{align}
    \cals X_{\kappa \tilde \kappa, \kappa'\tilde \kappa'}^{L,\lambda} = & (-1)^LC_{\tilde l_u' 0 \tilde l_u 0} ^{\lambda0} C_{l_u'0 l_u0}^{\lambda0}\Lrb{\begin{matrix} l_u' & l_u & \lambda\\ \tilde l_u & \tilde l_u' & L  \end{matrix}}, & \cals S_{m\kappa \tilde \kappa}^{JLS} = &  (-1)^{\tilde j-m} \hat j\hat{\tilde j}\hat l_u\hat{\tilde l}_u\hat L\hat S \begin{Bmatrix} j &\tilde j & J \\ l_u & \tilde l_u & L \\ \ff2 & \ff2 & S \end{Bmatrix} C_{\tilde j-m jm }^{J0}.
\end{align}
In the $6j$ and $9j$ symbols, $l_u$ corresponds to the orbital angular momentum of upper component of the spherical Dirac spinor (\ref{eq:nkappam}), and the one of the lower component reads as $l_d$. That is to say, for the terms with $-\kappa$, $l_u$ shall be replaced by $l_d$. In the symbols $\Pi$, the terms $A_\lambda(r,r')$ and $ D_{\lambda}(r,r')$ are defined as,
\begin{align}
    A_\lambda(r,r') \equiv & \sum_{i=1}^2 V_\lambda^i(r,r') A_i, & D_\lambda(r,r')\equiv & \sum_{i=1}^2 V_{\lambda}^i(r,r') D_i,
\end{align}
where $A_i = W_i - H_i P^\tau$, $D_i = B_i - M_i P^\tau$, and the radial expansion terms of the Gogny force $V_\lambda^i$ read as,
\begin{equation}
  V_\lambda^i(r,r') =  e^{-(r^2+r'^2)/\mu_i^2} \sqrt{2\pi\frac{\mu_i^2}{2rr'}} I_{\lambda+1/2}(\frac{2rr'}{\mu_i^2}),
\end{equation}
with the parameters of Gogny force $W_i$, $H_i$, $B_i$, $M_i$ and $\mu_i$, see Eq. (\ref{eq:Gogny}).

\section{Cent-of-mass correction under Bogoliubov scheme}\label{sec:APP-C}

As we did before, the cent-of-mass (CoM) correction on the binding energy of nuclei are evaluated in a microscopic way by considering the expectation with respect to the Bogoliubov ground state $\lrlc{\text{HFB}}$. Applying the quantization of the Dirac spinor field (\ref{eq:expansionHFB}), the expectation of $\svec P_{\text{c.m.}}^2$ can be derived as,
\begin{equation}
  \lrcl{\text{HFB}}\svec P_{\text{c.m.}}^2 \lrlc{\text{HFB}} =  \sum_{i}\svec p_{ii}^2 - \sum_{ii'} \svec p_{ii'}^{VV}\cdot\svec p_{i'i}^{VV} + \sum_{ii'} \svec p_{ ii'}^{VV}\cdot\svec p_{\bar i\bar i'}^{UU} .
\end{equation}

The second derivative term $\svec p_{ii}^2$ can be expressed as,
\begin{equation}
  \svec p_{ii}^2 =  -\sum_{i\kappa}v_i^2 \int dr \Big\{\cals G_{i\kappa}^V\Big[\frac{d^2}{dr^2} - \frac{l_u(l_u+1)}{r^2}\Big]\cals G_{i\kappa}^V + \cals F_{i\kappa}^V\Big[\frac{d^2}{dr^2} - \frac{l_d(l_d+1)}{r^2}\Big]\cals G_{i\kappa}^V\Big\},
\end{equation}
where $v_i^2$ denotes the degeneracy of the orbit $i$. For the first derivative term $\svec p_{ii'}^{VV}\cdot\svec p_{i'i}^{VV}$, one should be careful in dealing with the couplings between the orbits $i = (\nu\pi m)$ and $i' = (\nu'\pi'm')$, which can have opposite sign on the $m$-values. Here we use $\bar i$/$\bar i'$ to denote the orbits with negative $m$-values, accordingly the time-reversal partners. Thus, the term $\svec p_{ii'}^{VV}\cdot\svec p_{i'i}^{VV}$ can be derived as,
\begin{equation}
  \frac{v_i^2v_{i'}^2}{2}\Big[\svec p_{ii'}^{VV}\cdot\svec p_{i'i}^{VV} + \svec p_{i\bar i'}^{VV}\cdot\svec p_{\bar i'i}^{VV}\Big],
\end{equation}
where the detailed expressions of both terms read as,
\begin{align}
  \svec p_{ii'}^{VV}\cdot\svec p_{i'i}^{VV} = & - \sum_{\kappa_1\kappa_1'}\frac{\hat j_1\hat j'_1}{3}C_{j_1\ff2 j_1'-\ff2}^{10} C_{j_1'-m' j_1 m}^{1\mu }A_{i\kappa_1;i'\kappa_1'}^V \times \sum_{\kappa_2\kappa_2'}\frac{\hat j_2 \hat j_2'}{3}C_{j_2\ff2 j_2'-\ff2}^{10}C_{j'_2 -m' j_2 m}^{1\mu} A_{i'\kappa_2';i\kappa_2}^V \\
  \svec p_{i\bar i'}^{VV}\cdot\svec p_{\bar i'i}^{VV} = & - \sum_{\kappa_1\kappa_1'}\frac{\hat j_1\hat j'_1}{3}C_{j_1\ff2 j_1'-\ff2}^{10} C_{j_1'm' j_1 m}^{1\mu }A_{i\kappa_1;i'\kappa_1'}^V \times \sum_{\kappa_2\kappa_2'} \frac{\hat j_2 \hat j_2'}{3}C_{j_2\ff2 j_2'-\ff2}^{10}C_{j'_2 m' j_2m}^{1\mu} A_{i'\kappa_2';i\kappa_2}^V(-1)^{j_1'+j_2'+1},
\end{align}
with
\begin{equation}
  A_{i\kappa;i'\kappa'}^V =  \int r^2 dr \Big\{\frac{\cals G_{i\kappa}^V}{r}\Big[\frac{d}{dr} + \frac{\cals A_{l_u'l_u}}{r}\Big]\frac{\cals G_{i'\kappa'}^V}{r} + \frac{\cals F_{i\kappa}^V}{r}\Big[\frac{d}{dr} + \frac{\cals A_{l_d'l_d}}{r}\Big] \frac{\cals F_{i'\kappa'}^V}{r}\Big\}, \hspace{2em} \cals A_{l'l} =  \Big\{ \begin{array}{ll} -l' & l=l'+1\\ l'+1 & l = l'-1  \end{array}.
\end{equation}

Similar as the term $\svec p_{ii'}^{VV}\cdot\svec p_{i'i}^{VV}$, the contribution of the term $\svec p_{ ii'}^{VV}\cdot\svec p_{\bar i\bar i'}^{UU}$ also contains two parts as,
\begin{equation}
  \frac{v_i^2v_{i'}^2}{2}\Big[\svec p_{ii'}^{VV}\cdot\svec p_{\bar i\bar i'}^{UU} + \svec p_{i\bar i'}^{VV}\cdot\svec p_{\bar ii'}^{UU}\Big],
\end{equation}
where the detailed expressions of both terms read as,
\begin{align}
  \svec p_{ii'}^{VV}\cdot\svec p_{\bar i \bar i'}^{UU} = & - \sum_{\kappa_1\kappa_1'} \frac{\hat j_1\hat j_1'}{3} C_{j_1\ff2 j_1'-\ff2}^{10} C_{j_1'-m'j_1m}^{1\mu} A_{i\kappa_1;i'\kappa_1'}^V\times \sum_{\kappa_2\kappa_2'}\frac{\hat j_2\hat j_2'}{3} C_{j_2\ff2 j_2'-\ff2}^{10} C_{j_2'-m'j_2m}^{1\mu} A_{i\kappa_2;i'\kappa_2'}^U, \\
  \svec p_{i \bar i'}^{VV}\cdot\svec p_{\bar i i'}^{UU} = & - \sum_{\kappa_1\kappa_1'} \frac{\hat j_1\hat j_1'}{3} C_{j_1\ff2 j_1'-\ff2}^{10} C_{j_1'm'j_1m}^{1\mu} A_{i\kappa_1;i'\kappa_1'}^V\times \sum_{\kappa_2\kappa_2'}\frac{\hat j_2\hat j_2'}{3} C_{j_2\ff2 j_2'-\ff2}^{10} C_{j_2'm'j_2m}^{1\mu} A_{i\kappa_2;i'\kappa_2'}^U(-1)^{j_1'+j_2'+1}.
\end{align}
with
\begin{equation}
  A_{i\kappa;i'\kappa'}^U =  \int r^2 dr \Big\{\frac{\cals G_{i\kappa}^U}{r}\Big[\frac{d}{dr} + \frac{\cals A_{l_u'l_u}}{r}\Big]\frac{\cals G_{i'\kappa'}^U}{r} + \frac{\cals F_{i\kappa}^U}{r}\Big[\frac{d}{dr} + \frac{\cals A_{l_d'l_d}}{r}\Big] \frac{\cals F_{i'\kappa'}^U}{r}\Big\}.
\end{equation}
In all above expressions, $l_u$ and $l_d$ correspond to the orbital angular momenta of the upper and lower component of the spherical Dirac spinor (\ref{eq:nkappam}), respectively.

\end{widetext}


\end{document}